\documentclass{aastex701}
\usepackage{graphicx}
\usepackage{amsmath}
\usepackage{amssymb}
\usepackage{subcaption} 
\makeatletter
\IfFileExists{latexml.sty}{\usepackage{latexml}}{} 
\@ifundefined{iflatexml}{\newif\iflatexml\latexmlfalse}{} 
\makeatother

\newcommand{\EmailArxiv}[1]{%
  \iflatexml
    \email{#1}
  \else
    \email[show]{#1}
  \fi
}

\begin{document}

\title{Blowouts of Nascent Wind Bubbles in Pulsar-Driven Supernovae}

\author[orcid=0009-0004-9403-6019,gname=Mingxi, sname='Chen']{Mingxi Chen}
\affiliation{Astronomical Institute, Tohoku University, Sendai, Miyagi 980-8578, Japan}
\EmailArxiv{m.chen@astr.tohoku.ac.jp}

\author[orcid=0000-0003-4299-8799,gname=Kazumi, sname='Kashiyama']{Kazumi Kashiyama} 
\affiliation{Astronomical Institute, Tohoku University, Sendai, Miyagi 980-8578, Japan}
\email{kashiyama@astr.tohoku.ac.jp}

\author[orcid=0009-0008-0189-863X,gname=Masato, sname='Sato']{Masato Sato}
\affiliation{Department of Earth Science and Astronomy, The University of Tokyo, Tokyo 153-8902, Japan}
\email{satoms@g.ecc.u-tokyo.ac.jp}  

\begin{abstract}
Formation of a rapidly spinning, strongly magnetized neutron star (NS) may occur in various classes of core-collapse events. If the NS injects an amount of energy comparable to the explosion energy of the accompanying supernova (SN) before the SN ejecta becomes transparent, the nascent NS wind bubble can overtake the outer ejecta and undergo a blowout driven by hydrodynamic instabilities. Based on multidimensional numerical studies, we construct a minimal semi-analytic framework to follow the post-blowout dynamics and radiative evolution, map the blowout conditions by scanning the ejecta and NS parameters, and compute survey-ready multi-band light curves. For stripped-envelope SNe with an ejecta mass of $M_\mathrm{ej} \sim 10\,M_\odot$ and an explosion energy of $E_\mathrm{sn} \sim 10^{51}\,\mathrm{erg}$, blowout occurs for NSs with magnetic field strengths of $B_{\mathrm{dip}} \gtrsim 10^{13}\,\mathrm{G}$ and spin periods of $P_\mathrm{NS} \lesssim \mathrm{a\ few}\,\mathrm{ms}$. Relatively weak-field cases with $B_\mathrm{dip} \lesssim 10^{14}\,\mathrm{G}$ produce luminous double-peaked UV/optical light curves, as observed in the superluminous SN LSQ14bdq, while stronger-field cases with $B_\mathrm{dip} \gtrsim 10^{14}\,\mathrm{G}$ result in hypernovae preceded by X-ray blowout precursors. We also examine weaker and lower-mass SN explosions representing ultra-stripped SNe and accretion- or merger-induced collapse events, in which blowout is more readily achieved over a broader range of NS parameters, producing fast X-ray transients with durations of $ 10^{2\mbox{--}4}\,\mathrm{s}$ and peak luminosities of $10^{42\mbox{--}48}\,\mathrm{erg\,s^{-1}}$. Our results encourage coordinated UV, optical, and X-ray observations which constrain the formation of the most energetic NSs in the universe.

\end{abstract}


\section{Introduction}

Over the past decade, wide-field, high-cadence time-domain surveys such as the Asteroid Terrestrial-impact Last Alert System (ATLAS) and the Zwicky Transient Facility (ZTF) have enabled early discovery and monitoring of supernovae (SNe), revealing a diversity of early-time photometric behavior that was previously difficult to capture systematically \citep{Tonry2018,Bellm2019,Graham2019}. Within the stripped-envelope supernovae (SESNe) family, broad-lined Type Ic supernovae (SNe Ic-BL) occupy the high-velocity end \citep[e.g.,][]{Galama1998, Mazzali2002, Taddia2019}, while Type I superluminous supernovae (SLSNe-I) occupy the high-luminosity, fast-rising end \citep[e.g.,][]{Quimby2007, Pastorello2010, Nicholl2015, Ho2019}. Purely $^{56}$Ni-powered interpretations can imply $^{56}$Ni masses approaching those of the pair-instability supernova (PISN) channel; however, the large ejecta masses and long diffusion timescales expected for PISNe are generally in tension with many of these events
\citep{GalYam&Avishay2019,Kasen2011,Dessart2013}. These extreme explosions therefore motivate a systematic consideration of additional power sources beyond radioactive heating.

A widely discussed scenario is a newborn, millisecond-period, strongly magnetized neutron star (NS) that injects rotational energy into the ejecta and powers the light curve \citep{Kasen2010,Woosley2010,GalYam&Avishay2019}. This pulsar-driven SN framework quantitatively accounts for SLSNe-I and is also invoked for a subset of engine-driven Ic-BL explosions, including events with early luminous components such as SN~2006oz \citep{Leloudas2012}, LSQ14bdq \citep{Nicholl2015}, and DES14X3taz \citep{Smith2016}. Further motivation comes from the well-established associations between long-duration gamma-ray bursts (GRBs) and Ic-BL SNe (e.g., GRB~980425/SN~1998bw, GRB~060218/SN~2006aj, GRB~100316D/SN~2010bh, and GRB~171205A/SN~2017iuk; \citealt{Galama1998,Pian2006,Chornock2010,Wang2018}). In this context, a nascent magnetar has been discussed as a viable central engine in at least some cases, alongside accreting black-hole models \citep{Woosley2006,Metzger2011}. Moreover, FRB~200428 from the Galactic magnetar SGR~1935+2154 provides direct evidence that at least a fraction of fast radio bursts (FRBs) originate from magnetars \citep{Andersen2020,Bochenek2020}. This, in turn, motivates a potential temporal connection between magnetar-powered SNe and FRBs \citep{Murase2016,Metzger2017,Kashiyama2017} and highlights the importance of capturing observational signatures of magnetar birth.


Identifying pulsar-embedded SNe, and using their accompanying observables to constrain the birth environment and channel of the central engine, is increasingly important. Proposed diagnostics include searches for pulsar wind nebula (PWN)-powered non-thermal emission and ionization break-out \citep[e.g.,][]{Kotera2013,Metzger2014,Murase2016,Omand2018}, as well as hydrodynamic signatures of engine activity\citep[e.g.,][]{Blondin&Chevalier2017,Suzuki&Maeda2017,Chen2016}. 

Here we focus on the latter one. The engine wind inflates an overpressured bubble and sweeps up a thin shell; in the steep outer density gradient, hydrodynamic instabilities fragment the shell and trigger blowout. The resulting near-surface shock heating produces a short-lived ultraviolet (UV)/optical first peak, followed by a diffusion-powered main peak. This double-peaked light curve offers a complementary diagnostic of a newborn, rapidly rotating magnetar \citep{Kasen2016,Liu2021}. Direct radiation-hydrodynamic simulations of blowout remain computationally expensive because they must capture multidimensional instabilities and radiation transport \citep{Chen2016,Suzuki&Maeda2017,Suzuki&Maeda2019,Suzuki&Maeda2021}. Motivated by relativistic hydrodynamics (RHD) results that post-blowout mixing can establish an approximately radius-independent kinetic-energy flux, $\rho r^{2} v^{3} \approx \mathrm{const}$ \citep{Suzuki&Maeda2019,Suzuki&Maeda2021}, we adopt this relation as a guiding constraint and develop a semi-analytic model to explore the resulting observables.
This framework enables systematic parameter surveys that link blowout emission and double-peaked morphology to supernova ejecta properties and neutron-star engine parameters, and may help identify otherwise elusive channels such as ultra-stripped explosions and accretion- or merger-induced collapse (AIC/MIC) events.

This paper is organized as follows. In Section~\ref{sec:blowout_model} we present the theoretical model. In Section~\ref{sec:calculation_results} we report simulation results for SESNe, as well as ultra-stripped SNe and AIC/MIC SNe, and we discuss the resulting observational implications. In Section~\ref{sec:discussion} we delineate the regime of applicability of the model and discuss non-thermal emission. We summarize our conclusions in Section~\ref{sec:conclusion}.

\section{Blowouts in Pulsar-Driven Supernovae}
\label{sec:blowout_model}

In the following section, we detail our theoretical model for pulsar-driven SNe hosting a nascent wind bubble, motivated by early emission in some double-peaked light curves and anchored by the near radius-invariance of the kinetic-energy flux found in RHD simulations \citep{Kasen2016, Suzuki&Maeda2019, Suzuki&Maeda2021, Liu2021}.

\subsection{Model Setup}
\label{sec:model-setup}

\begin{figure}[ht!]
\epsscale{1.1} 
\plotone{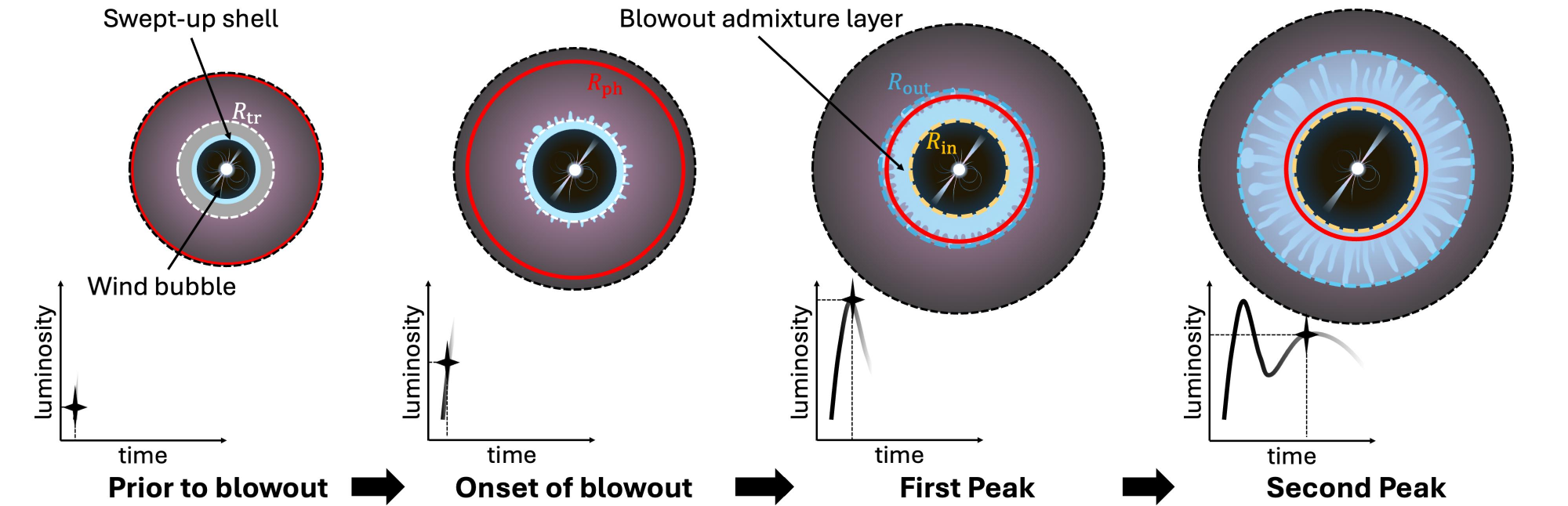}
\caption{
Schematic illustration of the blowout model and the resulting double-peaked light curve. From left to right, four evolutionary stages are shown. (i) \textit{Prior to blowout}: a central engine inflates a wind bubble and drives a thin swept-up shell that expands inside the dense ejecta at $r<R_{\mathrm{tr}}$. (ii) \textit{Onset of blowout}: the swept-up shell reaches the density-transition radius $R_{\mathrm{tr}}$, since the outer ejecta at $r>R_{\mathrm{tr}}$ is tenuous, the upstream ram pressure drops rapidly, and hydrodynamic instabilities broaden the shell. (iii) \textit{First peak}: the thin shell is converted into a blowout admixture layer between $R_{\mathrm{in}}$ and $R_{\mathrm{out}}$, where kinetic energy flux is invariant with radius: $\rho r^{2} v^{3} \approx \mathrm{const}$. The admixture layer rapidly expands until the diffusion front $R_{\mathrm{ph}}$ penetrates into $R_{\mathrm{out}}$. Energy released by shock heating and thermal photon diffusion leakage from $R_{\mathrm{out}}$ supported the fast first peak by that. (iv) \textit{Second peak}: $R_{\mathrm{ph}}$ recedes deep into the admixture layer as the ejecta become optically thin, and the release of trapped internal energy produces the second, broader peak in the light curve.
}
\label{fig:blowout-model}
\end{figure}

The schematic illustration of our blowout model has been summarized in Fig.~\ref{fig:blowout-model}. Soon after the explosion, the SN ejecta are assumed to undergo homologous expansion and are approximated by a broken power-law density profile \citep{Chevalier1992},
\begin{equation}
    \rho(r,t) = \zeta_{\rho}\frac{M_{\rm ej}}{R_{\rm tr}^3}
    \begin{cases}
        \left(r/R_{\rm tr}\right)^{-\delta}, & r < R_{\rm tr},\\[4pt]
        \left(r/R_{\rm tr}\right)^{-n},      & r \ge R_{\rm tr},
    \end{cases}
    \label{eq:rho_profile}
\end{equation}
where the transition radius is defined as $R_{\rm tr}\equiv v_{\rm tr}t$. The indices $\delta$ and $n$ describe a relatively flat inner core and a steep outer envelope, respectively. Typical values are $\delta\simeq 0$–$1$ and $n\simeq 7$–$10$, with $\delta<3$ and $n>5$ required for the total mass and energy to remain finite \citep{Chevalier1982,Matzner&McKee1999}. The normalization factor is
\begin{equation}
    \zeta_{\rho}
    = \frac{(3-\delta)(n-3)}{4\pi\,(n-\delta)}.
    \label{eq:zeta_rho}
\end{equation}
The transition velocity $v_{\rm tr}$ is related to the SN explosion energy by
\begin{equation}
    v_{\rm tr}
    = \zeta_{\rm v} \left( \frac{E_{\rm sn}}{M_{\rm ej}} \right)^{1/2},
    \label{eq:vtr_def}
\end{equation}
where
\begin{equation}
    \zeta_{\rm v}
    = \left[
        \frac{2(5-\delta)(n-5)}{(3-\delta)(n-3)}
      \right]^{1/2}
    \label{eq:zeta_v}
\end{equation}
is a dimensionless coefficient of order unity.

Following the SN explosion, a rapidly rotating, strongly magnetized pulsar may be born at the center. Its spin-down power emerges as a Poynting-flux–dominated outflow that drives a relativistic $e^\pm$ wind. This wind inflates a hot central bubble and launches a forward shock into the ambient ejecta. The subsequent evolution of the bubble is therefore tightly coupled to the pulsar's physical parameters, which determine its spin-down luminosity.

We characterize the central engine by a dipole magnetic field strength $B_{\rm dip}$, an initial spin period $P_{\rm NS}$, and an obliquity angle $\chi$ between the rotation and magnetic axes. Force-free simulations \citep[e.g.,][]{Gruzinov2005, Spitkovsky2006, Tchekhovskoy2013} relate the pulsar spin-down luminosity $L_{\rm sp}$ to these quantities as
\begin{equation}
    L_\mathrm{sd}
    = \frac{B_{\rm dip}^2 \left(2\pi/P_{\rm NS}\right)^4 R_{\rm NS}^6}{4c^3}
      \left(1 + C\sin^2\chi\right)
    \;\;\sim\;\;
    2.17\times10^{45}\ \mathrm{erg\,s^{-1}}\,
    \left(\frac{B_{\rm dip}}{10^{13}\ \mathrm{G}}\right)^{2}
    \left(\frac{P_{\rm NS}}{1\ \mathrm{ms}}\right)^{-4},
    \label{eq:Lsd}
\end{equation}
where $R_{\rm NS}$ is the neutron-star radius, $c$ is the speed of light, and we adopt $C\simeq 1$ and $\chi=45^\circ$ for fiducial estimates.

As the wind bubble expands within the ejecta, it sweeps up a dense shell. If, before the ejecta become transparent, this swept-up shell receives from the pulsar an energy comparable to the original SN explosion energy, it can be rapidly accelerated out to radii $r\gtrsim R_{\rm tr}$, where it encounters the steep outer density profile \citep{Chevalier1992}. In this regime, the swept-up shell contains substantial internal energy, while the confining ram pressure from the overlying ejecta becomes weak. Hydrodynamical instabilities at the contact discontinuity then grow nonlinearly, disrupting the thin shell and mixing the bubble material with the outer ejecta \citep[e.g.,][]{Jun1998, Blondin2001}. This process produces a broadened ``blowout admixture layer''. The disruption of the pulsar wind bubble in this manner is what we refer to in this work as a \emph{blowout} \citep[e.g.,][]{Blondin&Chevalier2017}.

Thermal photons begin to escape once the local photon diffusion speed exceeds the local expansion speed of the matter. The radius at which this condition is satisfied defines a diffusion front, which moves inward in mass coordinate as the ejecta expand \citep[e.g.,][]{Kisaka2015, Kashiyama2015}. Before the diffusion front reaches the admixture layer, the internal energy stored in the bubble can be efficiently converted into kinetic energy of the blowout material, sustaining its rapid expansion. Two important consequences follow: (i) the forward shock driven by the expanding bubble heats the surrounding ejecta, producing luminous shock-powered emission \citep{Kasen2016}; and (ii) the optically thick inner layers advect photons outward more rapidly than radiative diffusion alone, transporting energy to the outer ejecta. Together, these effects give rise to a fast, luminous early peak in the light curve.

Once the diffusion front penetrates the admixture layer, further acceleration of this material is no longer supported: the kinetic energy required to maintain rapid expansion is no longer efficiently replenished by newly injected internal energy, and the expansion velocity of the admixture layer gradually approaches that of the outer ejecta. The first peak thus terminates at this stage. Subsequently, as the ejecta continue to expand and the diffusion front recedes deeper in mass, the remaining internal energy is released, which is higher in the denser inner regions, causing the luminosity to rise again and form the second, main peak. This ``main'' peak corresponds to the conventional SN fireball phase, but with an energy budget dominated by the pulsar's rotational energy rather than radioactive decay \citep{Kasen2016}.

Throughout this paper, we refer to the disruption of the pulsar wind bubble as a ``blowout'', and to the associated early luminosity feature as the ``first peak''. Many previous works instead describe this early bump as a ``breakout'' \citep[e.g.,][]{Kasen2016}. We retain their original terminology when discussing those studies, but adopt ``first peak'' for our own models to avoid ambiguity.

\subsection{Blowout Condition}
\label{sec:blowout-condition}

A blowout occurs when the forward shock driven by the pulsar wind bubble reaches the transition radius $R_{\rm tr}$ \citep{Blondin&Chevalier2017}. For this to happen, the pulsar must deposit a sufficiently large amount of energy into the ejecta \emph{before} they become transparent, so that the energy injection is efficient. This requirement translates into constraints on both the pulsar's rotational-energy budget and its spin-down luminosity.

From the energy-budget perspective, by the time of blowout the total energy injected by the pulsar is
\begin{equation}
E_{\rm bo} = \zeta_{\rm bo}\,E_{\rm sn},
\label{eq:E_bo}
\end{equation}
where the dimensionless factor
\begin{equation}
\zeta_{\rm bo} =
\frac{2\,(n-5)\,(9-2\delta)\,(11-2\delta)}
{(5-\delta)^{2}\,(n-\delta)\,(3-\delta)}
\end{equation}
depends only on the density slopes. For $\delta=1$ and $n=10$, one obtains $\zeta_{\rm bo}=2.19$ \citep{Chevalier1992, Kasen2016}. Thus, the pulsar must spin rapidly enough that its rotational-energy reservoir exceeds $E_{\rm bo}$. Equating the available rotational energy of a pulsar of mass $M_{\rm NS}$, radius $R_{\rm NS}$ that initially rotating at a period of $P_{\rm NS}$ yields the constraint
\begin{equation}
P_{\rm NS} <
(2\pi)\,5^{-1/2}\,
\zeta_{\rm bo}^{-1/2}\,
M_{\rm NS}^{1/2}\,R_{\rm NS}\,E_{\rm sn}^{-1/2}.
\label{eq:energy_constraint}
\end{equation}

A second constraint arises from the timescale of energy deposition. The pulsar wind nebula must inject $E_{\rm bo}$ before radiation from the nebula begins to diffuse through the ejecta on a timescale \citep{Arnett1982, Kasen2016}
\begin{equation}
t_{\rm diff} =
\left(\frac{\kappa\,\zeta_{\rho}\,M_{\rm ej}}{v_{\rm tr}\,c}\right)^{1/2},
\end{equation}
where $\kappa$ is the opacity. Otherwise, one expects a standard pulsar-powered SN light curve without a distinct blowout-like first peak \citep[e.g.,][]{Kashiyama2016}. The characteristic time required to deposit the blowout energy is
\begin{equation}
t_{\rm bo} \simeq \frac{E_{\rm bo}}{L_{\rm sp}},
\label{eq:blowout_timescale}
\end{equation}
assuming that $L_{\rm sp}$ remains approximately constant for $t \lesssim t_{\rm bo}$ \citep{Chevalier1992, Kasen2016}. 
We do not consider cases in which $L_{\rm sp}$ has already decayed strongly (i.e., times later than the spin-down timescale), because by then most of the pulsar's rotational energy has been exhausted; if a blowout has not occurred by that stage, the initial spin is simply too slow to satisfy the energy requirement in Eq.~\eqref{eq:energy_constraint}.

Requiring that the energy be injected before photons can diffuse out, $t_{\rm bo} < t_{\rm diff}$, yields a constraint on the spin-down luminosity or, equivalently, on the combination of $B_{\rm dip}$ and $P_{\rm NS}$,
\begin{equation}
B_{\rm dip}\,P_{\rm NS}^{-2} >
\zeta_{\rm spl}\,
c^{7/4}\,
\left(1+C\sin^2\chi\right)^{-1/2}
R_{\rm NS}^{-3}\,
\kappa^{-1/4}\,
E_{\rm sn}^{5/8}\,
M_{\rm ej}^{-3/8},
\label{eq:luminosity_constraint}
\end{equation}
with
\begin{equation}
\zeta_{\rm spl} =
\frac{1}{2\pi^{2}}\,
\zeta_{\rm bo}^{1/2}\,
\zeta_\rho^{-1/4}\,
\zeta_v^{1/4},
\end{equation}
for which $\zeta_{\rm spl}\simeq 0.145$ when $\delta=1$ and $n=10$.

\begin{figure}[ht!]
\epsscale{0.618} 
\plotone{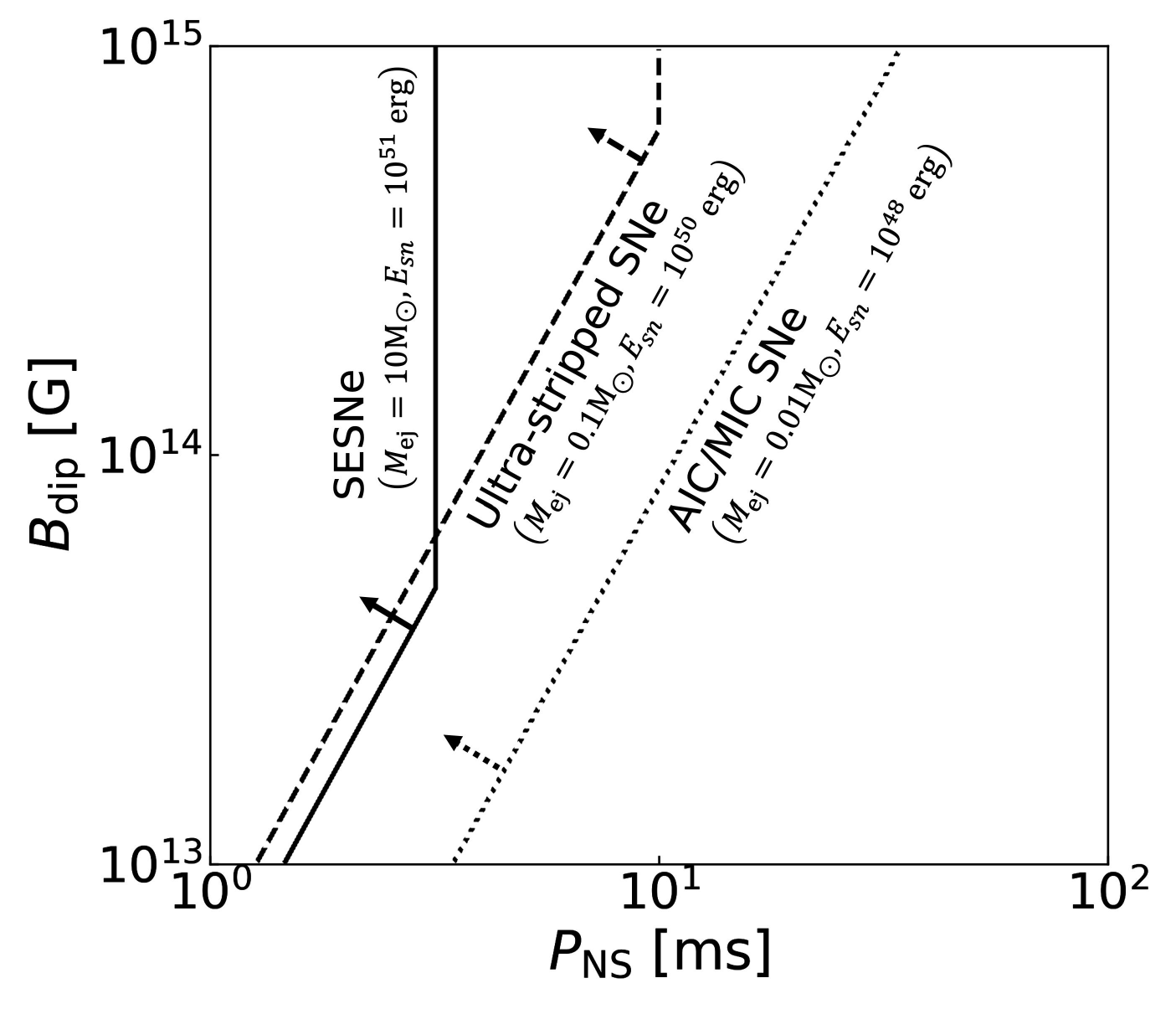}
\caption{
Phase diagram for the occurrence of blowout in different classes of SNe. We consider representative parameters for SESNe, ultra-stripped SNe, and AIC/MIC SNe. The lines correspond to specific SN types and parameter choices, as indicated by the labels next to each curve. Upper left region of each boundary line (i.e., the region pointed to by the arrows) corresponds to pulsar parameters for which a blowout can occur under the SN condition.
}
\label{fig:blowout-phase-diagram}
\end{figure}

For representative SESN parameters, these inequalities become
\begin{align}
\left(\frac{P_{\rm NS}}{1\ \mathrm{ms}}\right)
&< 3.18\,
\left(\frac{E_{\rm sn}}{10^{51}\ \mathrm{erg}}\right)^{-1/2},
&& \text{\makebox[3.8cm][r]{\footnotesize (energy constraint)}} \\
\left(\frac{B_{\rm dip}}{10^{13}\ \mathrm{G}}\right)
\left(\frac{P_{\rm NS}}{1\ \mathrm{ms}}\right)^{-2}
&> 0.468\,
\left(\frac{\kappa}{0.1\ \mathrm{cm^2\,g^{-1}}}\right)^{-1/4}
\left(\frac{E_{\rm sn}}{10^{51}\ \mathrm{erg}}\right)^{5/8}
\left(\frac{M_{\rm ej}}{10\ M_\odot}\right)^{-3/8}
\label{eq:L_sp_blowout_constraint}
&& \text{\makebox[3.8cm][r]{\footnotesize (spin-down luminosity constraint)}} 
\end{align}
where we have adopted $M_{\rm NS}=1.4\,M_\odot$ and $R_{\rm NS}=10\ \mathrm{km}$ for the numerical coefficients. Taken together, these two conditions define the blowout boundary in the phase diagram shown in Fig.~\ref{fig:blowout-phase-diagram}: the vertical segment is set by the energy constraint, whereas the inclined segment is determined by the spin-down luminosity constraint. Only the region in the upper left, where both inequalities are satisfied simultaneously, corresponds to pulsar parameters for which a blowout can occur.

These inequalities imply that, for canonical SESNe, a blowout generally requires a millisecond newborn NS and a magnetic field near or above the quantum critical value ($B_Q\simeq 4.4\times10^{13}\ {\rm G}$) \citep{Harding2006,Blondin&Chevalier2017}, so that a rotational-energy budget comparable to the explosion energy is deposited before the thermal-diffusion peak. Such conditions are most naturally realized in hydrogen-poor SLSNe-I, and possibly in some hyper-energetic broad-lined Type Ic events \citep{Campana2006, Kasen2016, Liu2021}. 
On the other hand, low-energy explosions, such as ultra-stripped SNe \citep{Tauris2013,Tauris2015} or AIC/MIC events \citep{Piro2012,Ablimit2022}, the requirements on the natal spin and magnetic field are substantially relaxed (see Fig.~\ref{fig:blowout-phase-diagram}). A larger fraction of these transients is therefore expected to exhibit a rapidly evolving blowout-powered first peak (i.e., a double-peaked light curve), making the early first peak a potentially useful diagnostic of such channels.

The discussion above presumes that the swept-up shell remains quasi-adiabatic and geometrically thin until the forward shock reaches $R_{\rm tr}$ \citep{Chevalier1992}. Equivalently, the diffusion front, the radius where radiation and matter begin to decouple and diffusion becomes dynamically important, must not overtake the shell before it reaches $R_{\rm tr}$. In practice, this is captured by requiring $t_{\rm bo}<t_{\rm diff}$ (Eqs.~\eqref{eq:blowout_timescale}--\eqref{eq:luminosity_constraint}); otherwise the system transitions to an inner-breakout--like evolution \citep{Kasen2016} and the thin-shell approximation breaks down.

Across the parameter space enclosed by Eqs.~\eqref{eq:energy_constraint} and \eqref{eq:luminosity_constraint} in Fig.~\ref{fig:blowout-phase-diagram}, this additional consistency condition is satisfied except near the luminosity threshold $L_{\rm sp}\sim E_{\rm bo}/t_{\rm diff}$, where $t_{\rm bo}\sim t_{\rm diff}$ and diffusion becomes marginally important already during the pre-blowout phase. This corresponds to the marginal-blowout regime discussed in Section~\ref{subsec:marginal_blowout}.

\subsection{Blowout dynamics and emission}
\label{sec:blowout-dynamics-and-emssion}

\subsubsection{Geometry and Scaling}
We first lay out the geometric and kinematic framework of the blowout admixture layer. Once the bubble surface crosses the transition radius $R_{\rm tr}$, the external confining ram pressure drops rapidly while the shell retains comparatively high thermal pressure. Rayleigh–Taylor (RT) instabilities promptly fragment the shell and open channels that puncture the tenuous, preshock outer ejecta \citep{Blondin&Chevalier2017}. Numerical simulations \citep{Suzuki&Maeda2019,Suzuki&Maeda2021} show that these channels efficiently transport pulsar-injected energy to larger radii. In this blowout admixture layer, the kinetic–energy flux is approximately flat between an inner and an outer radius,
\begin{equation}
\rho(r)\, r^{2}\, v(r)^{3} \;\approx\; \mathrm{const},
\label{eq:admixture_Ek_flux_const}
\end{equation}
where $R_{\rm in}$ is the radius of the relativistic $e^\pm$ wind bubble, $R_{\rm out}$ is the maximum extent of the channels, $\rho(r)$ is the shell-averaged density at radius $r\in[R_{\rm in},R_{\rm out}]$, and $v(r)$ is the local expansion speed.

Immediately after blowout the bubble shell is thin; subsequent RT-driven mixing with the ambient ejecta drives the system toward an admixture layer that gradually approaches quasi-homologous expansion. To capture this transition with minimal assumptions, we adopt an affine radial-velocity profile for $r\in[R_{\rm in},R_{\rm out}]$,
\begin{equation}
v(r) \;=\; v_{\rm in} + \zeta_{\rm af}\,\bigl(r - R_{\rm in}\bigr),
\label{eq:affine_velocity_profile}
\end{equation}
with slope $\zeta_{\rm af} \equiv (v_{\rm out}-v_{\rm in})/(R_{\rm out}-R_{\rm in})$. Here $v_{\rm in}\equiv v(R_{\rm in})$ and $v_{\rm out}\equiv v(R_{\rm out})$. With a time-dependent $\zeta_{\rm af}(t)$, the affine profile connects the thin-shell regime to near-homologous expansion; over intervals where $\partial_t\zeta_{\rm af}\!\approx\!0$, it reduces to an $R_{\rm in}$–anchored homologous scaling. We use this profile as our working approximation.

Combining Eqs.~\eqref{eq:admixture_Ek_flux_const} and \eqref{eq:affine_velocity_profile}, the density profile in the admixture layer is
\begin{equation}
\rho(r)=
\frac{M_{\rm adm}}{2\pi R_{\rm in}^3}
\left(\frac{r}{R_{\rm in}}\right)^{-5}
\left\{
\left[(1-h)^{-2}-\left(\zeta_{\rm r}-h\right)^{-2}\right]^{-1}
\left[1-h\left(\frac{R_{\rm in}}{r}\right)\right]^{-3}
\right\},
\end{equation}
where $\zeta_{\rm r}\equiv R_{\rm out}/R_{\rm in}$ and
\begin{equation}
h \;\equiv\; \frac{v_{\rm out}-\zeta_{\rm r} v_{\rm in}}{v_{\rm out}-v_{\rm in}}
\label{eq:h}
\end{equation}
quantifies the departure from strictly homologous expansion ($h=0$ corresponds to $v\propto r$ anchored at $R_{\rm in}$). In the limit $h\to 0$, the profile asymptotically approaches $\rho\propto r^{-5}$, which follows directly from the constant kinetic–energy–flux condition in Eq.~\eqref{eq:admixture_Ek_flux_const}. The total mass in the admixture layer is
\begin{equation}
M_{\rm adm} \;=\; M_{\rm ej} \,
\left[
1- \frac{(3-\delta)}{(n-\delta)}\,
\left(\frac{R_{\rm out}}{R_{\rm tr}}\right)^{3-n}
\right].
\label{eq:Mass_conservation}
\end{equation}

Given the velocity and density profiles, the kinetic energy is
\begin{equation}
E_{\rm kin,adm} =
\int_{R_{\rm in}}^{R_{\rm out}}
2\pi\,r^2\, \rho\, v^2\, dr
\;=\;
M_{\rm adm} \, 
\frac{v_{\rm in}^2 v_{\rm out}^2}{v_{\rm out}^2 - v_{\rm in}^2} 
\ln\!\left(\frac{v_{\rm out}}{v_{\rm in}}\right),
\label{eq:phaseI_Ekin_adm}
\end{equation}
and the radial momentum is
\begin{equation}
P_{\rm adm} 
=
\int_{R_{\rm in}}^{R_{\rm out}}
4\pi\,r^2\, \rho\, v\, dr
\;=\;
2M_{\rm adm} \, 
\frac{v_{\rm in} v_{\rm out}}{v_{\rm in} + v_{\rm out}}.
\label{eq:phaseI_P_adm}
\end{equation}

While the ejecta remains optically thick radiation and gas are tightly coupled (near LTE) \citep{Suzuki&Maeda2021}; the expansion is then well approximated as adiabatic. Here $p$ denotes the radiation pressure, implying $p\propto\rho^{4/3}$ and an internal–energy density $u=3p$. Accordingly, the internal energy stored in the admixture layer is
\begin{equation}
E_{\rm int,adm} 
=
\int_{R_{\rm in}}^{R_{\rm out}}
12\pi\,p\,r^2\, dr
\;=\;
12\pi\,p_{\rm in}\, R_{\rm in}^3\,
\mathcal{G}(\zeta_{\rm r}, h),
\label{eq:phaseI_Eint_adm}
\end{equation}
where $p_{\rm in}\equiv p(R_{\rm in})$ and $\mathcal{G}$ is the Gauss hypergeometric function
\begin{equation}
\mathcal{G}(\zeta_{\rm r}, h)
=\frac{3}{11}(1-h)^4
\left[
{_2}F_1\!\left(
\frac{11}{3},\,4;\,\frac{14}{3};\,h
\right)
-{\zeta_{\rm r}}^{-11/3}\,
{_2}F_1\!\left(
\frac{11}{3},\,4;\,\frac{14}{3};\,\frac{h}{\zeta_{\rm r}}
\right)
\right].
\end{equation}

Equations~\eqref{eq:Mass_conservation}–\eqref{eq:phaseI_Eint_adm} constitute the main relations that specify the state of the admixture layer before diffusion front penetrating into it. Obtaining a closed-form solution to this coupled system is generally intractable, particularly when the model is required to describe a continuous transition from a swept-up thin shell to the developing admixture layer, which introduces hypergeometric functions into the integrals. Once the time evolution of the background quantities and the initial conditions at blowout are specified, however, the system can be solved numerically in a self-consistent manner.

To close this subsection and for subsequent calculations, we summarize the energy budget of the swept-up shell at the blowout moment, using the thin-shell approximation of \citet{Chevalier1992}. We define $E_{\rm kin,shell}$ as the shell’s bulk kinetic energy, $E_{\rm int,shell}$ as its internal energy, and $E_{\rm rad,shell}$ as the cumulative energy radiated by the shell up to $t_{\rm bo}$. These quantities provide the initial conditions for our post-blowout model:
\begin{equation}
    E_{\rm kin,shell}=
    \frac{(5-\delta)(6-\delta)^2}{2(11-2\delta)(9-2\delta)}\,
    E_{\rm bo}
    =\frac{(n-5)(6-\delta)^2}{(5-\delta)(3-\delta)(n-\delta)}\,
    E_{\rm sn},
\label{eq:E_kin_initial}
\end{equation}
\begin{equation}
    E_{\rm int,shell}=
    \frac{5-\delta}{11-2\delta}\,
    E_{\rm bo}
    =\frac{2(n-5)(9-2\delta)}{(5-\delta)(3-\delta)(n-\delta)}\,
    E_{\rm sn},
\label{eq:E_int_initial}
\end{equation}
\begin{equation}
    E_{\rm rad,shell}=
    \frac{3-\delta}{2(11-2\delta)(9-2\delta)}\,
    E_{\rm bo}
    =\frac{n-5}{(5-\delta)^2(n-\delta)}\,
    E_{\rm sn}.
\label{eq:E_rad_initial}
\end{equation}

\subsubsection{Evolution Equations}
We evolve the bulk kinetic energy of the admixture layer by accounting for (i) boundary work, (ii) conversion between internal and kinetic energy via adiabatic $p\,dV$ work, and (iii) kinetic-energy advection by upstream ejecta entrained through the outer boundary.

At the inner boundary, the pulsar-wind bubble overpressurizes the layer and exerts an outward thrust set by the interface thermal pressure $p_{\rm in}$,
\begin{equation}
F_{\rm in} \;\simeq\; 4\pi R_{\rm in}^{2}\, p_{\rm in}.
\end{equation}
At the outer boundary $R_{\rm out}$, the layer outruns the preshock homologous ejecta and drives a forward shock. The associated ram pressure acts as a drag, but post-blowout fragmentation can render the interaction front porous so that strong shocks cover only a fraction of the surface. We parameterize this reduction by a dimensionless shock covering fraction $\eta_{S}\!\in\![0,1]$ and write
\begin{equation}
F_{\rm out} \;=\; \eta_{S}\, 4\pi R_{\rm out}^{2}\, \rho_{\rm pre}\, (\Delta v)^{2},
\end{equation}
where $\rho_{\rm pre}\equiv \rho_{\rm ej}(R_{\rm out},t)$ is the preshock ejecta density from the broken power-law profile, and
\begin{equation}
\Delta v \;\equiv\; v_{\rm out}-v_{\rm ej}(R_{\rm out},t)
\end{equation}
is the velocity jump at the shock (for a homologous preshock flow, $v_{\rm ej}(R_{\rm out},t)=R_{\rm out}/t$). The kinetic-energy extraction at $R_{\rm out}$ defines the shock-heating power,
\begin{equation}
L_{\rm sh} \;=\; \frac{1}{2}\,F_{\rm out}\,\Delta v,
\end{equation}
i.e., the rate at which bulk kinetic energy is converted into heat (and ultimately radiation, depending on the optical depth ahead).

We denote the $p\,dV$ work rate by
\begin{equation}
\dot{W}_{\rm pdV}
\;\equiv\;
\int_{R_{\rm in}}^{R_{\rm out}}
p\,(\nabla\!\cdot\!\boldsymbol{v})\, 4\pi r^{2}\,dr,
\qquad
\nabla\!\cdot\!\boldsymbol{v}
=\frac{1}{r^{2}}\frac{\partial}{\partial r}\bigl(r^{2}v\bigr)
\;=\;\frac{\partial v}{\partial r}+\frac{2v}{r}.
\end{equation}
Assuming a radiation-dominated equation of state, $p\propto\rho^{4/3}$, and adopting the affine velocity profile of Eq.~\eqref{eq:affine_velocity_profile}, the integral evaluates to
\begin{equation}
\dot{W}_{\rm pdV}
\;=\;
4\pi\,p_{\rm in}\, R_{\rm in}^{2}\,
\frac{v_{\rm out}-v_{\rm in}}{\zeta_{\rm r}-1}
\left[
3\,\mathcal{G}\!\left(\zeta_{\rm r},h\right)
-2h\,\mathcal{J}(\zeta_{\rm r},h)
\right],
\end{equation}
where $\mathcal{J}$ is another Gauss hypergeometric function,
\begin{equation}
\mathcal{J}(\zeta_{\rm r}, h)
=\frac{3}{14}(1-h)^4
\left[
{_2}F_1\!\left(
\frac{14}{3},\,4;\,\frac{17}{3};\,h
\right)
-\zeta_{\rm r}^{-14/3}\,
{_2}F_1\!\left(
\frac{14}{3},\,4;\,\frac{17}{3};\,\frac{h}{\zeta_{\rm r}}
\right)
\right].
\end{equation}
With this convention, $\dot{W}_{\rm pdV}>0$ corresponds to a net transfer of internal energy into bulk kinetic energy during expansion.

Mass loading also imports upstream kinetic energy into the layer as preshock ejecta crosses $R_{\rm out}$. We write the associated kinetic luminosity as
\begin{equation}
L_{\rm adv}
\;\equiv\;
\frac{1}{2}\,4\pi R_{\rm out}^{2}\,
\rho_{\rm pre}\,\Delta v\,\bigl[v_{\rm ej}(R_{\rm out},t)\bigr]^{2},
\end{equation}
which counts the inflow of upstream kinetic energy irrespective of whether the entrained material subsequently shocks strongly or mixes quasi-continuously.

Collecting the above contributions, the bulk kinetic energy evolves as
\begin{equation}
\frac{dE_{\rm kin,adm}}{dt}
\;=\;
F_{\rm in}\,v_{\rm in}
\;-\;
L_{\rm sh}
\;+\;
\dot{W}_{\rm pdV}
\;+\;
L_{\rm adv},
\label{eq:PhaseI_Ekin_evolution}
\end{equation}
closing the kinetic-energy budget for the admixture layer under our sign conventions.

Turning to the momentum evolution, and retaining only mechanical stresses at the boundaries (neglecting subdominant radiative momentum losses), we obtain
\begin{equation}
\frac{dP_{\rm adm}}{dt}
\;=\;
F_{\rm in}-F_{\rm out}+4\pi\,R_{\rm out}^2 \,\rho_{\rm pre}\, \Delta v\,v_{\rm out}.
\label{eq:PhaseI_P_evolution}
\end{equation}

In Eq.~\eqref{eq:PhaseI_P_evolution}, $F_{\rm in}-F_{\rm out}$ is the net boundary thrust driving bulk acceleration. The last term is the post-shock momentum flux associated with mass loading: ejecta is swept into the layer at the rate $\dot{M}=4\pi R_{\rm out}^{2}\rho_{\rm pre}\Delta v$ and is rapidly brought to co-motion with the outer boundary, so the momentum added to the layer is $\dot{M}\,v_{\rm out}$.

We finally consider the internal-energy budget while the admixture layer remains optically thick. In addition to $p\,dV$ conversion, internal energy is supplied by thermalization of the pulsar spin-down power and is depleted by radiative leakage through $R_{\rm out}$. Denoting by $\epsilon_{\rm adm}$ the fraction of $L_{\rm sp}(t)$ thermalized within the layer, we write
\begin{equation}
\frac{dE_{\rm int,adm}}{dt}
\;=\;
-\dot{W}_{\rm pdV}
\;+\;
\epsilon_{\rm adm}\,L_{\rm sp}
\;-\;
L_{\rm diff}
\;+\;
\left(1-f_{\rm esc}\right)L_{\rm sh},
\label{eq:PhaseI_Eint_evolution}
\end{equation}
with
\begin{equation}
\epsilon_{\rm adm} = 1-\exp(-\tau_{\rm ab}),
\qquad
\tau_{\rm ab}= \int_{R_{\rm in}}^{R_{\rm out}}\kappa_{\rm ab}\,\rho\,dr,
\label{eq:thermalization_efficiency}
\end{equation}
which approaches unity while the layer is optically thick. In this regime we evaluate $\tau_{\rm ab}$ using a simplified effective absorption opacity $\kappa_{\rm ab}$, treated as a proxy for the depth over which injected power is absorbed and thermalized. The $(1-f_{\rm esc})L_{\rm sh}$ term accounts for the fraction of shock-generated radiation that is trapped because photon transport in the exterior is inefficient: photons produced near $R_{\rm out}$ can be overtaken by the advancing boundary and reprocessed within the layer. Technically, this term is $\simeq 0$ in most cases of our calculation; see Eq.~\eqref{eq:f_sec} and the accompanying discussion.

Eq.~\eqref{eq:PhaseI_Eint_evolution} neglects energy injection from $^{56}$Ni decay for simplicity. In the blowout branch considered here, the engine injects an energy comparable to the explosion energy and well above the radioactive contribution within short time, which justifies ignoring $^{56}$Ni heating in most cases. While, we note that $^{56}$Ni decay can still produce a visible effect in the marginal blowout regime and to the post main peak observable signatures. We summarize these effects in Section~\ref{subsec:Ni_decay} and will incorporate a self consistent treatment of $^{56}$Ni decay in future extensions of the model.

While the diffusion front remains beyond $R_{\rm out}$, expansion outpaces diffusion across the layer, so photon transport is predominantly advective and direct leakage is weak. To capture the gradual onset of leakage as the diffusion front approaches $R_{\rm out}$, we model escape through $R_{\rm out}$ as a competition between diffusion and boundary advection, characterized by an effective diffusion speed $v_{\rm diff}$ and the outward boundary speed $v_{\rm out}$. The probability that a photon outruns the moving boundary is parameterized as
\begin{equation}
f_{\rm esc}
\;\equiv\;
\min\!\left[\,1,\;
\frac{2\,v_{\rm diff}}{v_{\rm diff}+v_{\rm out}}\,\right],
\label{eq:f_sec}
\end{equation}
where the factor of 2 approximates the forward-hemisphere excess in an outward-biased radiation field. The diffusive luminosity through $R_{\rm out}$ is then
\begin{equation}
L_{\rm diff}
\;=\;
12\pi\, f_{\rm esc}\, R_{\rm out}^{2}\, p_{\rm in}
\left[\frac{\rho(R_{\rm out})}{\rho(R_{\rm in})}\right]^{4/3}
v_{\rm diff}(R_{\rm out}),
\end{equation}
where $p(R)\propto\rho(R)^{4/3}$. The local diffusion speed follows from radiative diffusion with $u=3p$,
\begin{equation}
v_{\rm diff}(r)
\;=\;
\frac{D(r)\,|\partial_r p(r)|}{p(r)}
\;=\;
\frac{4c}{9\,\kappa\,\rho(r)}\,\bigl|\partial_r\ln\rho(r)\bigr|,
\qquad
D(r)=\frac{c}{3\,\kappa\,\rho(r)}.
\end{equation}
This construction reproduces the standard scaling
$L_{\rm diff}\propto R_{\rm out}^{2}\,[c/(\kappa\rho(R_{\rm out}))]\,|\partial_r p|$,
while providing a smooth transition from the advection-dominated regime to the leakage-dominated regime as the diffusion front reaches $R_{\rm out}$.

An analogous consideration applies to radiation produced at the forward shock. Before the diffusion front penetrates into the admixture layer, the forward shock lies \emph{inside} the photosphere ($\tau\!\sim\!1$) so that the upstream remains optically thick ($\tau\gtrsim1$) and the shock is radiation-mediated. A non-negligible fraction of the shock-generated photons is therefore captured by the following optically thick, rapidly expanding blowout frontier; accordingly, we deposit a fraction $(1-f_{\rm esc})$ of $L_{\rm sh}$ into the internal-energy reservoir, as represented by the final term in Eq.~\eqref{eq:PhaseI_Eint_evolution}.

\subsubsection{Admixture Layer Transparency}

As the ejecta expand, a diffusion front recedes inward in mass coordinate and enters the admixture layer. Across this front the local photon–diffusion speed becomes comparable to, and then exceeds, the bulk expansion speed; photons begin to decouple from the gas, and thermal pressure no longer drives further thickening efficiently. Accordingly, the thickening relations derived earlier no longer hold throughout the layer. We evaluate the diffusive luminosity locally as
\begin{equation}
L_{\rm diff}
\;=\;
12\pi\,
R_{\rm diff}^{2}\, p_{\rm in}\,
\left[\frac{\rho\!\left(R_{\rm diff}\right)}{\rho\!\left(R_{\rm in}\right)}\right]^{4/3}
v_{\rm diff}\!\left(R_{\rm diff}\right),
\end{equation}
where $p_{\rm in}$ is the pressure at $R_{\rm in}$ and $v_{\rm diff}(R_{\rm diff})$ is the local diffusion speed at the front. This stage mainly releases the internal energy accumulated previously. By this time the early, fast peak has already been produced by the rapid conversion of internal energy into bulk kinetic energy via shock heating. While diffusion dominates, the ejecta are still globally optically thick yet not fully transparent; this mechanism therefore controls the luminosity in the inter–peak trough of double–peaked light curves.

Motivated by the contrast in optical depth and thermodynamic state across the advancing diffusion front, we treat the layer as an optically thick interior and an optically thin exterior separated by $R_{\rm diff}$. Radiative leakage ahead of the front weakens pressure support on a dynamical timescale, whereas the interior remains in near LTE and is well described by an approximately adiabatic $\gamma\simeq 4/3$ equation of state.
(i) The \emph{optically thick} inner zone, $R_{\rm in}\le r\le R_{\rm diff}$, is described by the same structural profiles and equation of state as in the previous subsection, with the replacement $R_{\rm out}\!\rightarrow\!R_{\rm diff}$ (equivalently, $\zeta_{\rm r}\!\rightarrow\!\zeta_{\rm d}\!\equiv\!R_{\rm diff}/R_{\rm in}$). (ii) The \emph{optically thin} outer zone, $R_{\rm diff}\le r\le R_{\rm out}$, cools rapidly by diffusion and exerts only a subdominant dynamical back–reaction prior to global transparency.

Once $R_{\rm diff}$ has penetrated into the admixture layer, a strong forward shock at $R_{\rm out}$ is no longer sustained. The structure between $R_{\rm diff}$ and $R_{\rm out}$ therefore should not rapidly deviate from that at the moment the diffusion front passes. In the absence of strong external forcing capable of rapidly reshaping the stratification, we thus treat the optically thin exterior as kinematically frozen at that time. For density-dependent calculations, such as optical-depth estimates, we approximate $R_{\rm diff}\le r\le R_{\rm out}$ by extending the instantaneous thick-region density profile beyond $R_{\rm diff}$, retaining $\rho\propto r^{-5}$. 

With this partition, we write the state evolution of the optically thin zone, using a one-zone approximation, since pressure and temperature gradients are dynamically unimportant once photons decouple.
Let $\epsilon_{\rm th,\,thick}$ and $\epsilon_{\rm th,\,thin}$ denote the thermalization efficiencies of the pulsar spin–down power $L_{\rm sp}(t)$ within the thick and thin zones, respectively. The thin–zone internal energy evolves as
\begin{equation}
\frac{dE_{\rm int,\,thin}}{dt}
\;=\;
\epsilon_{\rm th,\,thin}\,\bigl(1-\epsilon_{\rm th,\,thick}\bigr)\,L_{\rm sp}
\;-\;
\frac{E_{\rm int,\,thin}}{t_{\rm dyn,\,thin}}
\;-\;
L_{\rm thin} ,
\end{equation}
where the second term encodes conversion of internal energy into bulk motion via $p\,dV$ work on a dynamical timescale evaluated at the front,
\begin{equation}
t_{\rm dyn,\,thin}
\;\equiv\;
\frac{R_{\rm diff}}{v\!\left(R_{\rm diff}\right)} ,
\end{equation}
and
\begin{equation}
\frac{dE_{\rm kin,\,thin}}{dt}
\;=\;
\frac{E_{\rm int,\,thin}}{t_{\rm dyn,\,thin}} .
\end{equation}
The emergent luminosity of the thin zone is parameterized as
\begin{equation}
L_{\rm thin}
\;=\;
\frac{E_{\rm int,\,thin}}{t_{\rm esc,\,thin}} ,
\end{equation}
with an escape time
\begin{equation}
t_{\rm esc,\,thin}
\;\approx\;
\frac{\tau_{\rm thin}}{c}\,\bigl(R_{\rm out}-R_{\rm diff}\bigr) ,
\end{equation}
where
\begin{equation}
\tau_{\rm thin}
\;\equiv\;
\int_{R_{\rm diff}}^{R_{\rm out}}\!\kappa\,\rho(r)\,dr .
\end{equation}
This prescription smoothly connects diffusive escape to free streaming as $\tau_{\rm thin}$ declines, consistent with the diffusion–time treatment of \citet{Arnett1982}. 

In mass coordinate, the diffusion front continues to recede and eventually reaches $R_{\rm in}$. At that point the entire admixture layer has become transparent, and the thermal–peak luminosity is expected to coincide with this transition. By then the internal energy accumulated during blowout and earlier have been released, and the local diffusive power $L_{\rm diff}$ vanishes. Thereafter the region is described uniformly by the thin–zone relations with the replacement $R_{\rm diff}\!\rightarrow\!R_{\rm in}$, which now serves as the physical inner boundary.

\subsubsection{Observable Signatures}
\label{subsubsec:observable_signatures}
Having specified the evolution from blowout to global transparency, we now connect the model to observables. We have distinguished three contributing channels: (i) shock heating at the forward shock, (ii) photon diffusion from the admixture layer, and (iii) graybody emission from the thermalized post-diffusion cold region. Crucially, especially for the shock–heating and photon diffusion components our expressions quantify \emph{local} photon losses at the emission site; they do not yet include radiative transfer through the overlying ejecta to the observer. During the intervals when these channels dominate, the photosphere $R_{\rm ph}$ (where $\tau\!\sim\!1$) typically lies exterior to the local radiating surface. The region beyond emission sites, although it may be relatively optically thin, still possesses a finite optical depth. Photons that leave the shock region or decouple at the diffusion front can therefore undergo additional scatterings in this residual scattering envelope.

In the residual-scattering envelope, photons have already largely decoupled from the gas and the transport is near the free-streaming limit ($\tau\!\lesssim\!{\rm few}$). The finite optical depth between a local emission radius $R_{\rm emit}$ and the last-scattering surface $R_{\rm ph}$ therefore acts primarily as a modest scattering-induced hold-up, delaying and temporally smoothing photons after they are released locally but before they free-stream from the photosphere. We model this effect with a one-zone residual-scattering reservoir, introduced solely to capture this small propagation delay, while leaving the stratification-dependent diffusion that sets $R_{\rm diff}$ and $L_{\rm diff}$ inside the optically thick ejecta unchanged.

In our setup the relevant local sources are the forward-shock heating luminosity $L_{\rm sh}$ (emitted at $R_{\rm out}$) and the diffusive leakage from the hot admixture layer $L_{\rm diff}$ (emitted at $R_{\rm diff}$); we evolve separate reservoirs because their emission radii differ and their scattering delays cannot, in general, be collapsed into a single effective one. By contrast, emission from the post-diffusion cold region is released effectively at $R_{\rm ph}$, so there is no overlying scattering envelope to traverse; any additional scattering delay is therefore negligible and we do not apply this treatment there. Importantly, near the first-peak maximum the outer layers are already tenuous and $R_{\rm ph}\simeq R_{\rm out}$, so the residual-scattering layer is geometrically thin and of low inertia; the one-zone closure is therefore sufficient and does not distort the key first-peak signature, nor does it bias the timing or luminosity of the subsequent main peak.

The transient storage of radiation within $R_{\rm emit}\!\lesssim\! r\!\lesssim\! R_{\rm ph}$ is represented by a reservoir energy $E_{\rm rs}(t)$, which evolves as
\begin{equation}
\frac{dE_{\rm rs}}{dt}
\;=\;
L_{\rm inj}
\;-\;
L_{\rm scat},
\label{eq:pool_evolution}
\end{equation}
where $L_{\rm inj}$ is the local radiative power injected into the reservoir and
\begin{equation}
L_{\rm scat}
\;=\;
\frac{E_{\rm rs}}{t_{\rm scat}}
\end{equation}
is the emergent luminosity after residual scatterings. Here $t_{\rm scat}$ is the characteristic scattering (escape) timescale across the layer between $R_{\rm emit}$ and $R_{\rm ph}$. Because the residual-scattering region contains little inertia, $p\,dV$ work is negligible; its primary role is to delay and temporally broaden the emergent signal. Since the shock- and diffusion-powered emission sites are generally not co-spatial, we treat their residual-scattering reservoirs independently.

Within this approximation, the bolometric luminosity is
\begin{equation}
L_{\rm bol}
\;=\;
L_{\rm sh,\,scat}
\;+\;
L_{\rm diff,\,scat}
\;+\;
L_{\rm th},
\end{equation}
where the subscript `scat' denotes the luminosity after passage through the residual scattering envelope. The corresponding effective blackbody temperature is
\begin{equation}
T_{\rm eff}
\;=\;
\left(
\frac{L_{\rm bol}}{4\pi\,R_{\rm bol}^{2}\,\sigma}
\right)^{1/4},
\end{equation}
with
\begin{equation}
R_{\rm bol}
\;=\;
\max\!\left(R_{\rm ph},\,R_{\rm in}\right) .
\end{equation}
With the bolometric luminosity and graybody temperature derived above, we can readily predict the corresponding multi-band photometric light curves. This is particularly well motivated because the early-time emission of these events has been shown to be well described by blackbody spectra \citep[e.g.,][]{Leloudas2012,Nicholl2016,Moriya2018}.

Our blowout model is intentionally minimal: it relies on conservation of mass, momentum, and energy, together with a constant kinetic-energy flux in the admixture layer and an approximately affine velocity profile. Even so, a fully analytic solution is not available, and accurate light-curve evolution must be obtained by semi-analytic, time-dependent integration of the governing equations. Here we only summarize compact estimates for the characteristic timescales and luminosities of the two peaks, as a practical guide to the observable signatures.

We first focus on the first peak. Once the diffusion front enters the admixture layer, the stored internal energy is drained rapidly: the thermal pressure can no longer drive rapid expansion and thickening of the leading edge, the shock-heating luminosity fades, and the power source of the first peak is effectively shut off. At the same time, the material ahead of the front is nearly optically thin, so photons produced by shock heating and internal-energy leakage escape efficiently. We therefore identify the moment when the diffusion front first enters the admixture layer as a good proxy for the epoch of maximum luminosity, and adopt it as the characteristic timescale of the first peak. We decompose this timescale as
\begin{equation}
t_{\rm first} = t_{\rm bo} + t_{\rm bro},
\end{equation}
where $t_{\rm bo}$ is the time required to reach blowout (Eq.~\eqref{eq:blowout_timescale}) and $t_{\rm bro}$ is the post-blowout interval during which the leading edge of the admixture layer completes its rapid expansion.

To estimate $t_{\rm bro}$, we assume that by the time of the first peak the admixture layer has broadened from an initially thin shell and is nearly co-expanding with the outer ejecta in a homologous manner, $v_{\rm out}\simeq \zeta_r v_{\rm in}$. In the regime of interest we further take $M_{\rm adm}\simeq M_{\rm ej}$ (for $\delta=1$ and $n=10$, Eq.~\eqref{eq:Mass_conservation} shows that the layer already encloses $\simeq M_{\rm ej}$ by $R_{\rm out}\gtrsim 2R_{\rm tr}$). Under these assumptions, imposing $v_{\rm diff}(R_{\rm out})=v_{\rm out}$ at the time of the first peak gives
\begin{equation}
R_{\rm out}(t_{\rm first}) \simeq 
\left[
  \frac{
    81\,E_{\rm kin,adm}\,M_{\rm ej}\,\kappa^2\,(\zeta_{r}^2-1)
  }{
    1600\,\pi^2\,c^2\,\zeta_r^4\,\ln \zeta_{r}
  }
\right]^{1/4},
\label{eq:R_out_1st_peak_estimation}
\end{equation}
and
\begin{equation}
v_{\rm out}(t_{\rm first})  \simeq 
\left(
  \frac{E_{\rm kin,adm}}{M_{\rm ej}}\,
  \frac{\zeta_{r}^2-1}{\ln \zeta_{r}}
\right)^{1/2}.
\end{equation}

Because the detailed velocity evolution does not admit a simple analytic form, it remains difficult to obtain an exact expression for $t_{\rm bro}$ directly from these state variables. However, we can approximate $t_{\rm bro}$ by a dynamical timescale using
\begin{equation}
t_{\rm bro}\simeq \frac{R_{\rm out}(t_{\rm first})-R_{\rm bo}}{\tilde{{v}}_{\rm out}},
\end{equation}
where $\tilde{{v}}_{\rm out}$ is the average expansion speed of $R_{\rm out}$ between blowout and the time when the diffusion front penetrates the admixture layer. In our numerical solutions, $R_{\rm bo}\lesssim 0.1\,R_{\rm out}(t_{\rm first})$ in almost all cases, and most of the acceleration occurs early while the layer is still relatively thin-shell-like, so $\hat{v}_{\rm out}\approx v_{\rm out}(t_{\rm first})$ is an adequate approximation. This yields
\begin{equation}
t_{\rm bro} \simeq 
\left[
  \frac{
    81\,M_{\rm ej}^3\,\kappa^2\,\ln\zeta_r
  }{
    1600\,\pi^2\,E_{\rm kin,adm}\,c^2\,\zeta_r^4\,(\zeta_r^2-1)
  }
\right]^{1/4},
\label{eq:first_peak_broadening}
\end{equation}
and hence
\begin{equation}
t_{\rm first} \simeq 
\left[
  \frac{
    81\,M_{\rm ej}^3\,\kappa^2\,\ln\zeta_r
  }{
    1600\,\pi^2\,E_{\rm kin,adm}\,c^2\,\zeta_r^4\,(\zeta_r^2-1)
  }
\right]^{1/4}
+
\frac{\zeta_{\rm bo}E_{\rm sn}}{L_{\rm sp}}.
\label{eq:first_peak_timescale}
\end{equation}
Here $E_{\rm kin,adm}$ is the kinetic energy accumulated in the admixture layer prior to $t_{\rm first}$; most of the pulsar-injected energy before this epoch is converted into kinetic energy of the admixture layer and then partially reprocessed into radiation by shock heating. Thus, at similar SNe conditions the first-peak timescale therefore follows approximately constant $L_{\rm sp}$ contours in parameter space, with an effective dependence that can be viewed as a shifted $t_{\rm first}\propto L_{\rm sp}^{-1}$ relation arising from the sum of the diffusion and blowout terms.

When the first peak occurs before the pulsar has spun down significantly ($t_{\rm first} < t_{\rm sp}$), the spin-down luminosity can be approximated as constant, so that $E_{\rm kin,adm}\sim L_{\rm sp}\,t_{\rm first}$. Substituting this into Eq.~\eqref{eq:first_peak_timescale} gives
\begin{equation}
t_{\rm first}\simeq
\left[
  \frac{
    81\,M_{\rm ej}^3\,\kappa^2\,\ln\zeta_r
  }{
    1600\,\pi^2\,(\zeta_{\rm bo}E_{\rm sn})\,c^2\,\zeta_r^4\,(\zeta_r^2-1)
  }
\right]^{1/4}
+ \frac{\zeta_{\rm bo}E_{\rm sn}}{L_{\rm sp}}
\simeq 25.3~\mathrm{day}\,
\left( \frac{L_{\rm sp}}{10^{45}\,\mathrm{erg\,s^{-1}}} \right)^{-1}
\left( \frac{E_{\rm sn}}{10^{51}\,\mathrm{erg}} \right).
\label{eq:first_peak_timescale_est1}
\end{equation}
which is satisfied when $L_{\rm sp}\lesssim 10^{47} \mathrm{erg\, s^{-1}}$ with SESNe conditions.

Conversely, for stronger engines with spin-down luminosities above this threshold, $t_{\rm first}> t_{\rm sp}$ will be seen. The energy input effectively saturates at the pulsar’s rotational energy. In this regime, it is sufficient to replace $E_{\rm kin,adm}$ in Eq.~\eqref{eq:first_peak_timescale} by the neutron-star rotational energy $E_{\rm rot}$, which yields
\begin{equation}
t_{\rm first}
\simeq
0.44~\mathrm{day}\,
\left( \frac{M_{\rm ej}}{10\,M_\odot} \right)^{3/4}
\left( \frac{E_{\rm rot}}{10^{52}\,\mathrm{erg}} \right)^{-1/4}
\left( \frac{\kappa}{0.1\,\mathrm{cm^2\,g^{-1}}} \right)^{1/2} 
\label{eq:first_peak_timescale_est2}
\end{equation}

All numerical evaluations above have assumed a thickness factor $\zeta_r\simeq 20$, which is just a typical in our simulations and broadly consistent with values directly inferred from \citet{Suzuki&Maeda2019,Suzuki&Maeda2021}. We note, however, that $\zeta_r$ itself depends on $L_{\rm sp}$: for very large spin-down luminosities, the earlier blowout allows the admixture layer to broaden more while it is still optically thick, which can increase $\zeta_r$ and thereby shorten $t_{\rm first}$.

Another quantity of interest is the luminosity of the first peak. Since we focus on cases where the radiative efficiency before blowout is low \citep{Chevalier1992}, the relevant emission timescale is $t_{\rm bro}$, and the radiated energy $E_{\rm rel}$ during the first peak is of the same order as the kinetic energy previously stored in the admixture layer, $E_{\rm rel}\sim E_{\rm kin,adm}$. Thus, to order of magnitude,
\begin{equation}
 L_{\rm first}\sim \frac{E_{\rm kin,adm}}{t_{\rm bro}}.
\end{equation}

For relatively low spin-down luminosities, where $t_{\rm first}\simeq t_{\rm bo}>t_{\rm bro}$, the injected energy prior to the first peak is approximately the blowout energy $E_{\rm bo}$, so that $E_{\rm kin,adm}\sim t_{\rm bo}L_{\rm sp}\sim \zeta_{\rm bo}E_{\rm sn}$. This yields
\begin{equation}
\begin{aligned}
 L_{\rm first}\simeq \frac{t_{\rm bo}\,L_{\rm sp}}{t_{\rm bro}}
  =\frac{\zeta_{\rm bo}\,E_{\rm sn}}{t_{\rm bro}}
& =
  \left[
  \frac{
    1600\,\pi^2\,\zeta_{\rm bo}^5\,E_{\rm sn}^5\,c^2\,\zeta_r^4\,(\zeta_r^2-1)
  }{
    81\,M_{\rm ej}^3\,\kappa^2\,\ln\zeta_r
  }
\right]^{1/4}
\\
& =3.92\times10^{46} \,\mathrm{erg\, s^{-1}}\,
\left( \frac{M_{\rm ej}}{10\,M_\odot} \right)^{-3/4}
\left( \frac{E_{\rm sn}}{10^{51}\,\mathrm{erg}} \right)^{5/4},
\end{aligned}
\label{eq:first_peak_Lbol_1}
\end{equation}
so that, in this regime, the first-peak luminosity is primarily controlled by the explosion energy.

In contrast, for high spin-down luminosities the injected energy saturates at the pulsar rotational energy, $E_{\rm kin,adm}\sim E_{\rm rot}$, and we obtain
\begin{equation}
\begin{aligned}
 L_{\rm first}\simeq \frac{E_{\rm rot}}{t_{\rm bro}}
& =
  \left[
  \frac{
    1600\,\pi^2\,E_{\rm rot}^5\,c^2\,\zeta_r^4\,(\zeta_r^2-1)
  }{
    81\,M_{\rm ej}^3\,\kappa^2\,\ln\zeta_r
  }
\right]^{1/4}
\\
& =2.61\times10^{47} \,\mathrm{erg\, s^{-1}}\,
\left( \frac{M_{\rm ej}}{10\,M_\odot} \right)^{-3/4}
\left( \frac{E_{\rm rot}}{10^{52}\,\mathrm{erg}} \right)^{5/4}.
\end{aligned}
\label{eq:first_peak_Lbol_2}
\end{equation}
These relations indicate that, once a pulsar-driven blowout is realized, the bolometric luminosity of the first peak is only weakly sensitive to, though not completely independent of, the detailed value of $L_{\rm sp}$. Instead, it asymptotes to two limiting regimes across the critical spin-down luminosity: at low $L_{\rm sp}$ the first-peak luminosity is set mainly by the initial explosion energy $E_{\rm sn}$, whereas at high $L_{\rm sp}$ it is governed by the rotational energy reservoir $E_{\rm rot}$ of the pulsar.

With estimates for the admixture-layer outer radius at the first peak (Eq.~\eqref{eq:R_out_1st_peak_estimation}) and the corresponding bolometric luminosity (Eqs.~\eqref{eq:first_peak_Lbol_1}--\eqref{eq:first_peak_Lbol_2}), we can obtain a compact prediction for the characteristic color temperature of the first peak. By construction, the first maximum occurs when the diffusion front has just entered the admixture layer, i.e. when $R_{\rm diff}$ approaches the outer edge. At this epoch the ejecta exterior to the diffusion front is already close to optically thin, so the key radii are well approximated by $R_{\rm diff}\sim R_{\rm ph}\sim R_{\rm out}$. We therefore estimate the first-peak effective temperature using $R_{\rm out}(t_{\rm first})$ as the photospheric scale,

\begin{equation}
T_{\rm eff,1}\;\equiv\;
\left(\frac{L_{\rm first}}{4\pi \sigma R_{\rm out}^2(t_{\rm first})}\right)^{1/4}
\simeq
1.61\times 10^{5}\ {\rm K}\;
\left(\frac{E_{\rm kin,adm}}{10^{51}\ {\rm erg}}\right)^{3/16}
\left(\frac{M_{\rm ej}}{10\,M_\odot}\right)^{-5/16}
\left(\frac{\kappa}{0.1\ {\rm cm^{2}\,g^{-1}}}\right)^{-3/8},
\label{eq:Teff_first_numeric}
\end{equation}

Eq.~\eqref{eq:Teff_first_numeric} implies $T_{\rm eff,1}\sim 10^{5}\,$K for fiducial SESNe parameters, thus, the first peak should therefore be predominantly UV, and any contemporaneous optical emission is expected to be extremely blue. In the most extreme cases the optical bands probe the Rayleigh--Jeans tail of the spectrum, naturally yielding a comparatively muted or even visually inconspicuous first peak in optical light curves despite a large bolometric output.

As for the second, main peak, this corresponds to the light curve maximum in pulsar-powered SNe \citep[e.g.,][]{Kashiyama2016}, which has been studied extensively and will not be discussed further here.

\section{Results}
\label{sec:calculation_results}

\subsection{Blowouts in SESNe}
\label{subsec:results-SESNe}

Using the blowout model developed in Section~\ref{sec:blowout_model}, we now present representative calculation results and discuss their observational implications. We start from the SESNe regime, where double-peaked SLSNe-I may be viewed as an extreme subset, therefore illustrate the main features of our model and its primary target population.

\subsubsection{A Fiducial SESNe Blowout}
\label{subsubsec:SESN-fiducial}

As a representative setup, we consider Type Ib/c–like ejecta with $M_{\rm ej}=10\,M_\odot$, a transition radius $R_{\rm tr}\simeq 4.5\times10^{12}\ {\rm cm}$ at initialization, and an explosion energy $E_{\rm sn}=10^{51}\ {\rm erg}$, comparable to the parameters adopted by \citet{Kasen2016} and \citet{Suzuki&Maeda2019, Suzuki&Maeda2021}. A newborn pulsar with $B_{\rm dip}=1\times10^{14}\ {\rm G}$ and $P_{\rm NS}=2\ {\rm ms}$ acts as the central engine, providing a rotational-energy reservoir $E_{\rm rot}\simeq 5.5\times10^{51}\ {\rm erg}$, a substantial fraction of which is deposited over $t_{\rm sp}\sim10^{5}\ {\rm s}$. For the adopted density slopes ($\delta=1$, $n=10$), the blowout condition $E_{\rm bo}=\zeta_{\rm bo}E_{\rm sn}$ with $\zeta_{\rm bo}\simeq 2.19$ is satisfied. The artificial parameter $\eta_{S}$ and the initial relative thickness $\Delta R/R_{\rm in}$ are set to 1 and 0.1, respectively, in the following calculations unless otherwise stated. Unless explicitly specified, we adopt gray scattering and absorptive opacities of $\kappa_{\rm s}=\kappa_{\rm abs}=0.1\,\mathrm{cm^2\,g^{-1}}$.

\begin{figure}[ht!]
  \centering
  \begin{subfigure}{0.43\textwidth}
    \centering
    \includegraphics[width=\linewidth]{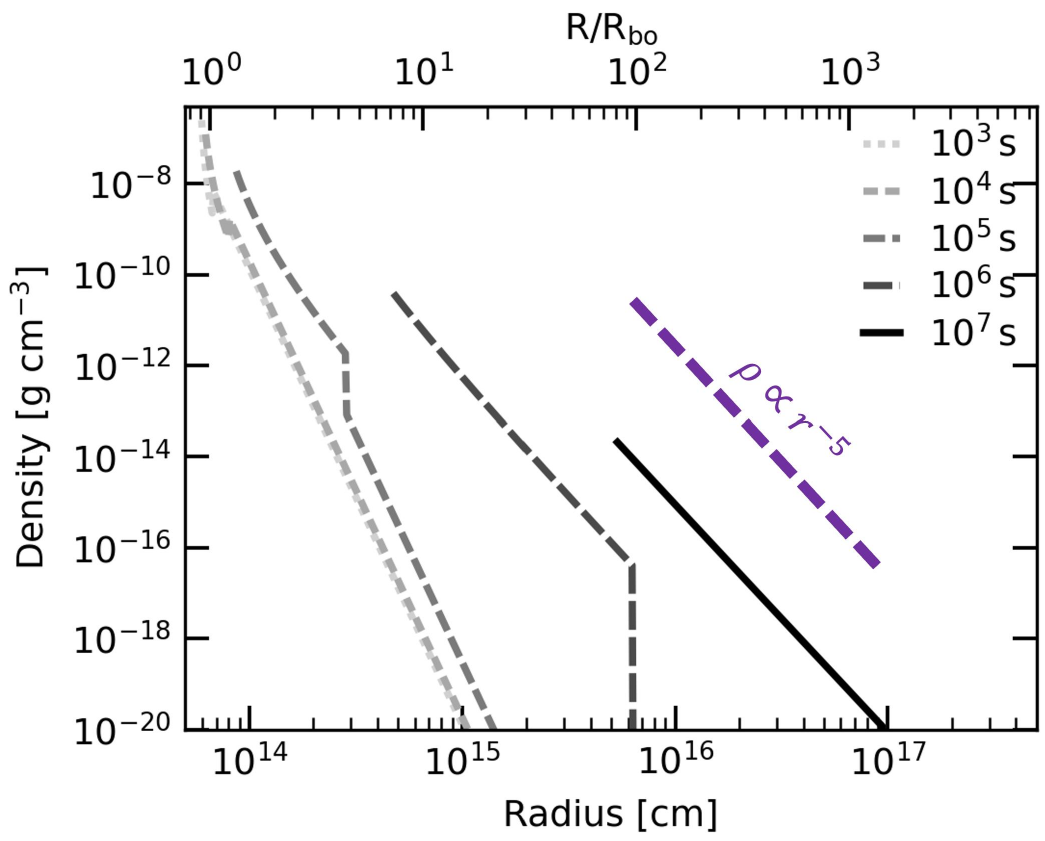}
  \end{subfigure}
  \begin{subfigure}{0.43\textwidth}
    \centering
    \includegraphics[width=\linewidth]{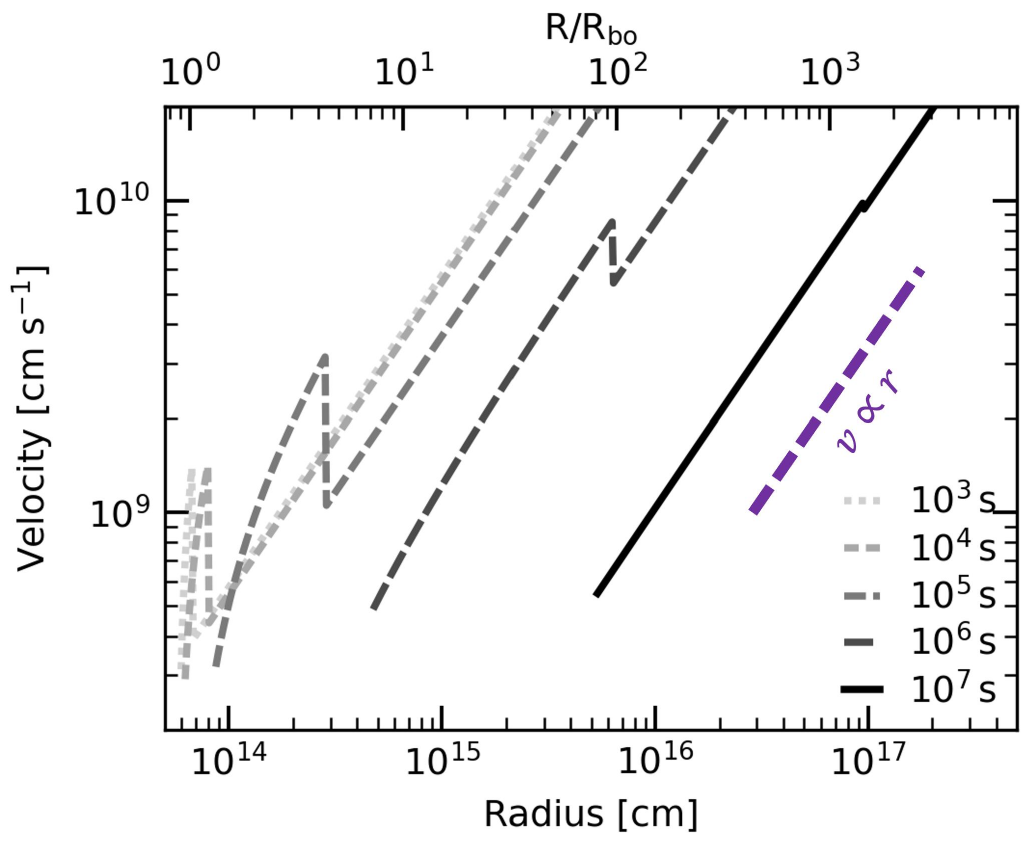}
  \end{subfigure}
  \begin{subfigure}{0.43\textwidth}
    \centering
    \includegraphics[width=\linewidth]{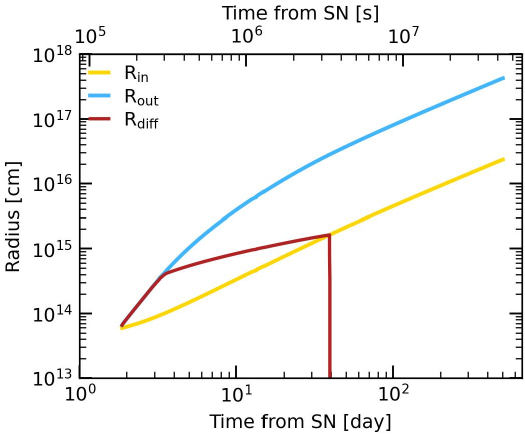}
  \end{subfigure}
  \caption{
    Post-blowout evolution of the density, expansion-velocity, and characteristic radii in a representative SESN blowout scenario. Profile curves are labeled by the time elapsed since blowout. Purple dashed lines show reference scalings $\rho\propto r^{-5}$ and $v\propto r$. At blowout, the swept-up ejecta form a thin shell and the inner profiles deviate markedly from homologous expansion. The shell then rapidly broadens, the density relaxes toward $\rho\propto r^{-5}$, and the velocity field approaches a nearly homologous profile. The right panel shows the time evolution of the inner and outer boundaries of the admixture layer, $R_{\rm in}$ and $R_{\rm out}$, together with the diffusion front $R_{\rm diff}$. For this case, $R_{\rm diff}$ lies close to $R_{\rm out}$ immediately after blowout but remains outside the dense admixture layer. Once the diffusion front penetrates into $R_{\rm out}$, the growth of its relative thickness slows, and the thickness factor $\zeta_{\rm r} = R_{\rm out}/R_{\rm in}$ asymptotes to $\simeq 20$.
  }
  \label{fig:SESN_fid_structural}
\end{figure}

\begin{figure}[ht!]
 \centering
  \begin{subfigure}{0.45\textwidth}
    \centering
    \includegraphics[width=\linewidth]{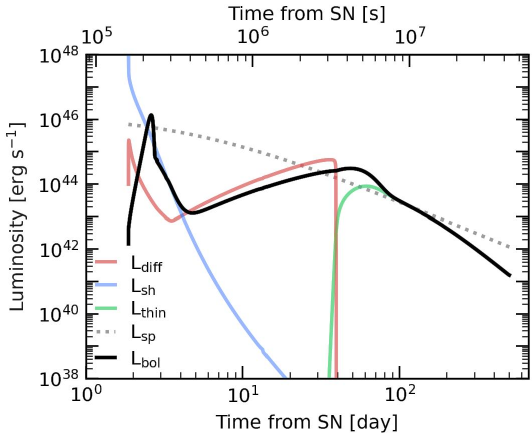}
  \end{subfigure}
  \begin{subfigure}{0.43\textwidth}
    \centering
    \includegraphics[width=\linewidth]{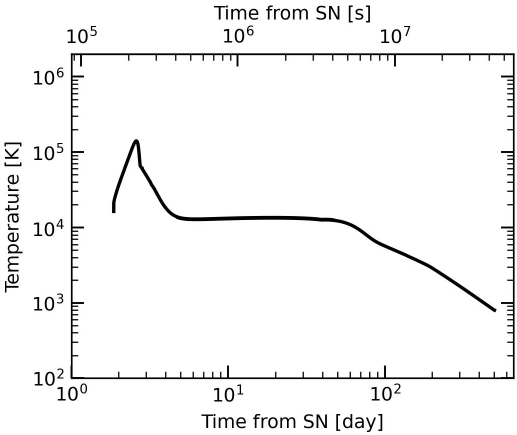}
  \end{subfigure}
  \caption{
    Bolometric luminosity (left) and graybody temperature (right) for the fiducial pulsar-driven SESN case with $M_{\rm ej}=10\,M_\odot$, $E_{\rm sn}=10^{51}\ {\rm erg}$, $B_{\rm dip}=1\times10^{14}\ {\rm G}$, and $P_{\rm NS}=2\ {\rm ms}$. In the left panel, we also show the individual contributions from shock heating at leading edge of the admixture layer ($L_{\rm sh}$), photon diffusive leakage through $R_{\rm out}$ or from $R_{\rm diff}$ ($L_{\rm diff}$), graybody emission from previously $R_{\rm diff}$-swept, optically thin ejecta ($L_{\rm thin}$), and the pulsar spin-down power ($L_{\rm sp}$). The double-peaked morphology arises from shock-powered and leakage-powered emission at early times (first peak), followed by the thermal-diffusion–dominated release of the residual internal energy (second, main peak).
  }
  \label{fig:SESN_fid_L_T}
\end{figure}

The corresponding structural evolution is shown in Fig.~\ref{fig:SESN_fid_structural}. Shortly after blowout, the swept-up bubble shell is geometrically thin 
and the inner region associated with the admixture layer still shows a pronounced deviation from homologous expansion, 
with both density and velocity exhibiting steep inner gradients. During this phase, thermal pressure in the pulsar wind bubble drives rapid thickening of the admixture layer and pushes its leading edge into the outer ejecta. Within a short time, this transient relaxes: the layer broadens, the density approaches a power law close to $\rho\propto r^{-5}$, and the velocity field tends toward a nearly homologous profile $v\propto r$. This evolution is consistent with the picture in which a thin shell at blowout gradually turns into a broadened admixture layer that co-expands with the outer ejecta. The key structural variables, in particular $R_{\rm in}$, $R_{\rm out}$, and the diffusion front radius $R_{\rm diff}$, fix the structure of the admixture layer and the locations of shock dissipation and photon escape, and thereby determine the bolometric luminosity and color-temperature evolution shown in Fig.~\ref{fig:SESN_fid_L_T}.

The resulting bolometric light curve indeed exhibits a clear double-peaked structure due to the blowout (Fig.~\ref{fig:SESN_fid_L_T}). The first peak is powered jointly by shock heating, $L_{\rm sh}$, and radiative leakage through the outer boundary of the admixture layer, $L_{\rm diff}$, with $L_{\rm sh}$ providing the dominant contribution. During this stage, the diffusion front $R_{\rm diff}$ lies just outside $R_{\rm out}$, so leakage is already appreciable. The admixture layer broadens most rapidly at this time: $p\,dV$ work efficiently converts internal energy into kinetic energy, causing both the internal-energy content and $L_{\rm diff}$ to decline quickly. This trend is reversed only once $R_{\rm diff}$ penetrates into the admixture layer at $t\gtrsim 2\times10^{5}\ {\rm s}$, which marks the end of the first peak.

As the ejecta continue to expand and dilute, the luminosity rises again. Around $\sim 40$ days after explosion, the diffusion front reaches $R_{\rm in}$ and disappears; by then the ejecta are globally close to transparent, and photon diffusion rather than advection by the flow fully controls the emergent radiation. Although the internal-energy density is highest near $R_{\rm in}$ and the localized $L_{\rm diff}$ reaches a maximum when $R_{\rm diff}$ arrives there, photons must still undergo scatterings in the overlying material before escaping. As a result, the second, main peak in the light curve lags slightly behind the disappearance of $R_{\rm diff}$, though the offset is modest.

The component $L_{\rm thin}$ denotes graybody emission from the optically thin region already swept by the diffusion front, where pulsar power has been absorbed and reprocessed. Thermalization is strongest near the density maximum at $R_{\rm in}$, so $L_{\rm thin}$ remains sub-dominant until that region becomes transparent. It then grows rapidly as the system transitions into the PWN phase. 

Motivated by observations that such SNe can be well described by blackbody emission, at least up to and around the main peak \citep[e.g.,][]{Leloudas2012,Nicholl2016,Moriya2018}, we use the bolometric luminosity and graybody temperature derived above to compute synthetic light curves in realistic survey bandpasses. In our fiducial SESN, the photospheric temperature remains above $\sim10^{4}\,$K for an extended period and reaches $\gtrsim10^{5}\,$K near the first peak, indicating that pulsar-driven SNe with wind-bubble blowout could, in some cases, produce detectable UV and soft X-ray emission. For additional illustrative cases and band-specific predictions, we refer to Section~\ref{subsubsec:SESN-diverse}.

\subsubsection{LSQ14bdq}
\label{subsubsec:LSQ14bdq}

Having completed a detailed case study of blowout in a representative blowout in pulsar–driven SESN, we now apply the same framework to an observed event. Specifically, we fit and interpret the well-studied, double-peaked SLSN LSQ14bdq using our model. LSQ14bdq \citep{Nicholl2015,Kasen2016} is widely treated as a benchmark hydrogen-poor SLSN-I with a likely central-engine origin: its slowly declining main peak and spectra closely resemble those of other SLSNe~I that are successfully modeled with magnetar spin-down, whereas its bright, relatively fast precursor bump (lasting $\sim 2$ weeks) is too luminous and narrow to be powered by ${}^{56}$Ni and is difficult to reproduce with simple ejecta–CSM interaction alone \citep[e.g.][]{Nicholl2015,Nicholl2015b,Nicholl2016}. In this sense, a highly magnetized fast rotating pulsar engine is often regarded as the most natural explanation for LSQ14bdq and similar double-peaked SLSNe, although fallback accretion onto a compact remnant and more complex CSM or ${}^{56}$Ni distributions have also been discussed as alternative power sources \citep[e.g.][]{Piro2011,Kasen2016,Orellana2022}.

Below we apply our blowout model to multi-band light curves of LSQ14bdq. In addition to the La Silla QUEST photometry converted to an SDSS $r$-like filter, we include the nearly contemporaneous $g$, $i$, and $z$-band measurements reported by \citet{Nicholl2021} in the fit. 

In the blowout framework, the two peaks are not fitted independently but are coupled by a small set of physical constraints that largely break parameter degeneracies. The main peak fixes the ejecta scale through $M_{\rm ej}$ and the total energy available up to peak, $E_{\rm tot}\simeq E_{\rm sn}+E_{\rm rot}$ \citep{Kashiyama2016}. The color evolution around the main peak provides an additional handle on the engine characteristics, because $E_{\rm rot}$ and $L_{\rm sp}$ imprint on $T_{\rm eff}(t)$ (e.g., Fig.~\ref{fig:SLSN_2nd_peak}; \citet{Kashiyama2016}). The first peak then constrains the early engine through the combination $E_{\rm sn}/L_{\rm sp}$ that sets its characteristic timescale and luminosity as described in Section~\ref{subsubsec:observable_signatures}, while the solution must also lie in the blowout-enabled region of parameter space.

\begin{figure}[ht!]
\epsscale{0.6}
\plotone{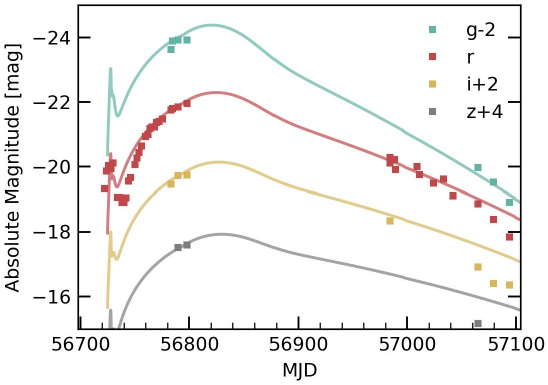}
\caption{Multi-band observational data and fitting light curves of LSQ14bdq using our blowout model.}
\label{fig:LSQ14bdq-Fitting}
\end{figure}

The resulting best-fit model is shown in Fig.~\ref{fig:LSQ14bdq-Fitting}, obtained for

$M_{\rm ej}\simeq 20\,M_{\odot}$,
$E_{\rm sn}\simeq 1\times 10^{51}\,{\rm erg}$,
$B_{\rm dip}\simeq 3.69\times 10^{13}\,{\rm G}$,
and $P_{\rm NS}\simeq 1.77\,{\rm ms}$. 
These parameters are broadly consistent with the data. The model first peak is modestly faster and brighter than observed, plausibly due to the $\sim$2-day cadence \citep{Nicholl2015} that could have missed a brief, brighter maximum, and because the inferred engine lies close to the marginal blowout boundary (Section~\ref{subsec:marginal_blowout}), where the assumptions underlying our initial conditions become marginal while the same underlying physical picture still applies.

\subsubsection{Diverse Blowout Signatures in SESNe}
\label{subsubsec:SESN-diverse}

\begin{figure}[ht!]
 \centering
  \begin{subfigure}{0.8\textwidth}
    \centering
    \includegraphics[width=\linewidth]{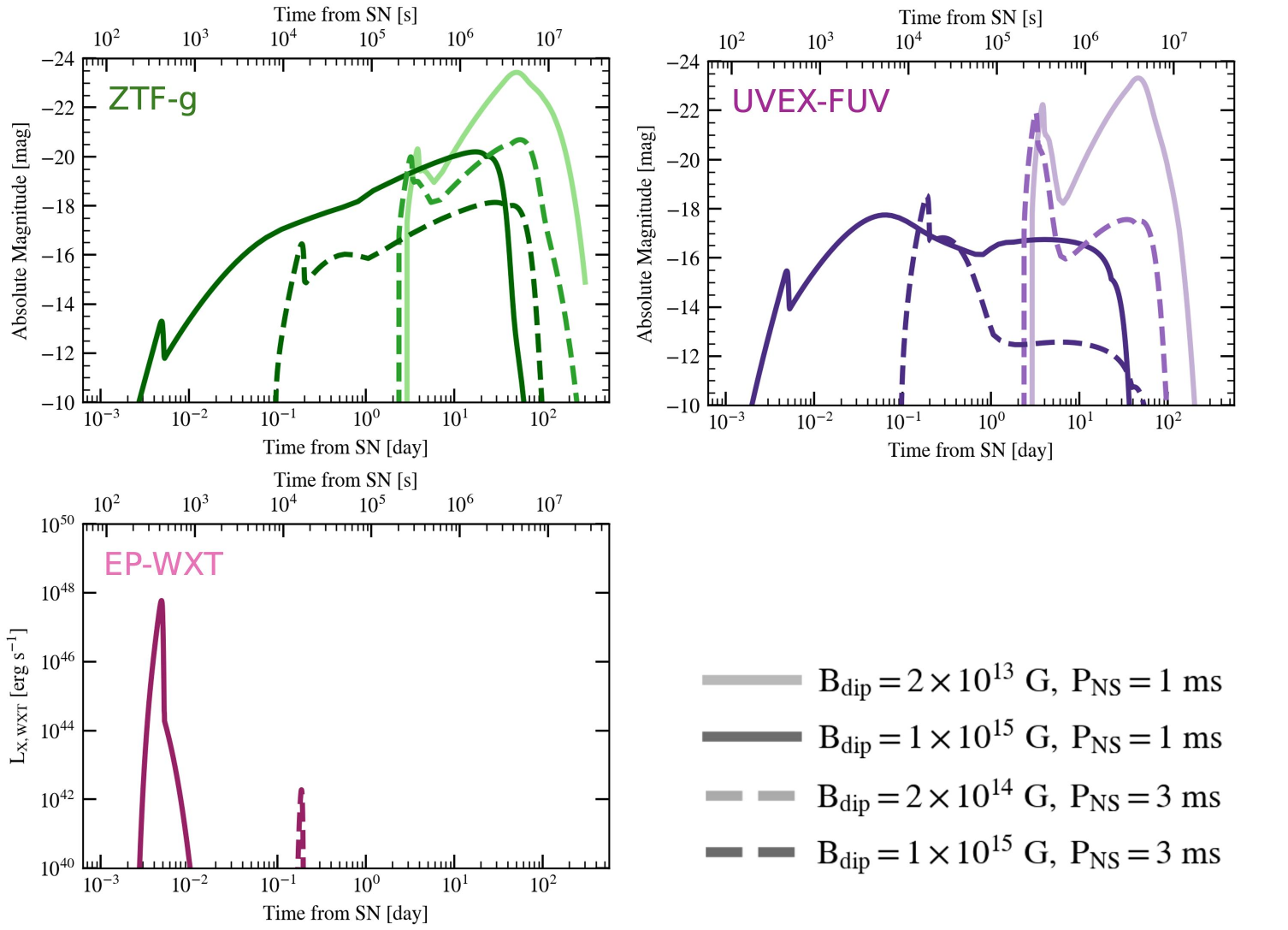}
  \end{subfigure}
  \caption{
    Survey-band light curves for four illustrative parameter choices that permit blowout in SESNe within our framework, spanning corner points in $(L_{\rm sp},E_{\rm rot})$. From left to right we show the ZTF $g$ band, the UVEX FUV band, and the EP-WXT soft X-ray band. The optical and UV light curves exhibit a double-peaked morphology, although the first peak can be faint and fast and may be missed in survey data. High-$L_{\rm sp}$ cases can also show a short-lived but luminous soft X-ray counterpart.
    }
  \label{fig:SESN_survey}
\end{figure}

We now examine how variations in the pulsar engine parameters imprint diverse signatures on SESNe, fixing the ejecta mass at $M_{\rm ej}=10\,M_\odot$ and the explosion energy at $E_{\rm sn}=10^{51}\,{\rm erg}$. 
To provide a more representative view beyond a single case, Fig.~\ref{fig:SESN_survey} presents survey-ready light curves for four corner points of the blowout-enabled parameter space with the considering SESNe condition. These models span the combinations of high/low spin-down power and high/low rotational energy: (i) high $L_{\rm sp}$ and high $E_{\rm rot}$ ($B_{\rm dip}=10^{15}\,{\rm G}$, $P_{\rm NS}=1\,{\rm ms}$), (ii) low $L_{\rm sp}$ and high $E_{\rm rot}$ ($B_{\rm dip}=2\times10^{13}\,{\rm G}$, $P_{\rm NS}=1\,{\rm ms}$), (iii) high $L_{\rm sp}$ and low $E_{\rm rot}$ ($B_{\rm dip}=10^{15}\,{\rm G}$, $P_{\rm NS}=3\,{\rm ms}$), and (iv) low $L_{\rm sp}$ and low $E_{\rm rot}$ ($B_{\rm dip}=2\times10^{14}\,{\rm G}$, $P_{\rm NS}=3\,{\rm ms}$). 
As shown in the figures, the optical (ZTF) and UV (UVEX) bands display clear double-peaked morphologies in most samples, except in the UV for the $B_{\rm dip}=10^{15},$G, $P_{\rm NS}=3,$ms case; as we will show below in Fig.~\ref{fig:SLSN_2nd_peak}, this high-$B_{\rm dip}$, short-$P_{\rm NS}$ engine is more hypernova-like, with a cooler main peak that can suppress the second UV peak.
In most cases, although the bolometric luminosity at the first peak is very high, the photospheric temperature is also high, placing the optical bands on the Rayleigh--Jeans tail of the spectral energy distribution (SED). Consequently, the first peak is typically far less prominent in the optical than in the UV. By contrast, models with lower $L_{\rm sp}$ are expected to have lower post-blowout graybody temperatures and therefore can display a more distinct optical double peak. In all cases, the early first peak evolves rapidly, so that at larger distances it is likely to be missed in purely optical searches. This in turn suggests that the intrinsic fraction of blowout events with double-peaked light curves among SLSNe-I may be higher than currently inferred from optical samples alone. Future UV-capable facilities such as UVEX could substantially increase the detection rate of such events and provide much richer diagnostic information.

In addition, in the soft X-ray band our model predicts that the high-$L_{\rm sp}$ cases can exhibit a short-lived counterpart near the time of the first peak, with a duration of $\sim 10^{2}$--$10^{4}\,{\rm s}$ that is, in principle, accessible to modern transient surveys \citep{Yuan2015, Cheng2025}. The two high-$L_{\rm sp}$ examples shown here, $B_{\rm dip}=10^{15}\,{\rm G}$ with $P_{\rm NS}=1\,{\rm ms}$ and $P_{\rm NS}=3\,{\rm ms}$, produce X-ray peaks as bright as $\sim 10^{48}\,{\rm erg\,s^{-1}}$ and $\sim 10^{42}\,{\rm erg\,s^{-1}}$, respectively. For the typical transient sensitivity of EP-WXT, $F_{\rm lim}\sim{\rm few}\times 10^{-11}\,{\rm erg\,cm^{-2}\,s^{-1}}$ \citep{Yuan2025}, these correspond to detection horizons of $\gtrsim 10\,{\rm Mpc}$. This implies that the hottest events, for which the first optical peak can be difficult to distinguish, may instead be accompanied by a luminous X-ray precursor, underscoring the value of coordinated multiwavelength searches that include X-rays. 

\begin{figure}[ht!]
  \centering
  \begin{subfigure}{0.45\textwidth}
    \centering
    \includegraphics[width=\linewidth]{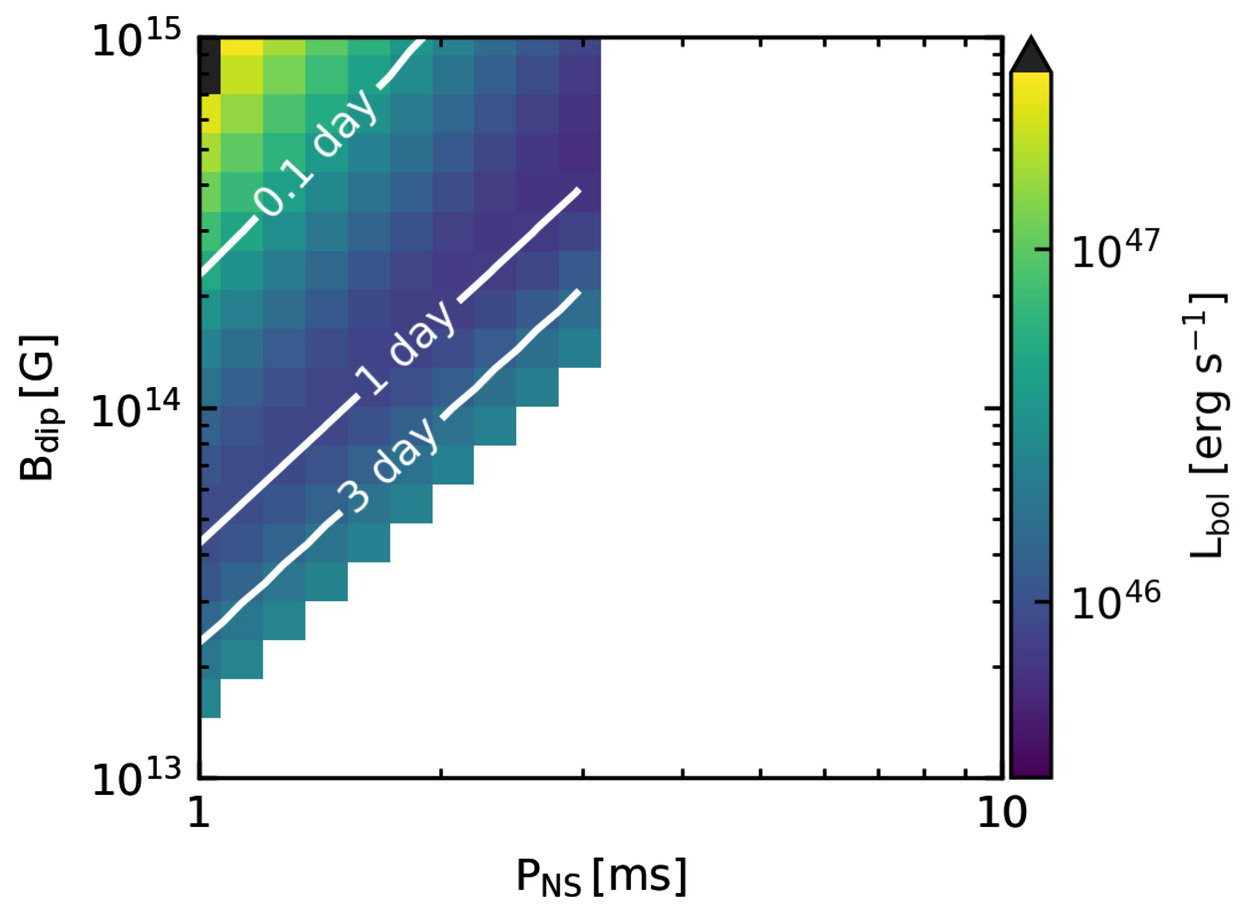}
  \end{subfigure}
  \begin{subfigure}{0.45\textwidth}
    \centering
    \includegraphics[width=\linewidth]{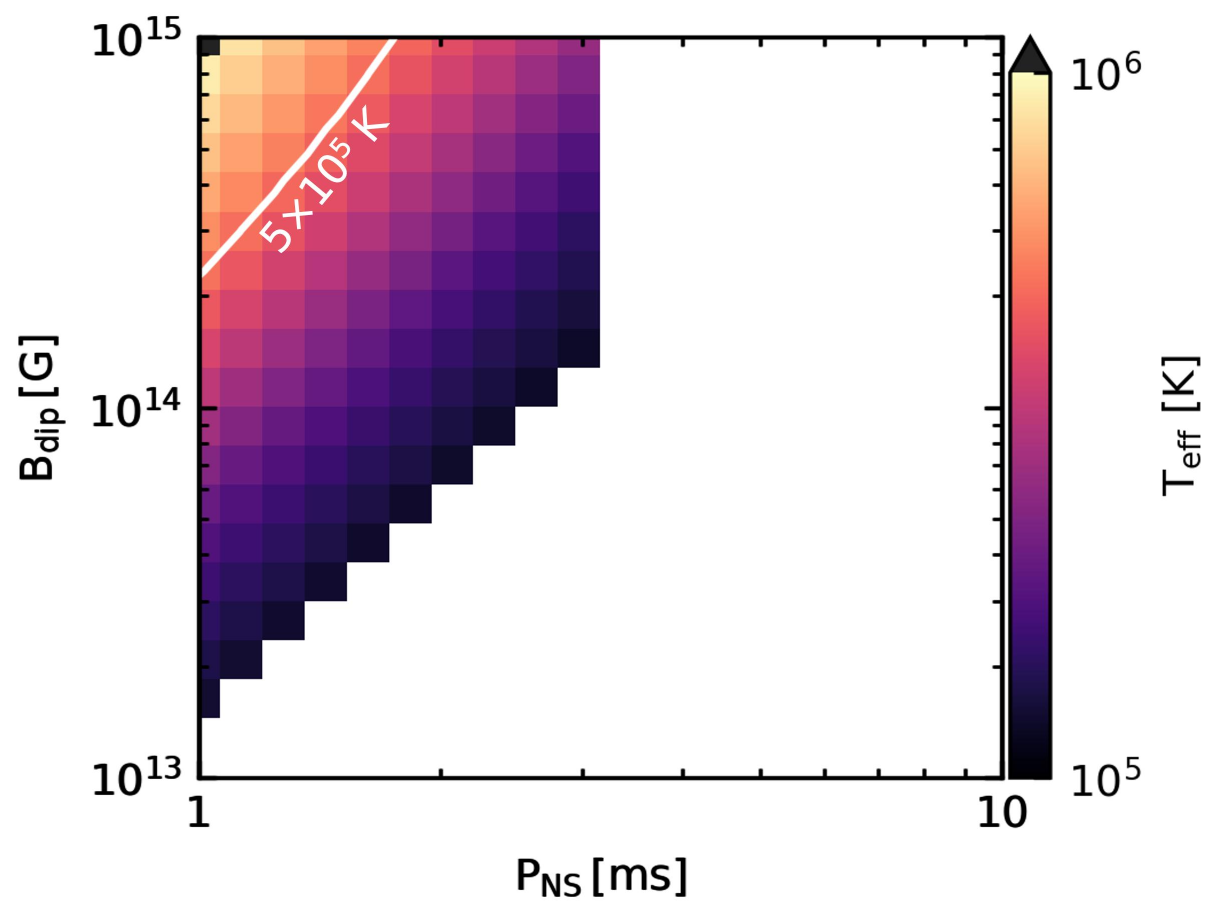}
  \end{subfigure}
  \caption{
    Bolometric first-peak properties for models with $M_{\rm ej}=10\,M_\odot$ and $E_{\rm sn}=10^{51}\,{\rm erg}$ and varying pulsar engine parameters. Colors denote the first-peak bolometric luminosity, and contours show the first-peak timescale in left panel, and the effective blackbody temperature information is shown in the right panel}.
  \label{fig:SESN_first_peaks_Lbol}
\end{figure}

A more systematic exploration of the diversity of these signals is presented below.
We first focus on the properties of the first peak. As a starting point, we investigate the bolometric luminosity and timescale of the first peak, as summarized in Fig.~\ref{fig:SESN_first_peaks_Lbol}.

The contours of the first-peak timescale in the $(B_{\rm dip},P_{\rm NS})$ plane approximately follow lines of constant $B_{\rm dip}P_{\rm NS}^{-2}$, i.e., they run nearly parallel to constant $L_{\rm sp}$ curves. From the $t_{\rm first}\simeq 0.1$ and $1$~day contours, we find that increasing $B_{\rm dip}$ by a factor of five shortens $t_{\rm first}$ by roughly an order of magnitude, corresponding to an effective scaling $t_{\rm first}\propto L_{\rm sp}^{-0.7}$. This behavior is broadly consistent with the analytic estimates in Eqs.~(\ref{eq:first_peak_timescale_est1},\ref{eq:first_peak_timescale_est2}), especially once we account for the tendency of higher $L_{\rm sp}$ to increase the thickness factor $\zeta_r$ of the admixture layer and for the broadening timescale $t_{\rm bro}$, both of which slightly harden the dependence relative to the idealized scalings.

\begin{figure}[ht!]
  \centering
  \begin{subfigure}{0.43\textwidth}
    \centering
    \includegraphics[width=\linewidth]{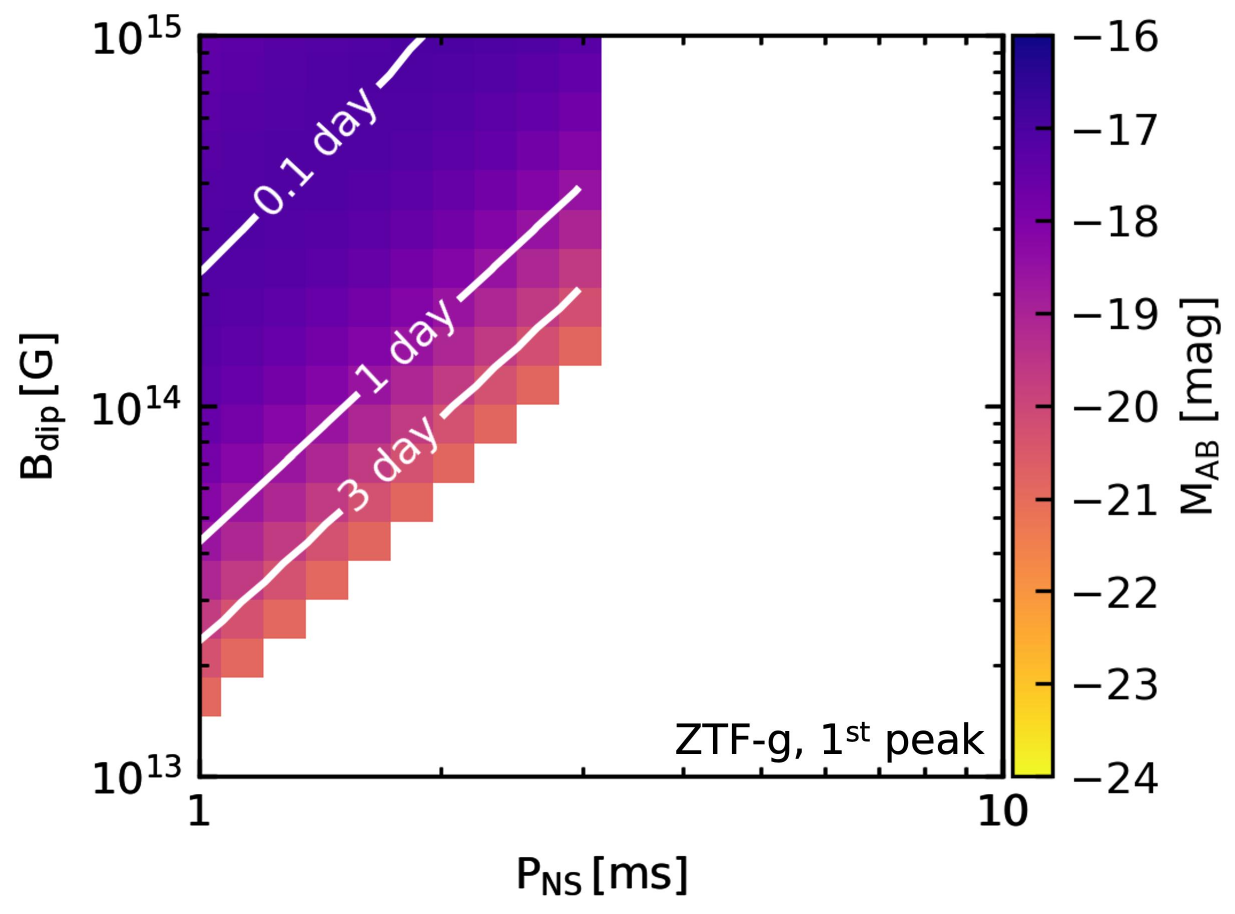}
  \end{subfigure}
  \begin{subfigure}{0.43\textwidth}
    \centering
    \includegraphics[width=\linewidth]{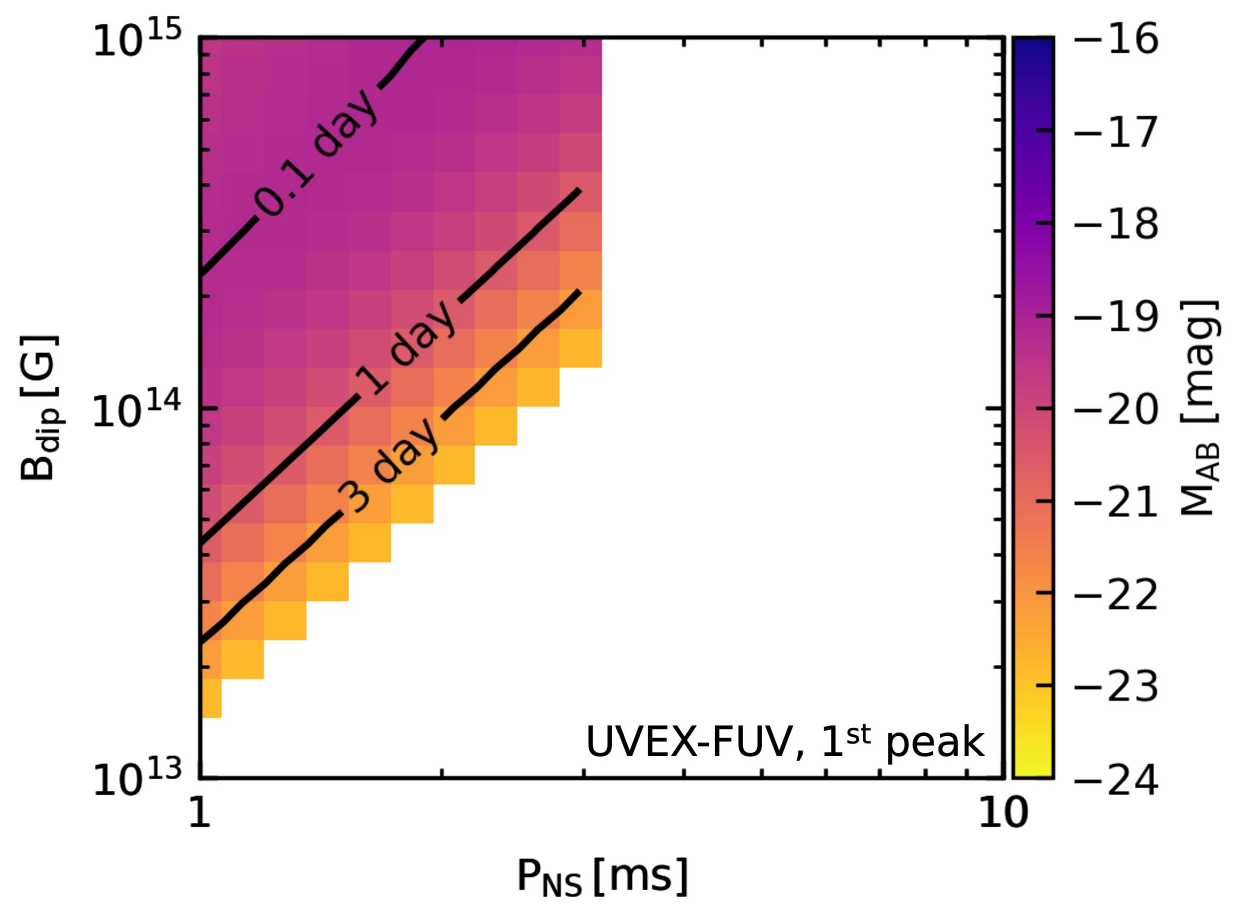}
  \end{subfigure}
  \begin{subfigure}{0.43\textwidth}
    \centering
    \includegraphics[width=\linewidth]{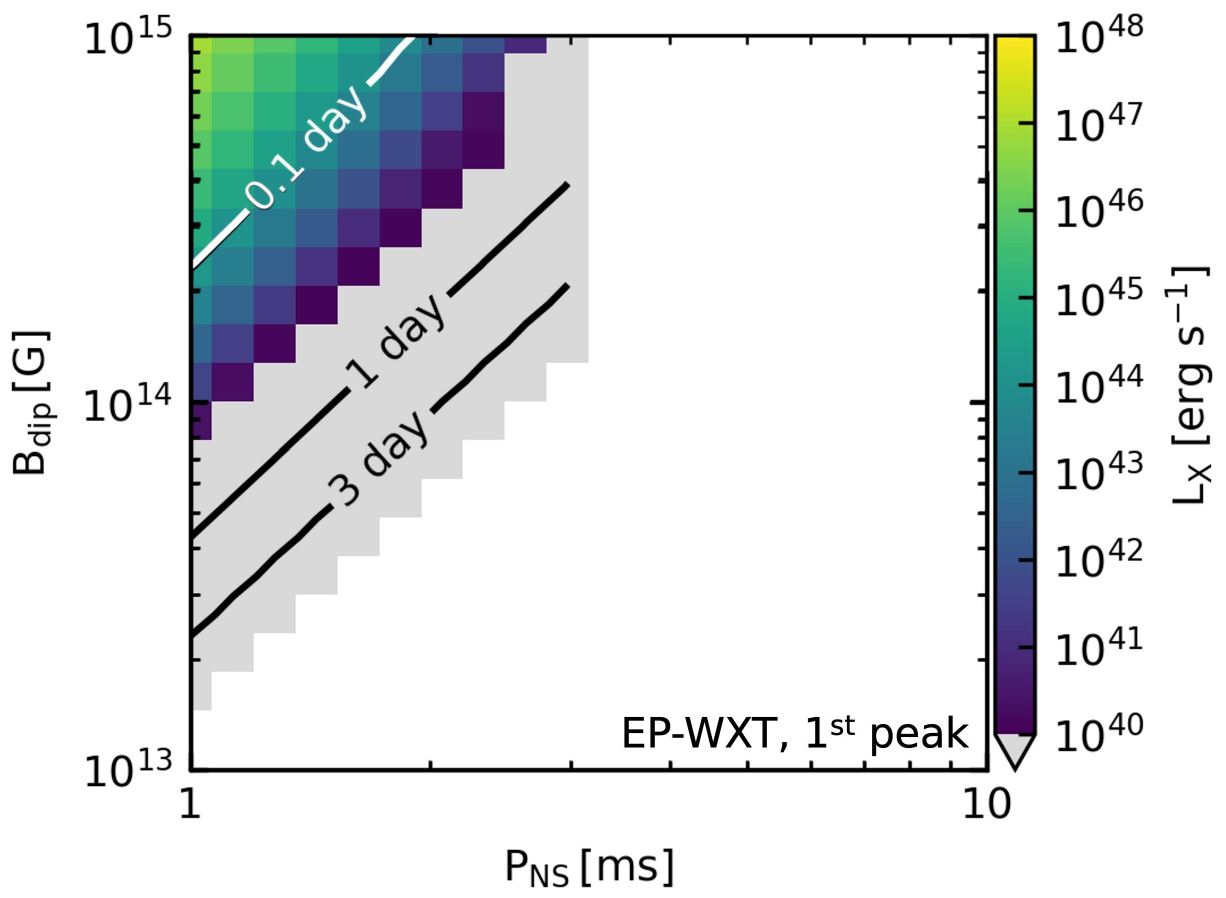}
  \end{subfigure}
  \caption{
    First-peak properties in the ZTF $g$, UVEX FUV, and EP--WXT bands for models with fixed ejecta mass and explosion energy ($M_{\rm ej}=10\,M_\odot$, $E_{\rm sn}=10^{51}\,\mathrm{erg}$) and varying pulsar engine parameters. In each panel, colors indicate the predicted peak luminosity (or flux) in the corresponding band, and contours show the time of maximum in that band.
  }
  \label{fig:SLSN_1st_peak}
\end{figure}

The first-peak bolometric luminosity varies only slowly with $L_{\rm sp}$ in the low–$L_{\rm sp}$ regime, remaining near $\sim 10^{46}\ {\rm erg\,s^{-1}}$. At higher $L_{\rm sp}$, corresponding to $E_{\rm rot}\sim 10^{52}\ {\rm erg}$ (or $P_{\rm NS}\sim 1\ {\rm ms}$), the first-peak luminosity can reach a few $\times 10^{47}\ {\rm erg\,s^{-1}}$, in agreement with the trends inferred from Eqs.~(\ref{eq:first_peak_Lbol_1},\ref{eq:first_peak_Lbol_2}). These results support the picture that, within the regime where our model applies, the interplay between the diffusion front $R_{\rm diff}$ and the outer boundary $R_{\rm out}$ indeed plays the dominant role in setting the properties of the first peak. To connect more directly with observables, we next consider the first-peak signatures in the optical (ZTF $g$), UV (UVEX FUV), and soft X-ray (EP--WXT) bands, as shown in Fig.~\ref{fig:SLSN_1st_peak}.

The diagnostics for the second, main peak are summarized in Fig.~\ref{fig:SLSN_2nd_peak}. We focus here on the ZTF $g$-band properties of the main peak.
\begin{figure}[ht!]
  \centering
  \begin{subfigure}{0.45\textwidth}
    \centering
    \includegraphics[width=\linewidth]{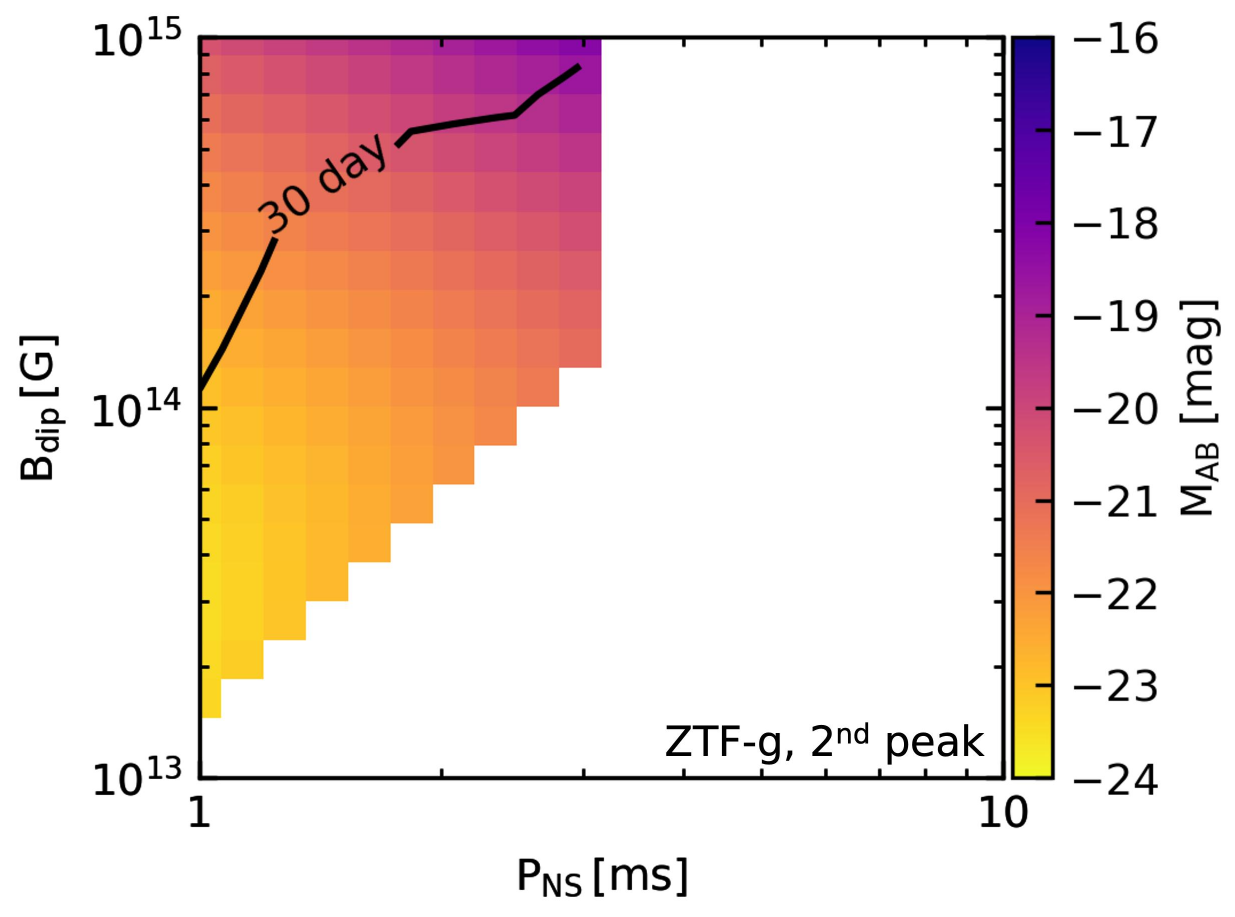}
  \end{subfigure}
  \begin{subfigure}{0.45\textwidth}
    \centering
    \includegraphics[width=\linewidth]{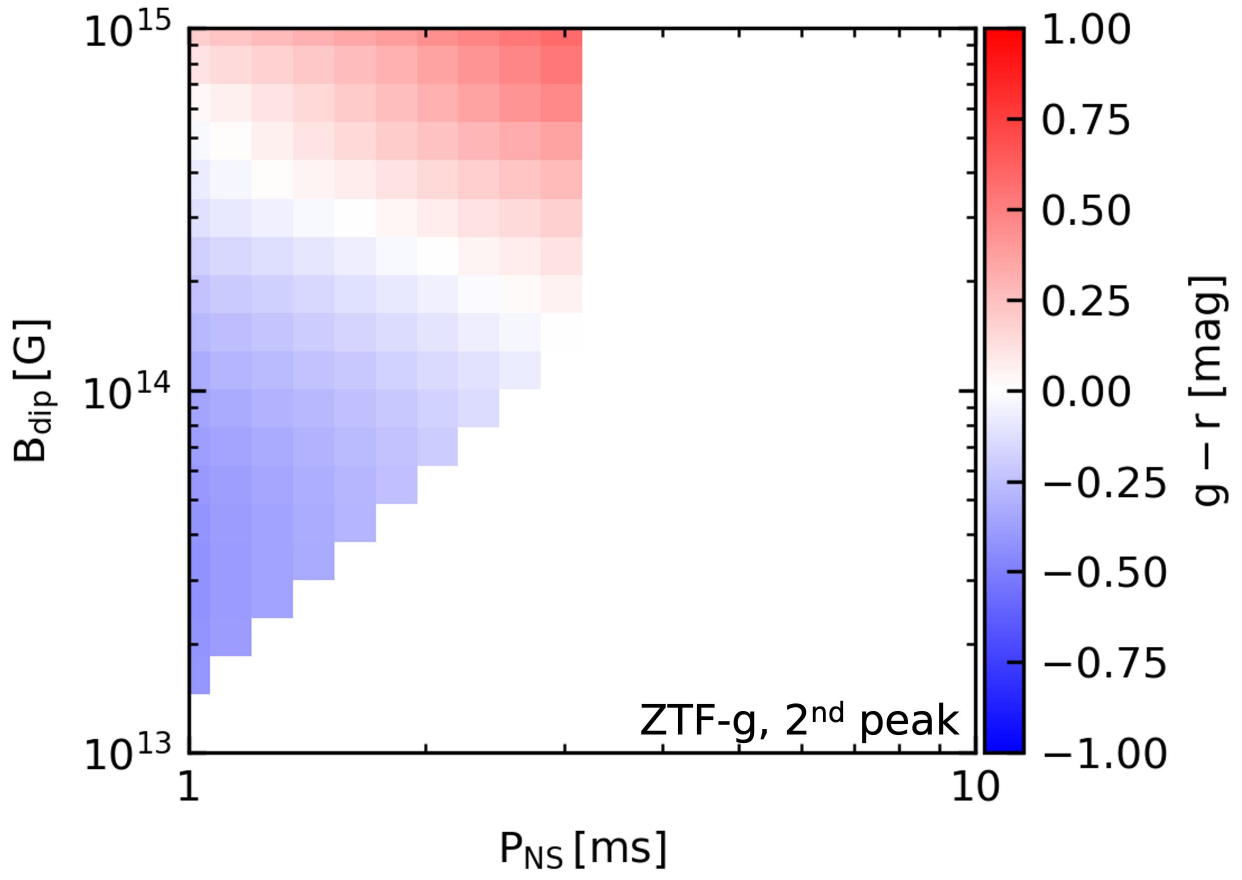}
  \end{subfigure}
  \caption{
    Second-peak properties in the ZTF $g$ band for models with varying pulsar engine parameters. Left: peak magnitude and time of maximum in $g$. Right: $g-r$ color at the $g$-band peak.
  }
  \label{fig:SLSN_2nd_peak}
\end{figure}
Comparing the optical-band results reveals two notable trends. First, although the bolometric first peak can be extremely luminous, its high temperature places the optical bands far from the SED maximum. As a result, the first peak in ZTF $g$ is typically up to $\sim 3$ mag fainter than the second, main peak. Combined with the short first-peak timescale (from sub-day to a few days) and the reduced effective time window at larger distances, this makes the first peak easy to miss in optical surveys. It is therefore plausible that the true fraction of SESNe with blowout and double-peaked light curves among SLSN-I–like events is higher than currently inferred from optical data alone, and that optical surveys may have a limited detection efficiency for the early first peak. However, our simulations predict that the UV first peak is generally comparable in brightness to the optical main peak. Future UV time-domain facilities may provide crucial new leverage, increasing the number of identified double-peaked SESNe and yielding a more complete view of their early-phase emission.

Second, a higher spin-down luminosity $L_{\rm sp}$ does not necessarily produce a bluer, brighter, and faster SN. Blowout SESNe can display both SLSN-like and hypernova-like characteristics. As seen in Fig.~\ref{fig:SLSN_2nd_peak}, contours of the $g$-band peak magnitude at $\sim 30$ days no longer simply follow lines of constant $L_{\rm sp}$. Models with a small rotational-energy reservoir (large $P_{\rm NS}$) but high $L_{\rm sp}$ (large $B_{\rm dip}$) tend to produce relatively faint ($M_g\gtrsim -20$ mag), redder main peaks, because the engine shuts off early while the ejecta are still relatively optically thick, so a larger fraction of the injected energy is converted into kinetic energy and the outcome resembles a hypernova. In contrast, models with a large $E_{\rm rot}$ (small $P_{\rm NS}$) but more modest $L_{\rm sp}$ (small $B_{\rm dip}$) inject energy over a longer thermalization timescale, allowing a larger fraction of the rotational energy to emerge directly as radiation. These events reach $M_g\lesssim -21$ mag and exhibit bluer colors, more closely resembling SLSN-I–like explosions, in qualitative agreement with previous studies such as \citet{Kashiyama2016}.

Soft X-rays provide an additional, complementary window, particularly in light of ongoing and planned high-cadence X-ray surveys such as Einstein Probe and EAGLE \citep{Yonetoku2025}. Following \citet{Frohmaier2021}, we adopt a local volumetric rate for SLSNe-I of $\mathcal{R}_{\rm SLSN-I}\sim 3\times10^{-8}\ {\rm Mpc^{-3}\,yr^{-1}}$. Requiring at least one detected blowout per year in an all-sky survey with duration $T$ and flux limit $F_{\rm WXT}$ in the EP--WXT band implies a minimum X-ray luminosity
\begin{equation}
L_{X,\min}
\approx
5\times10^{43}\ {\rm erg\,s^{-1}}\,
\left(\frac{\mathcal{R}_{\rm SLSN-I}}{3\times10^{-8}\ {\rm Mpc^{-3}\,yr^{-1}}}\right)^{-\frac{2}{3}}
\left(\frac{T}{1\ {\rm yr}}\right)^{-\frac{2}{3}}
\left(\frac{F_{\rm WXT}}{10^{-11}\ {\rm erg\,s^{-1}\,cm^{-2}}}\right).
\end{equation}
In our model, this threshold corresponds roughly to
$B_{\rm dip}P_{\rm NS}^{-2}\gtrsim 10^{14}\ {\rm G}\,(1\,{\rm ms})^{-2}$, thereby excluding a substantial fraction of the pulsar parameter space. X-ray surveys therefore have the potential to discover a valuable subset of the most energetic early peaks of SESNe and to constrain the upper end of the engine parameter space. In such cases, the events would appear as X-ray transients or precursors, but X-ray searches alone are unlikely to capture the bulk of double-peaked precursors.

\subsection{Blowouts in Ultra-stripped and AIC/MIC SNe}
\label{subsec:Results_USSNe_AICSNe}

In addition to canonical SESNe, ultra-stripped and AIC/MIC SNe constitute two further channels that warrant particular attention. Both arise from mass transfer onto, or mergers involving, compact objects in close binaries. Although rare, these events are important for forming double NS systems that merge within a Hubble time and for producing apparently young pulsars in old stellar populations. Observationally, only a small number of transients have been discussed as ultra-stripped SN candidates. iPTF14gqr \citep{De2018} remains the most convincing and best-observed case to date, while a few additional candidates have been reported in recent years \citep[e.g.,][]{Yao2020,Sawada2022,Agudo2023}. No AIC/MIC SN has yet been unambiguously identified. Because these explosions are expected to have relatively low kinetic energies, blowout is particularly easy to realize in such events (see Fig.~\ref{fig:blowout-phase-diagram}), making blowout signatures a promising diagnostic of these channels.

\begin{figure}[ht!]
  \centering
  \begin{subfigure}{0.8\textwidth}
    \centering
    \includegraphics[width=\linewidth]{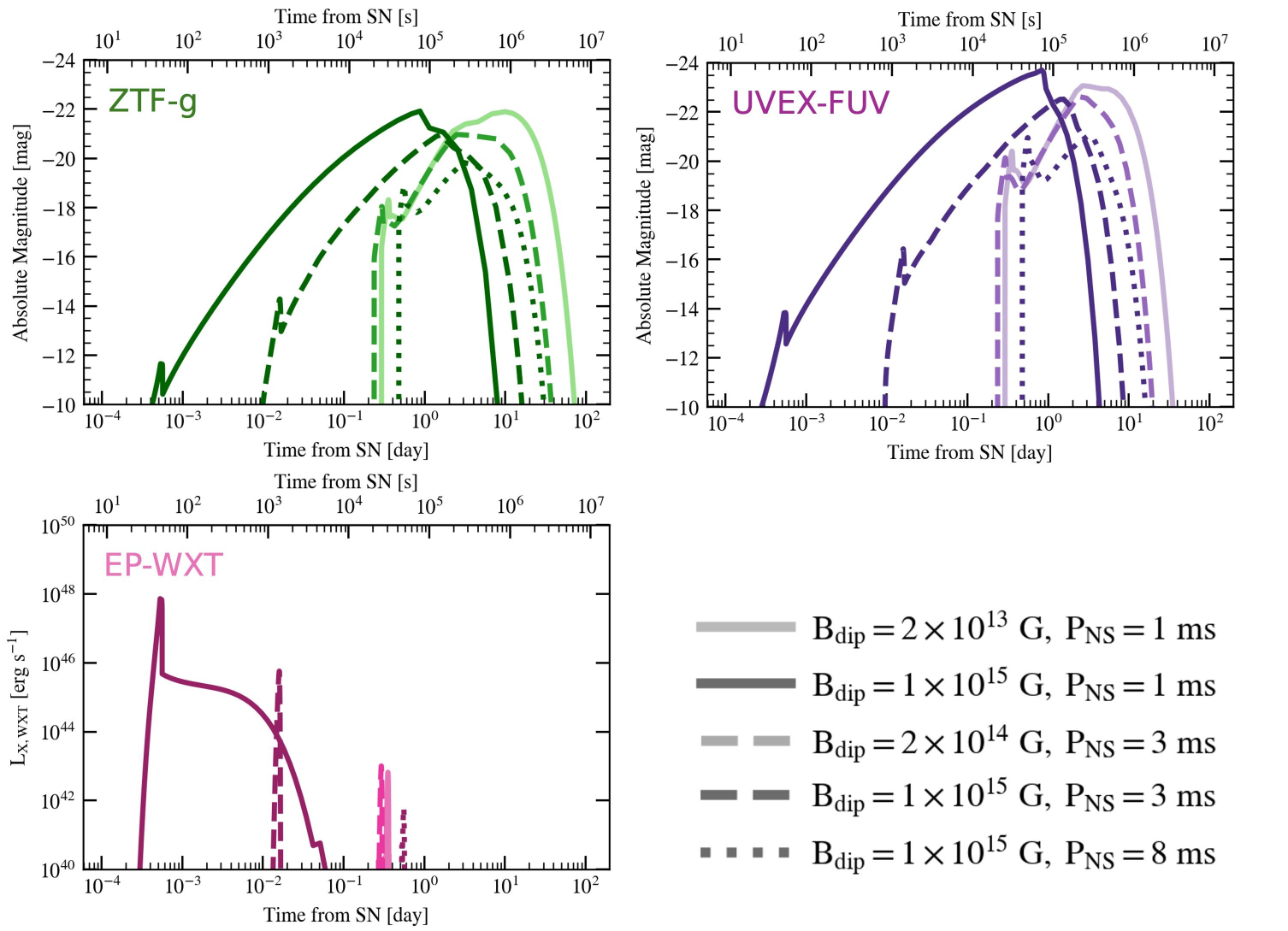}
  \end{subfigure}
  \caption{
    Survey-band light curves for ultra-stripped SNe for several illustrative pulsar-engine parameter choices (labeled in the panels). From left to right we show the ZTF $g$ band, the UVEX FUV band, and the EP--WXT soft X-ray band.
  }
  \label{fig:US_representives}
\end{figure}

In this subsection we present a systematic parameter study of pulsar-driven wind-bubble blowouts in ultra-stripped and AIC/MIC SNe. We begin by focusing on ultra-stripped SNe with $M_{\rm ej}=0.1\,M_{\odot}$ and $E_{\rm sn}=10^{50}\,{\rm erg}$. Before presenting the full set of contour and color maps, we first show several illustrative engine parameter choices and the corresponding survey-ready light-curve predictions in the ZTF $g$, UVEX FUV, and EP--WXT bands as Fig.~\ref{fig:US_representives}. In addition to the parameter choices used in Section~\ref{subsubsec:SESN-fiducial} for direct comparison, we include an additional case with $B_{\rm dip}=10^{15}\,{\rm G}$ and $P_{\rm NS}=8\,{\rm ms}$. This is motivated by the fact that, because ultra-stripped SNe have a much smaller $E_{\rm sn}$ than SESNe, the blowout-permitted region extends to substantially lower $L_{\rm sp}$ (i.e., to longer $P_{\rm NS}$) at fixed $B_{\rm dip}$.

From these illustrative cases, we find that ultra-stripped SNe evolve substantially faster than SESNe for the same engine parameters, owing to their much smaller ejecta mass. They also reach higher graybody temperatures and hence exhibit bluer light curves: the first peak is typically much fainter than the main peak in both the optical and UV bands, and the UV emission often remains brighter than the optical through the second peak, in some cases by up to $\sim 2$ mag. In addition, all cases shown here produce a luminous soft X-ray counterpart associated with the first peak.

\begin{figure}[ht!]
  \centering
  \begin{subfigure}{0.43\textwidth}
    \centering
    \includegraphics[width=\linewidth]{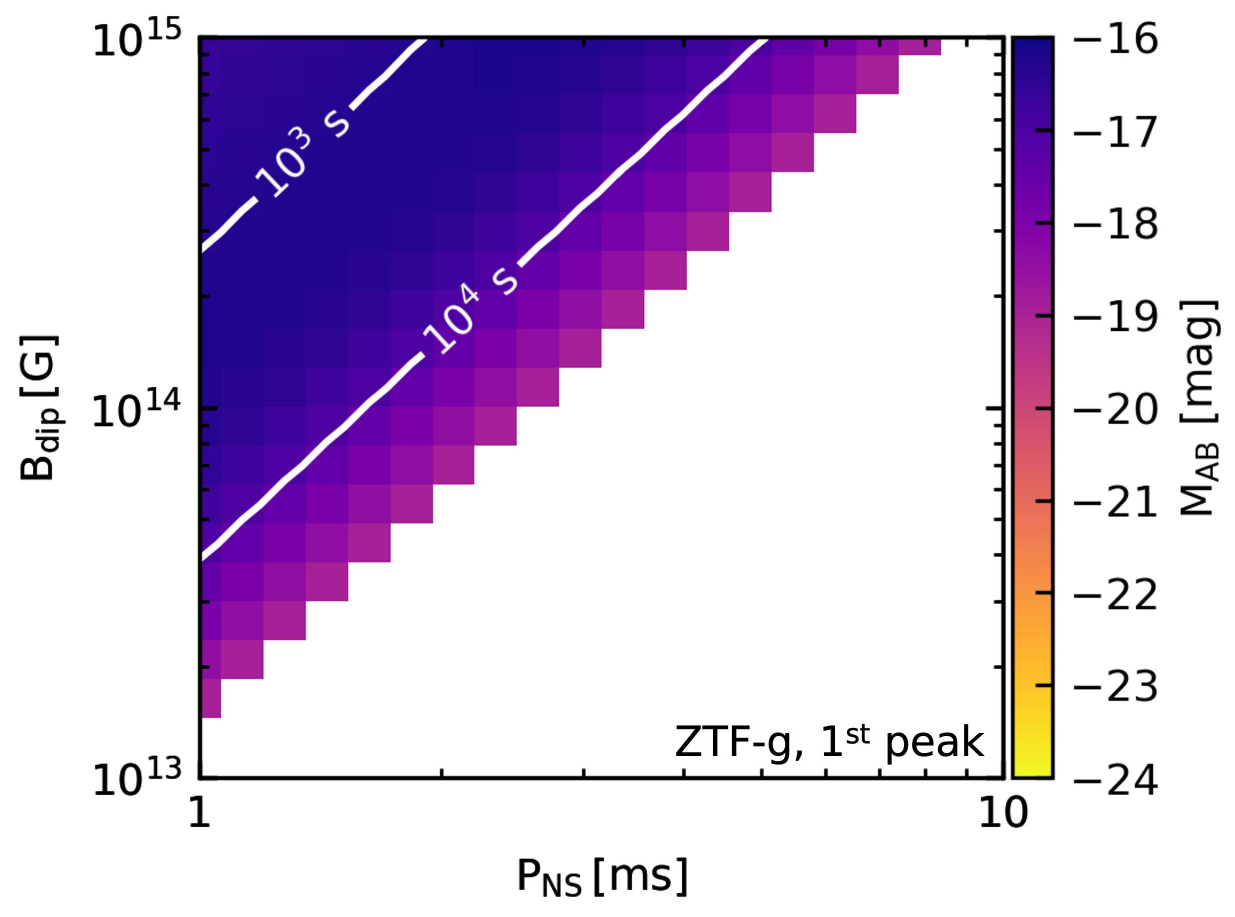}
  \end{subfigure}
  \begin{subfigure}{0.43\textwidth}
    \centering
    \includegraphics[width=\linewidth]{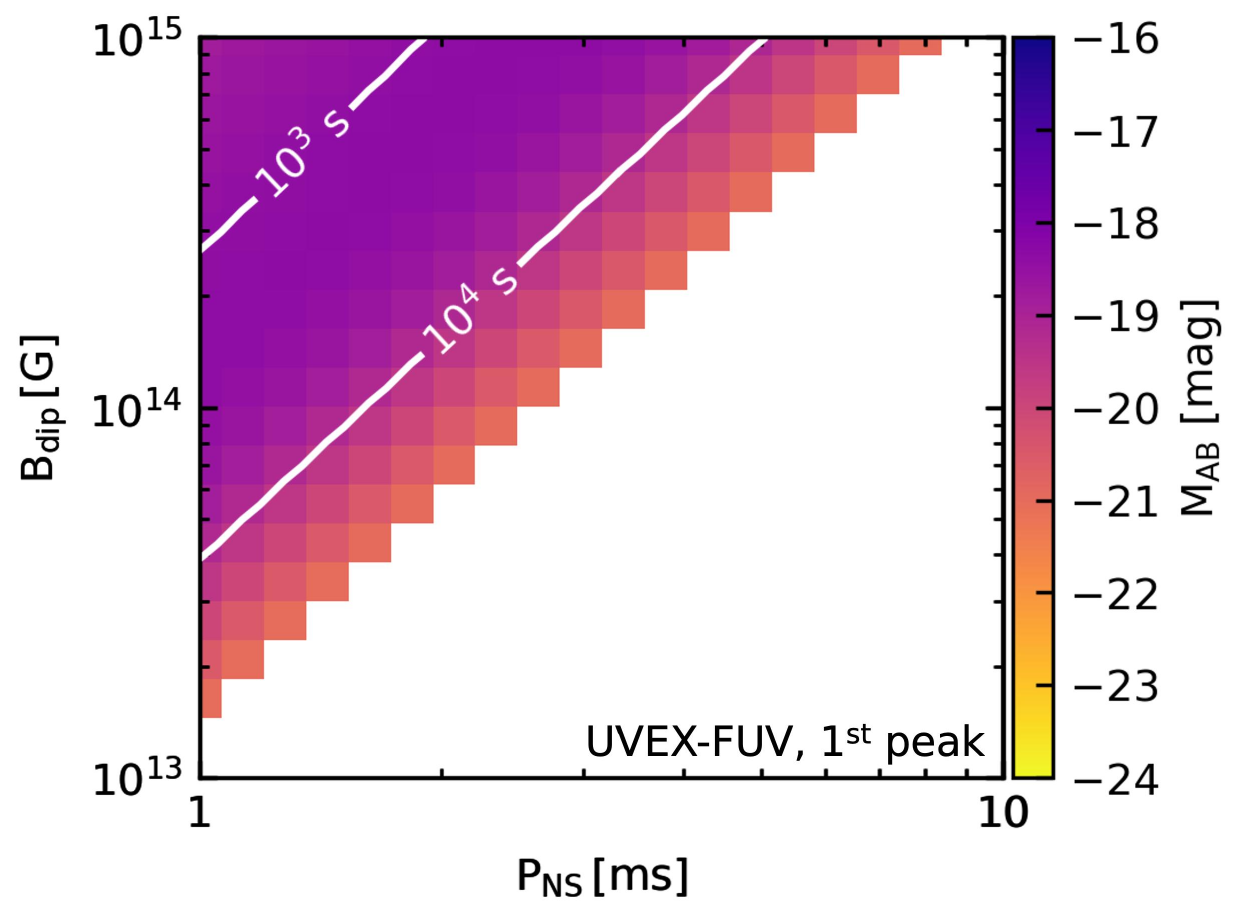}
  \end{subfigure}
  \begin{subfigure}{0.43\textwidth}
    \centering
    \includegraphics[width=\linewidth]{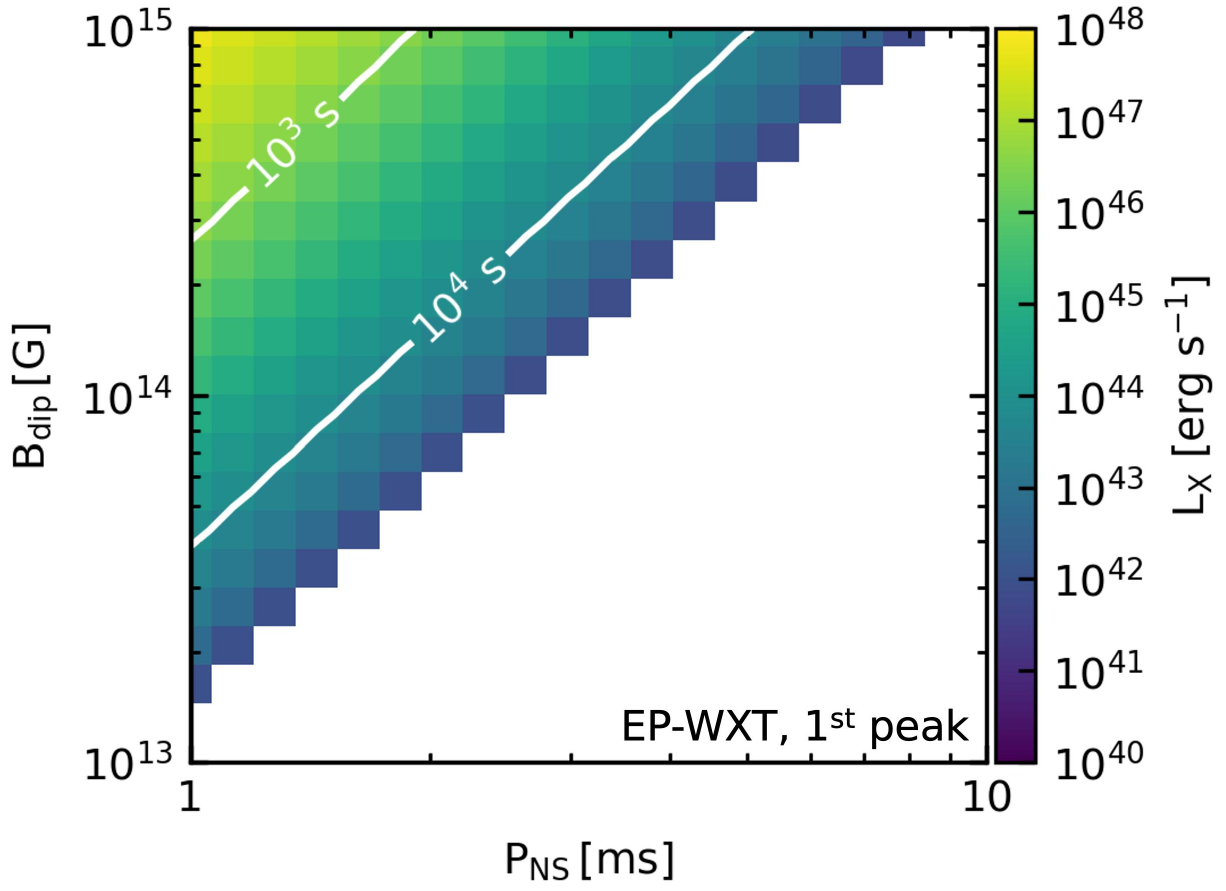}
  \end{subfigure}
  \caption{
    First peaks of the light curves for ultra-stripped SNe in the ZTF $g$, UVEX FUV, and EP--WXT bands for a grid of pulsar engine parameters. In each panel, colors indicate the predicted peak luminosity (or flux) in the corresponding band, and contours show the time of maximum in that band.
  }
  \label{fig:US_SNe_breaks}
\end{figure}

Having established the qualitative behavior with these illustrative cases, we next present a systematic survey over the $(B_{\rm dip},P_{\rm NS})$ parameter space. Fig.~\ref{fig:US_SNe_breaks} summarizes survey-ready observables for the first peaks, including the peak luminosities (or fluxes) and the times of maximum in the optical (ZTF $g$), UV (UVEX FUV), and soft X-ray (EP--WXT) bands for a grid of pulsar-engine parameters.

\begin{figure}[ht!]
  \centering
  \begin{subfigure}{0.45\textwidth}
    \centering
    \includegraphics[width=\linewidth]{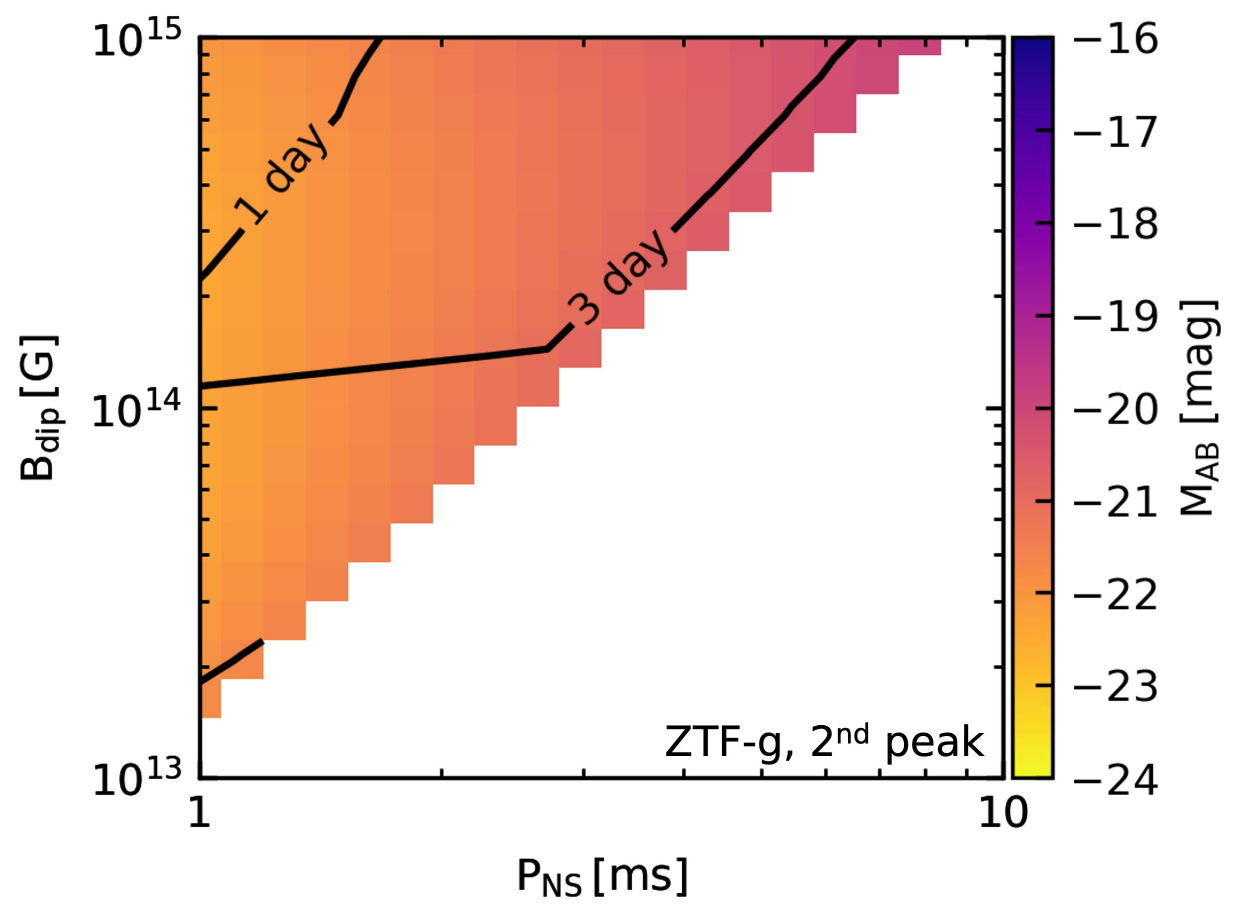}
  \end{subfigure}
  \centering
  \begin{subfigure}{0.45\textwidth}
    \centering
    \includegraphics[width=\linewidth]{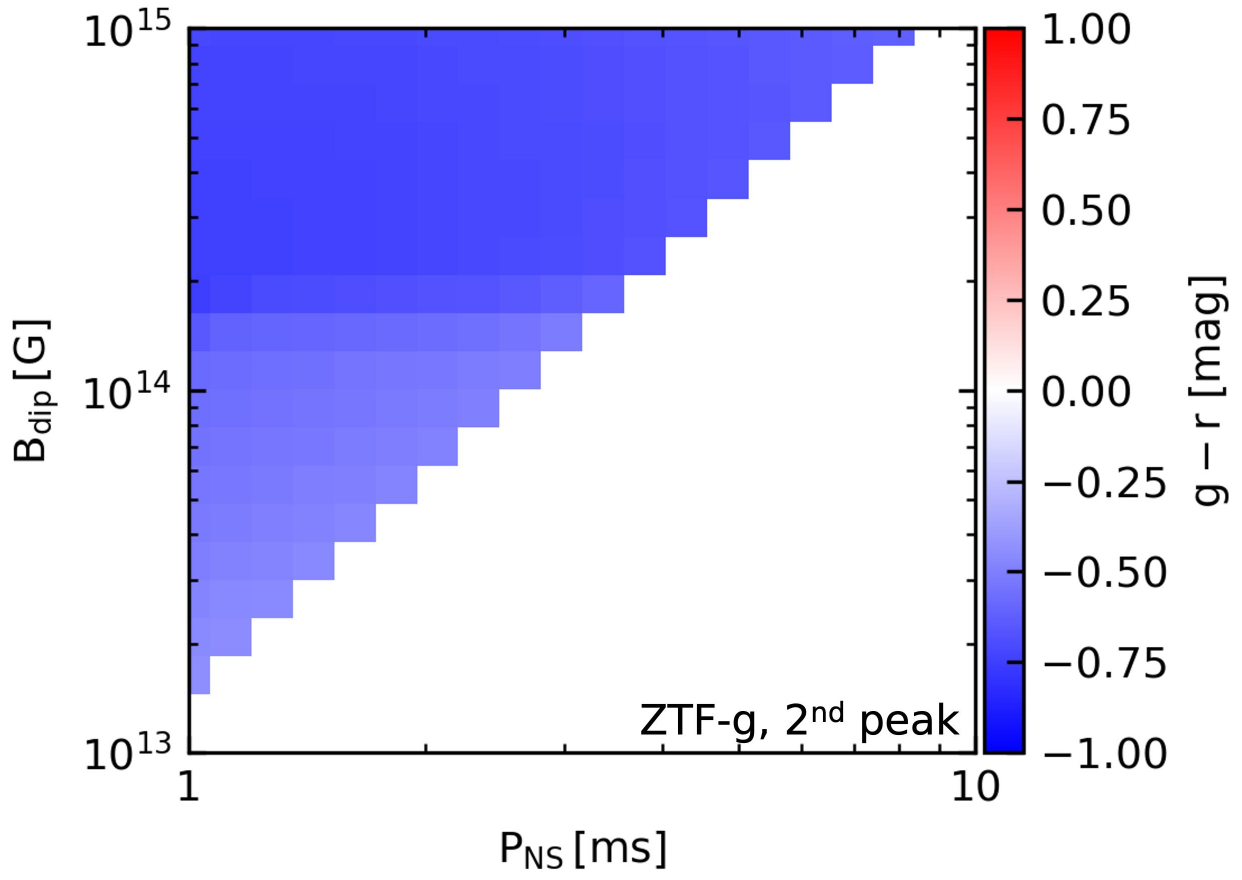}
  \end{subfigure}
  \caption{
    Second-peak properties of ultra-stripped SNe predicted by pulsar-driven blowout models in the ZTF $g$ band. \textit{Left:} peak $g$-band magnitude (color scale) and the time of maximum (contours). \textit{Right:} the predicted $g-r$ color at $g$-band maximum.
  }
  \label{fig:US_SNe_thermal}
\end{figure}

Despite the small ejecta mass and the correspondingly rapid evolution of the second, main peak, such events remain accessible and scientifically valuable targets for high-cadence optical surveys. In particular, the ZTF partnership high-cadence survey (HC) has already been collecting data for several years \citep{Bellm2019, Ho2023}. 
Fig.~\ref{fig:US_SNe_thermal} shows the ZTF $g$-band properties of the main peak for ultra-stripped SNe with pulsar-driven blowout.

Because of the smaller ejecta mass and explosion energy, the early first peak in ultra-stripped SNe is even shorter-lived and hotter than in SESNe as we already seen in Fig.~\ref{fig:US_representives}. This further suppresses the detectability of the first peak in optical bands. By contrast, the region of parameter space that produces bright soft X-ray emission is substantially enlarged, covering nearly the entire domain in which our model solutions exist. This suggests that, although identifying the early blowout 
peaks optically is challenging, soft X-ray observations offer a promising route, especially in light of high-performance soft X-ray time-domain facilities such as EP--WXT.

The smaller ejecta mass and the resulting short peak-luminosity timescale also tend to produce a hotter main peak, so that the optical bands lie farther from the SED maximum. As a result, even for similar neutron-star spin periods $P_{\rm NS}$ and comparable total injected energies, ultra-stripped events are systematically fainter at optical maximum than canonical SESNe, although they can still reach $M_g\lesssim -20$ mag. We further note that the reduced ejecta mass causes the ejecta to become optically thin earlier, making radiative losses more effective prior to the main peak. This reduces the fraction of the deposited energy that is retained and subsequently converted into bulk kinetic energy. For the same reason, the efficiency of converting early-time engine deposition into kinetic energy is limited, so the diversity of second-peak luminosities is markedly smaller than in the SESN case, and the color evolution is relatively muted: across the parameter space we explore, the main peak remains stably blue with $g-r\lesssim -0.2$ mag. In terms of their main-peak properties, such events are therefore more reminiscent of the brighter fast blue optical transients (FBOTs) described by \citet{Ho2023}.

\begin{figure}[ht!]
  \centering
  \begin{subfigure}{0.43\textwidth}
    \centering
    \includegraphics[width=\linewidth]{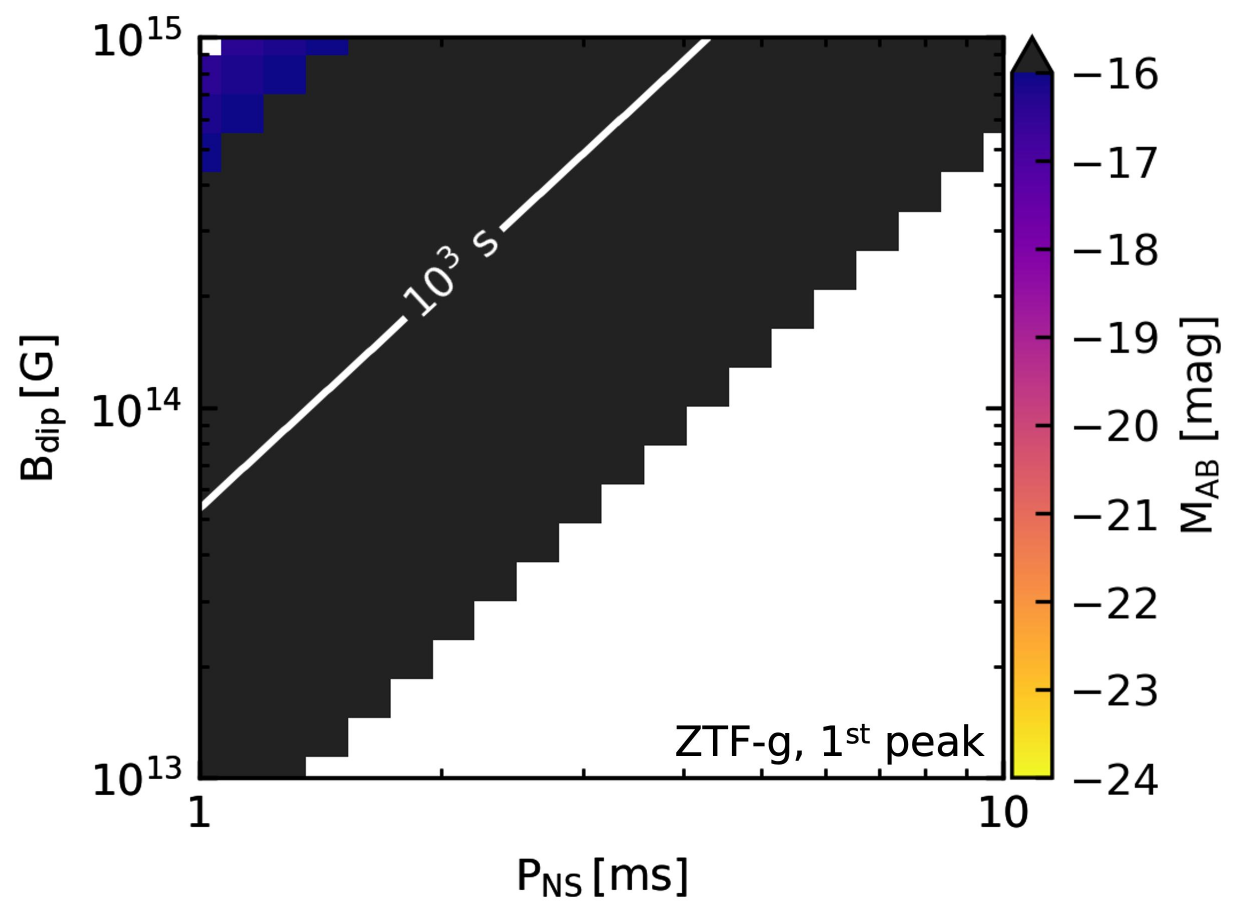}
  \end{subfigure}
  \begin{subfigure}{0.43\textwidth}
    \centering
    \includegraphics[width=\linewidth]{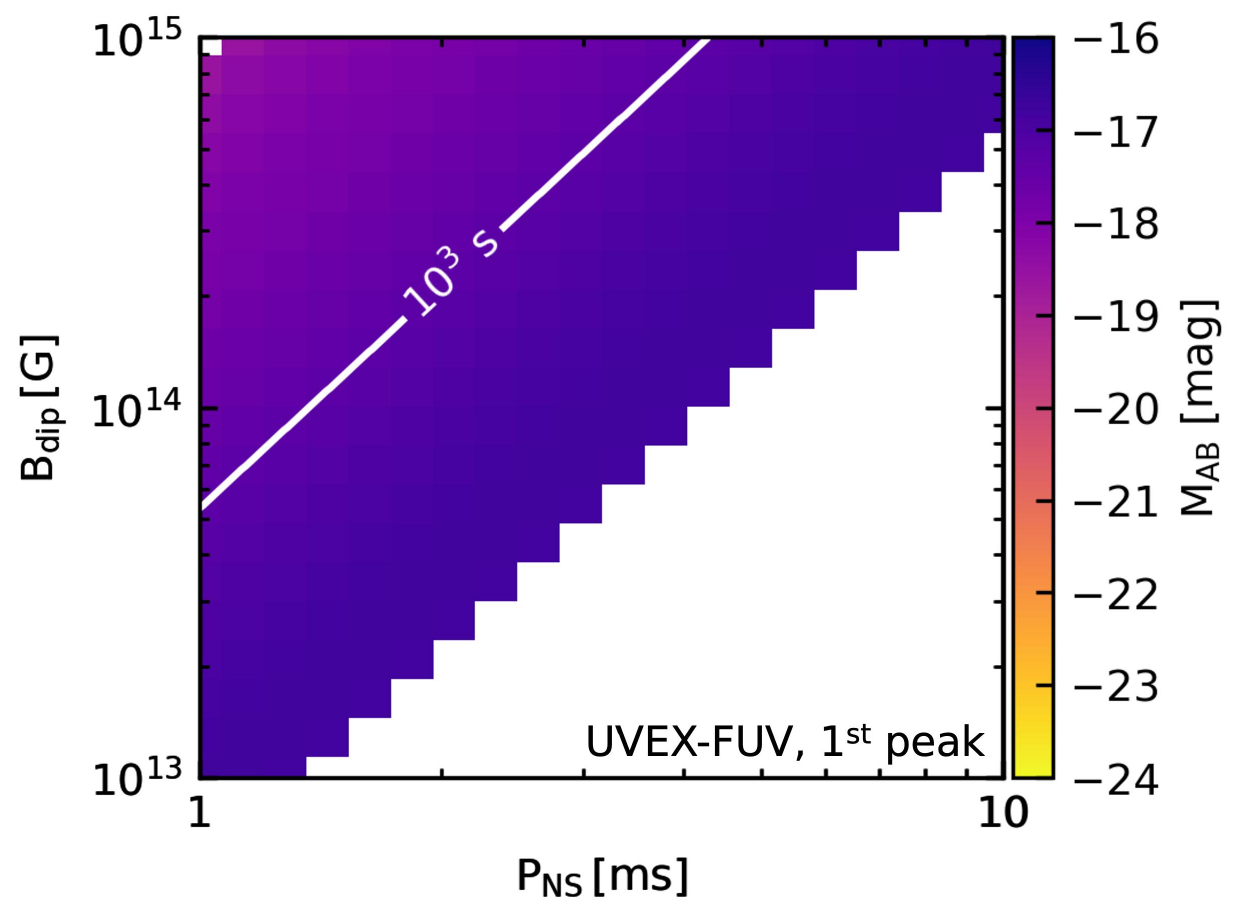}
  \end{subfigure}
  \begin{subfigure}{0.43\textwidth}
    \centering
    \includegraphics[width=\linewidth]{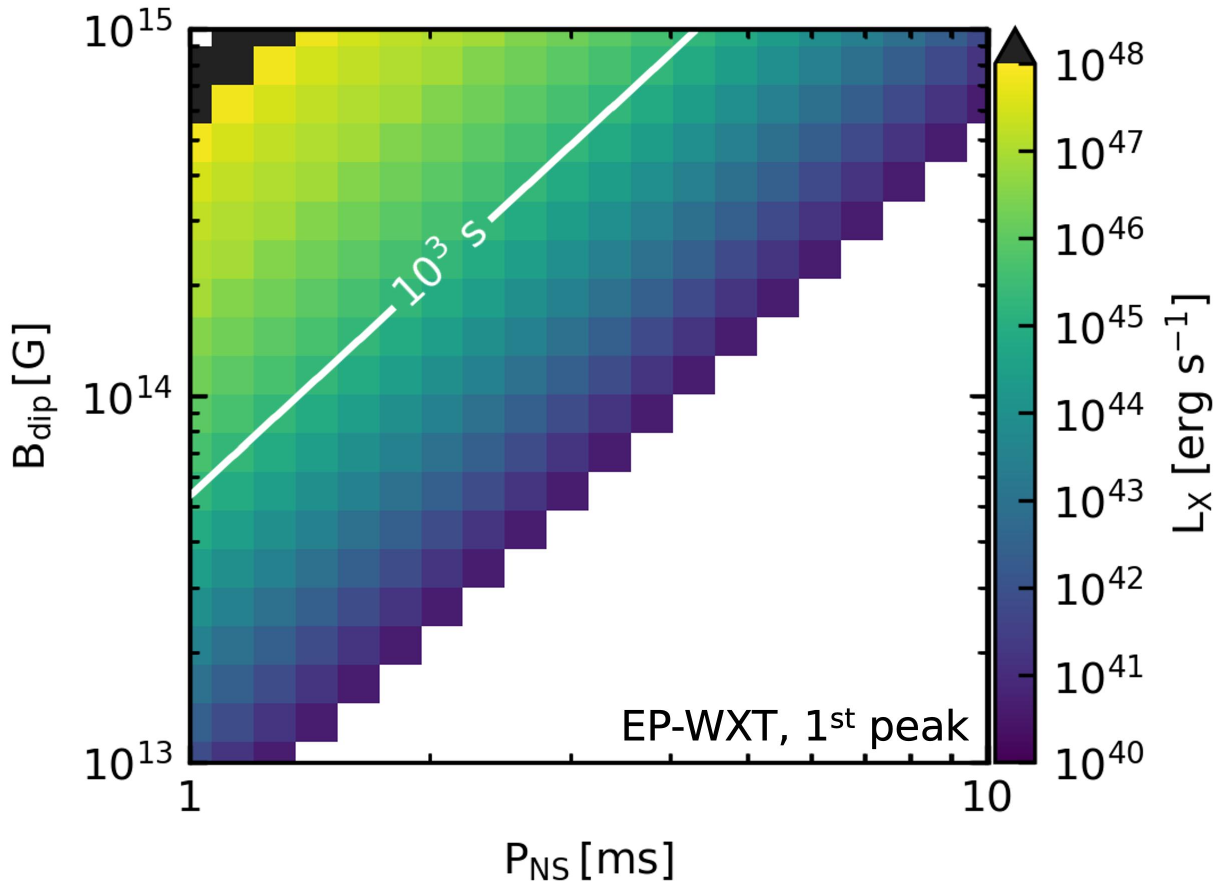}
  \end{subfigure}
  \caption{
    blowout-powered first-peak observables for AIC/MIC SNe in the ZTF $g$, UVEX FUV, and EP--WXT bands, for a grid of pulsar engine parameters. In each panel, the color scale shows the predicted peak luminosity (or flux) in the corresponding band, and contours trace the time of maximum in that band.
  }
  \label{fig:AIC_SNe_breaks}
\end{figure}

\begin{figure}[ht!]
  \centering
  \begin{subfigure}{0.45\textwidth}
    \centering
    \includegraphics[width=\linewidth]{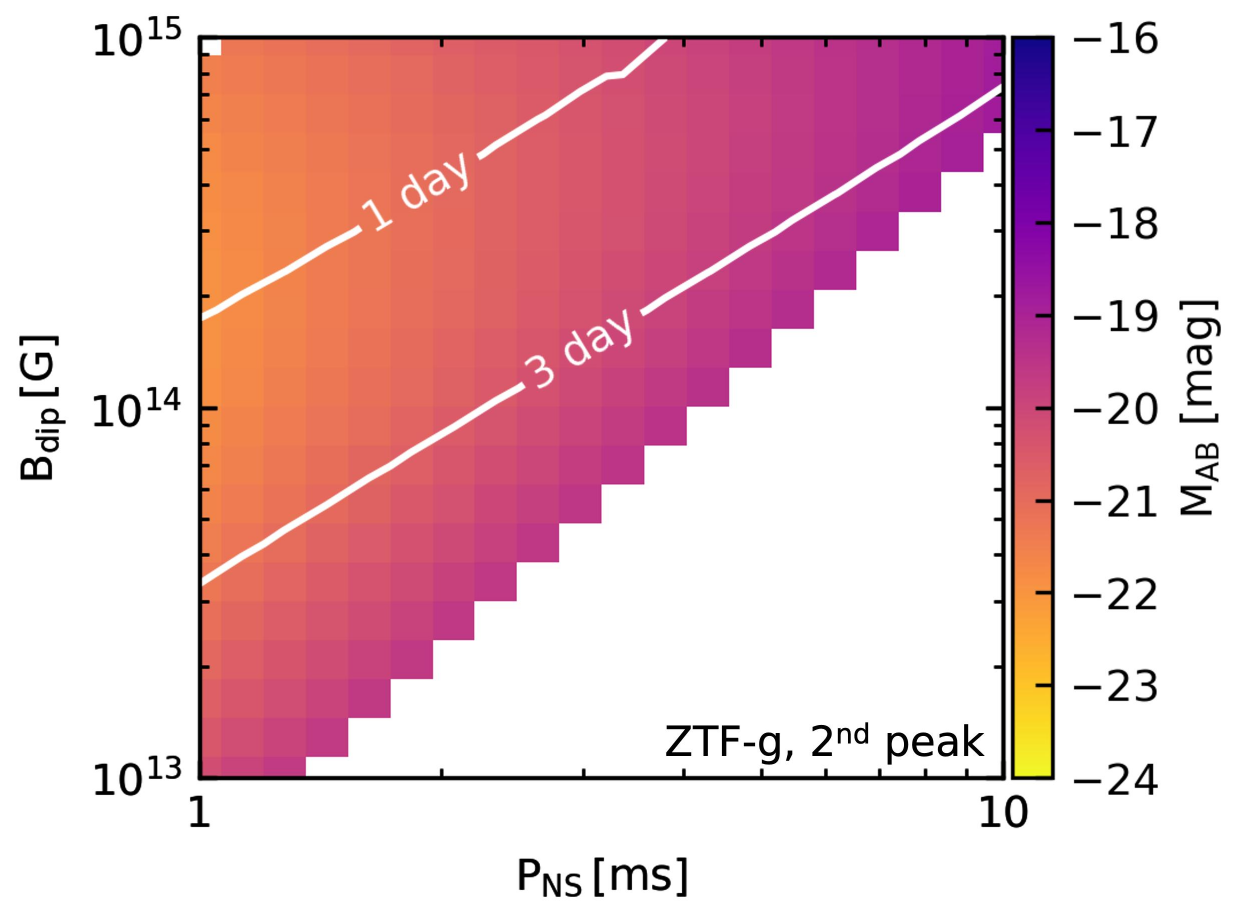}
  \end{subfigure}
  \begin{subfigure}{0.45\textwidth}
    \centering
    \includegraphics[width=\linewidth]{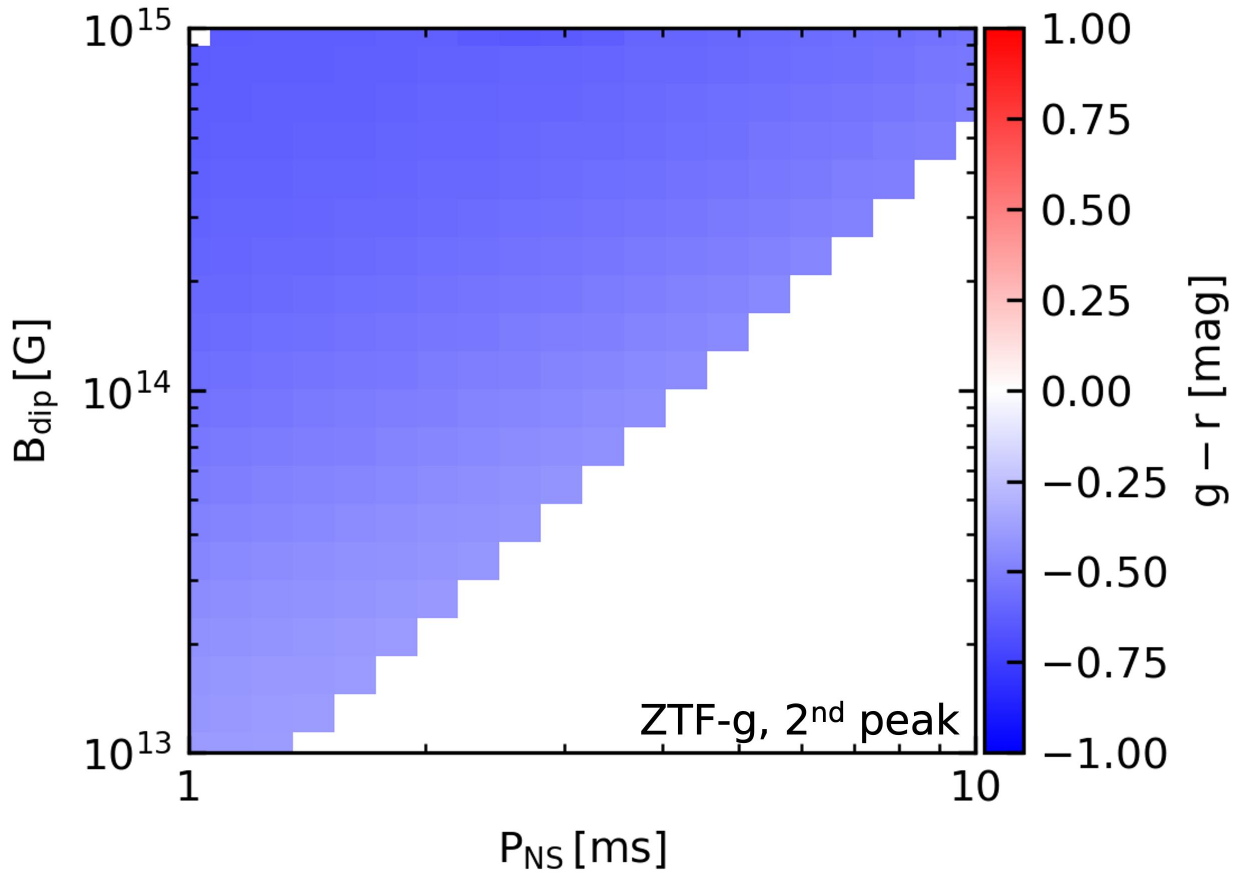}
  \end{subfigure}
  \caption{
    Main-peak properties of AIC/MIC SNe in the ZTF $g$ band for models with pulsar-driven blowout. \textit{Left:} peak $g$-band magnitude (color scale) and the time of maximum (contours). \textit{Right:} the predicted $g-r$ color at $g$-band maximum.
  }
  \label{fig:AIC_SNe_thermal}
\end{figure}

We now turn to AIC/MIC SNe. For this class of explosions we adopt
$M_{\rm ej}=0.01\,M_{\odot}$ and $E_{\rm sn}=10^{48}\,\mathrm{erg}$ as representative parameters. The overall trends resemble those found for ultra-stripped SNe, but in the most extreme case (high $L_{\rm sp}$ and high $E_{\rm rot}$; $B_{\rm dip}=10^{15}\,{\rm G}$, $P_{\rm NS}=1\,{\rm ms}$) a double-peaked morphology can appear even in the soft X-ray band.
A systematic summary of the multi-band properties of the blowout-powered first peak are summarized in Fig.~\ref{fig:AIC_SNe_breaks}.
The ZTF $g$-band properties of the second, main peak for AIC/MIC SNe with pulsar-driven blowout are shown in Fig.~\ref{fig:AIC_SNe_thermal}.

\begin{figure}[ht!]
  \centering
  \begin{subfigure}{0.8\textwidth}
    \centering
    \includegraphics[width=\linewidth]{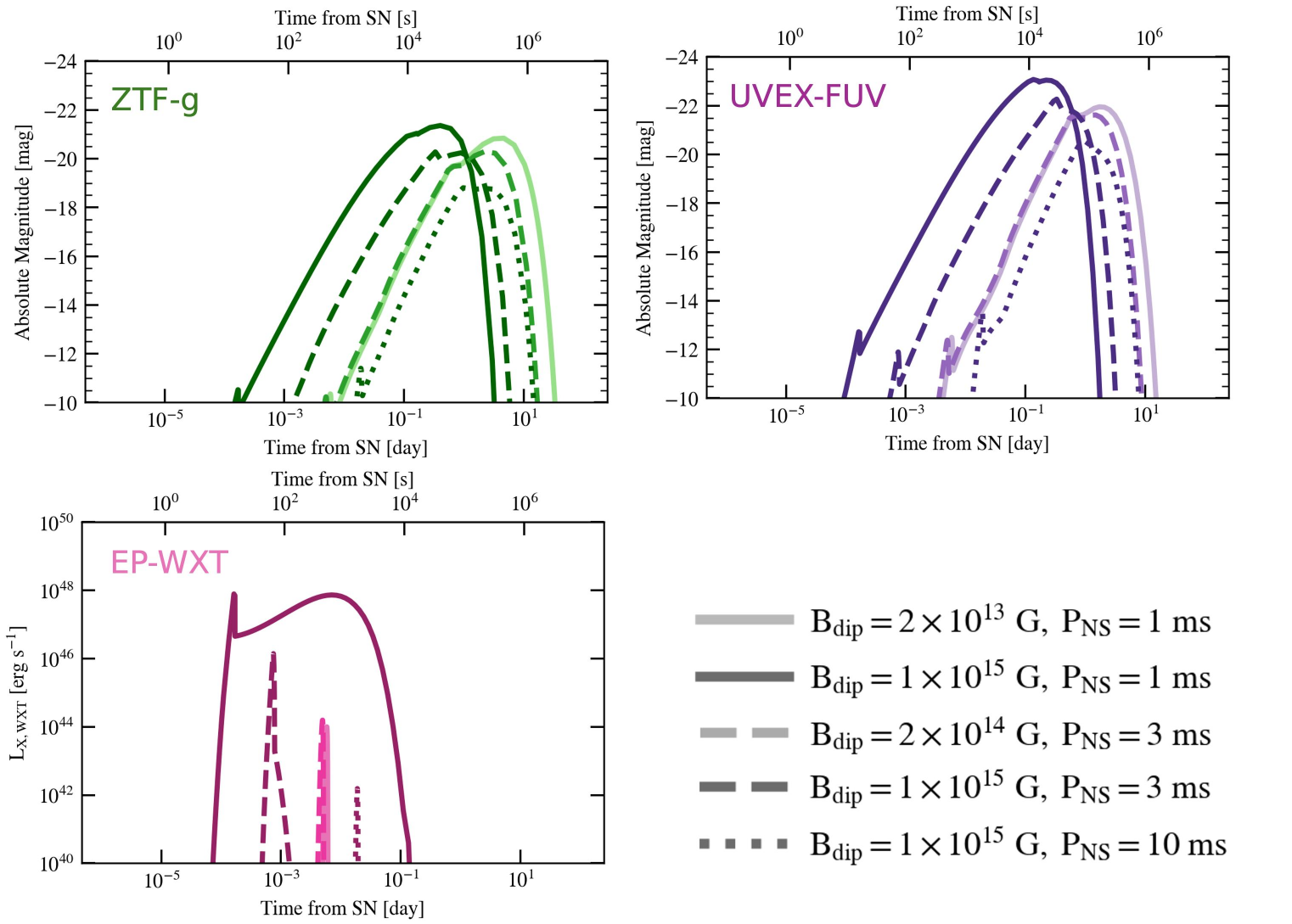}
  \end{subfigure}
  \caption{
    Survey-band light curves for AIC/MIC SNe for several illustrative pulsar-engine parameter choices (labeled in the panels). From left to right we show the ZTF $g$ band, the UVEX FUV band, and the EP--WXT soft X-ray band.
  }
  \label{fig:AIC_representives}
\end{figure}

For AIC/MIC SNe, which have even smaller ejecta masses and explosion energies than ultra-stripped SNe, the trends discussed above become more extreme. The blowout-powered first peak is faster and hotter, making it even more difficult to capture in optical and UV surveys. The main peak is also intrinsically fainter: within our explored parameter space, the optical peak magnitudes are generally fainter than $M_g\simeq -20$~mag. As in the ultra-stripped case, the diversity in main-peak luminosity is modest and the color evolution is weak, with the main peak remaining persistently blue ($g-r\lesssim -0.2$~mag).

Having discussed ultra-stripped and AIC/MIC SNe separately, we now compare them across subclasses; despite their similarly low ejecta masses and explosion energies, several observational trends are worth noting.
One notable point is that, in our fiducial setup, an earlier and more rapid blowout does \emph{not} imply brighter soft X-ray emission throughout the blowout-permitted parameter space. At low $L_{\rm sp}$, the injected energy at blowout is of order $\zeta_{\rm bo}E_{\rm sn}$, so reducing from the ultra-stripped case to an AIC/MIC-like explosion ($E_{\rm sn}\rightarrow 0.01E_{\rm sn}$, $M_{\rm ej}\rightarrow 0.1M_{\rm ej}$) lowers the first-peak luminosity by a factor
$(0.01)^{5/4}/(0.1)^{3/4}\sim 0.02$, as implied by Eq.~\eqref{eq:first_peak_Lbol_1}. Conversely, when $L_{\rm sp}$ is sufficiently large that the engine deposits essentially all of its rotational energy before the first peak, the first-peak luminosity depends mainly on $E_{\rm rot}$ and $M_{\rm ej}$ through Eq.~\eqref{eq:first_peak_Lbol_2}; at fixed engine, reducing the ejecta mass by a factor of 0.1 enhances the peak luminosity by $0.1^{-3/4}\simeq 6$. These scalings should, however, be viewed as illustrative for our chosen parameter set: in realistic AIC/MIC SNe, both $M_{\rm ej}$ and $E_{\rm sn}$ are expected to span a broad range \citep[e.g.,][]{Dessart2006,Mori2025}, and the soft X-ray first peak is sensitive to their combination (through the $E_{\rm sn}^5/M_{\rm ej}^3$ dependence in Eq.~\eqref{eq:first_peak_Lbol_1}, rather than being determined by either quantity separately.

\begin{figure}[ht!]
  \centering
  \begin{subfigure}{0.43\textwidth}
    \centering
    \includegraphics[width=\linewidth]{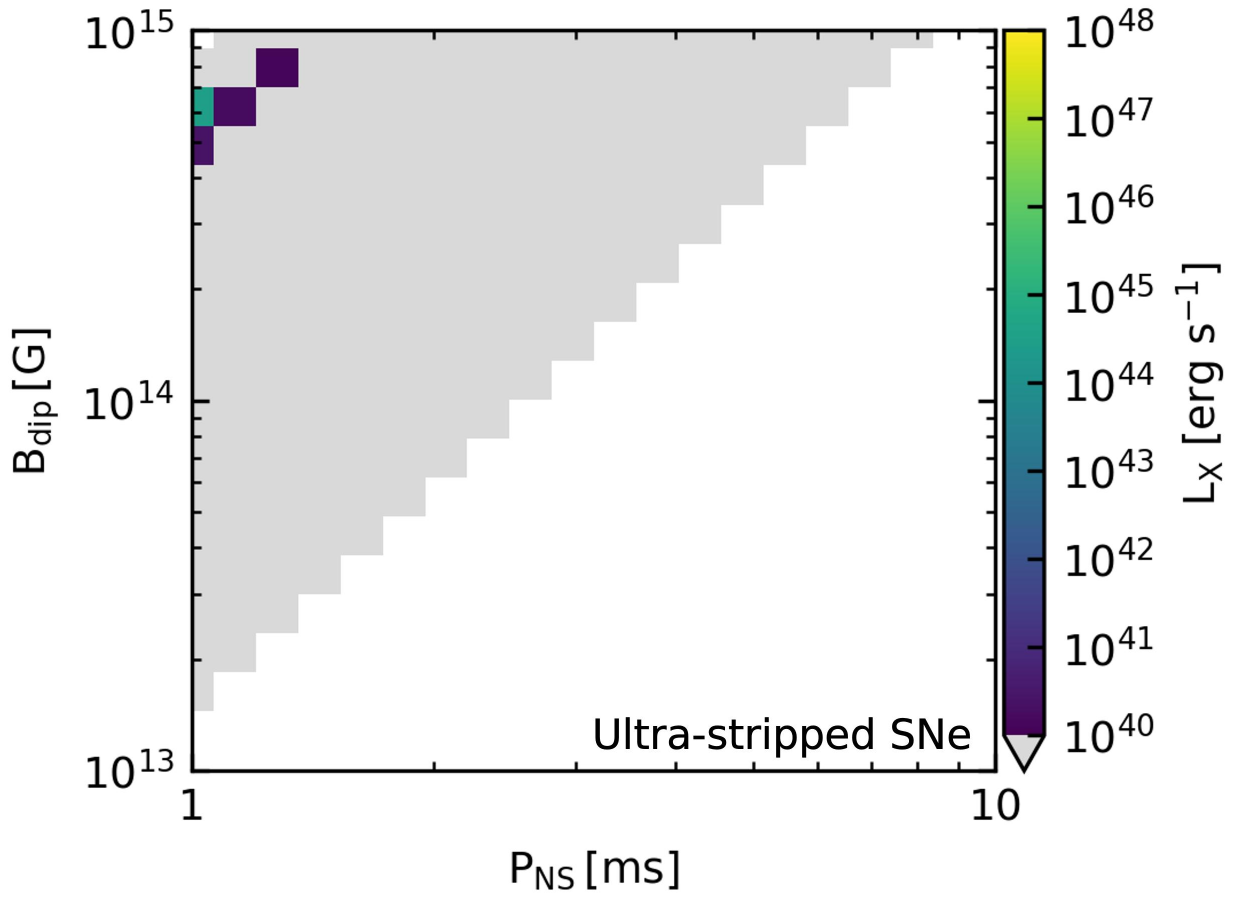}
  \end{subfigure}
  \begin{subfigure}{0.43\textwidth}
    \centering
    \includegraphics[width=\linewidth]{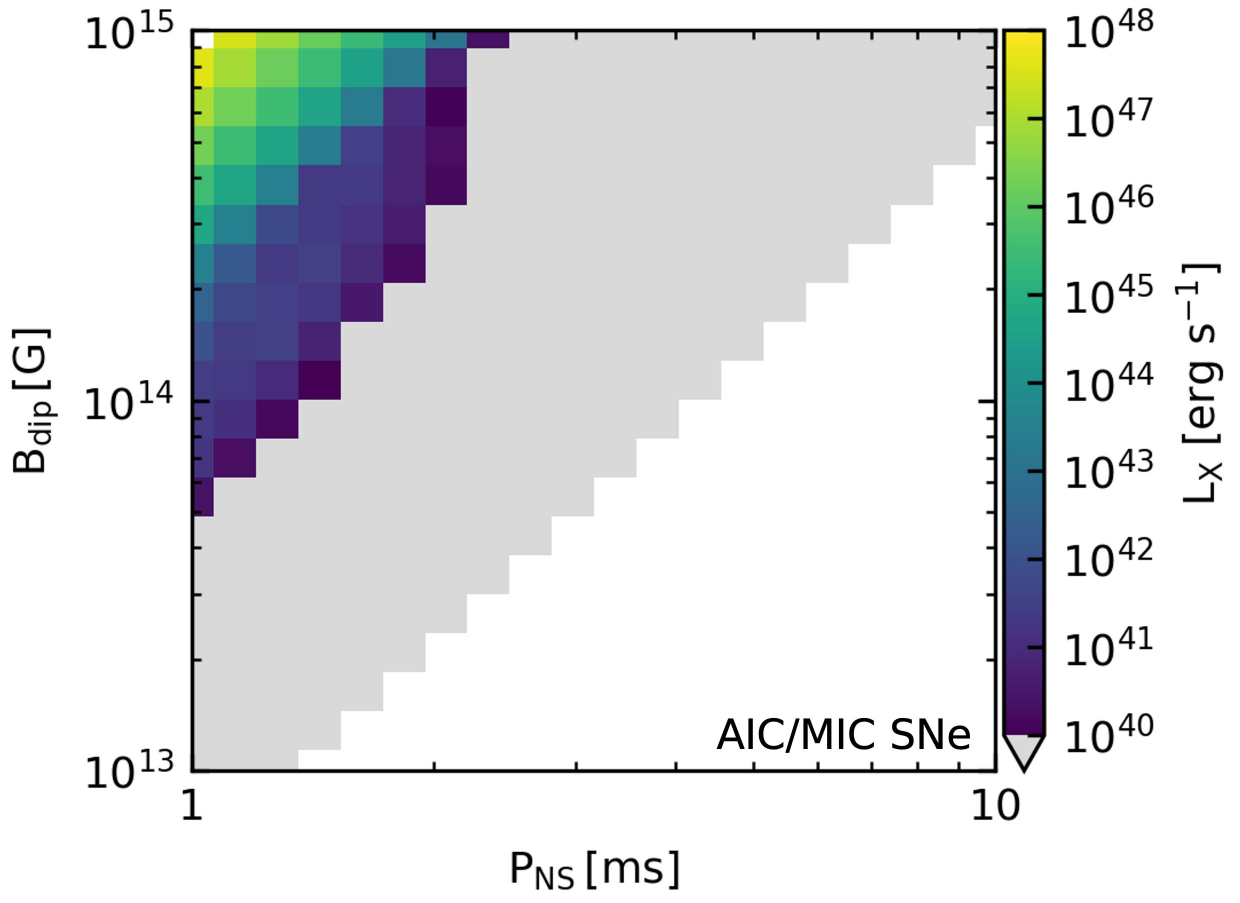}
  \end{subfigure}
  \caption{
    Soft X-ray main-peak diagnostics for low-mass ejecta with strong central energy injection. Left: EP--WXT peak brightness for ultra-stripped SNe. Right: the corresponding EP--WXT peak brightness for AIC/MIC SNe. While the ultra-stripped models show no parameter region with a soft-X-ray--bright main peak, the AIC/MIC models can exhibit an X-ray--luminous main peak for strongly magnetized, millisecond pulsars.
  }
  \label{fig:WXT_mainpeak_US_AIC}
\end{figure}

In addition to the diagnostics in the same bands discussed for SESNe, low-mass ejecta introduce an important qualitative difference in the second (main) peak. The higher main-peak temperatures can shift a substantial fraction of the emission into the soft X-ray band, as suggested by Fig.~\ref{fig:AIC_representives}, a behavior not seen in the SESN parameter regime explored above. As illustrated in Fig.~\ref{fig:WXT_mainpeak_US_AIC}, the AIC/MIC models exhibit a region with a soft-X-ray--bright main peak at $B_{\rm dip}\gtrsim 2\times 10^{14}\,{\rm G}$ and $P_{\rm NS}\lesssim 2\,{\rm ms}$, whereas no comparable region is present for the ultra-stripped models. The apparent discontinuity in the upper-left corner of the ultra-stripped panel arises because the light curves there do not show a clear double-peaked morphology in soft X-rays; after the first peak they more closely resemble a short-lived plateau, so a second peak is not robustly identified. This distinction may provide an observational clue for identifying AIC/MIC SNe, which have not yet been unambiguously detected.

Building on the results above, despite the rapid optical evolution expected for ultra-stripped and AIC/MIC SNe, the predicted rise and decline timescales remain well matched to modern high-cadence optical programs such as ZTF-HC\citep{Ho2023}. In the optical, the first peak may still be challenging when it is both faint and fast, but the depth of the Vera C. Rubin Observatory Legacy Survey of Space and Time, together with rolling-cadence strategies and high-cadence mini-survey modes, may extend sensitivity to such short-lived features \citep{Ivezic2019,Bianco2022,Gezari2018}; nevertheless, the extremely brief duration of the first peak can still make its secure identification difficult and may require favorable sampling and cadence. Moreover, our models predict luminous soft X-ray emission in both channels, offering an additional, highly discriminating signature of these engine-powered explosions. Looking ahead, rapid-response programs such as the Rubin Observatory Target-of-Opportunity framework could further expand discovery space by enabling triggered, high-cadence follow-up during the earliest phases \citep{Andreoni2024}. With Einstein Probe now delivering substantial survey exposure, systematic searches for such fast transients are both feasible and timely. A forthcoming companion study will leverage existing archival data to search for these events and, if no detection is achieved, place robust rate limits on explosions hosting an embedded, rapidly rotating, strongly magnetized pulsar.

\section{Discussion}
\label{sec:discussion}

In Sections~\ref{sec:blowout_model} and \ref{sec:calculation_results} we modelled the nascent wind bubble blowout within a one-dimensional (1D) framework and presented the resulting calculations. Here we place those results in a broader context and discuss implications and limitations that complement our theoretical model and main findings. 

\subsection{Multidimensional Effects}

We emphasize that multidimensional effects can be important in some cases. Here we represent the post-blowout region as a spherically symmetric admixture layer, implicitly assuming that fingers produced by hydrodynamical instabilities rapidly fragment and mix; this provides an effective description of a strongly mixed branch suggested by multidimensional simulations \citep[e.g.,][]{Blondin&Chevalier2017,Chen2016,Suzuki&Maeda2017,Suzuki&Maeda2019}.

In multidimensions, fingers need not be erased. Long-lived structures can leave low-density gaps that open localized, aspherical leakage pathways that vent the bubble \citep[e.g.,][]{Blondin&Chevalier2017,Suzuki&Maeda2017}. Engine power may then escape preferentially through these channels, instead of being deposited and thermalized in the admixture layer. Reduced thermalization would alter the flow, the bolometric luminosity, and the inferred blackbody temperature; in the most extreme cases, the blowout-driven peak could be muted or absent.

Magnetic fields can further steer the outcome. Ordered fields at the interface suppress small-scale shear and entrainment via magnetic tension, making fingers harder to shred and mixing less efficient \citep[e.g.,][]{Stone&Gardiner2007}. Magnetized bubbles also tend to be dominated by toroidal fields; the associated hoop stress can enhance anisotropy and collimate the energy flux, promoting and sustaining low-density channels \citep[e.g.,][]{Bucciantini2008,Bucciantini2009}.

Assessing the stability of the pre- and post-blowout structure, the degree of mixing, and the radiative imprint requires (relativistic) multidimensional radiation-hydrodynamics and magnetohydrodynamics (MHD) calculations. We therefore interpret our 1D admixture-layer treatment as an effective description of the fully mixed, quasi-spherical branch. A systematic exploration of long-lived fingers and venting-dominated configurations, and their observational consequences, is deferred to future work.

\subsection{Marginal Blowout}
\label{subsec:marginal_blowout}

In Section~\ref{subsubsec:LSQ14bdq}, our fit yields a first peak that is slightly faster and brighter than inferred from the data. The present framework also struggles to reproduce events with markedly slower first peaks. We do not view this as a failure of the blowout picture, but rather as a limitation of our current initialization of the pre-blowout phase. We assume that the swept-up shell evolves nearly adiabatically prior to blowout, with only minor energy losses through shock heating \citep[e.g.,][]{Chevalier1992}. This approximation works best for strongly driven cases (high $B_{\rm dip}$ and short $P_{\rm NS}$), but it can bias solutions in the marginal-blowout regime, where radiative leakage may already be non-negligible before $t_{\rm bo}$ (Eq.~\ref{eq:f_sec}). This is related to the ``inner-breakout'' regime discussed by \citet{Kasen2016}, in which diffusion begins to affect the dynamics and the emerging emission while the swept-up shell is still within the inner ejecta.

More specifically, to keep the adopted initial conditions self-consistent, we impose an additional admissibility cut: at $t_{\rm bo}$ we require the diffusion radius to lie outside the outer boundary, $R_{\rm diff}\geq R_{\rm out}$. This criterion is stricter than the physical blowout requirement. It excludes realizations that lie within the nominal blowout-allowed region of Fig.~\ref{fig:blowout-phase-diagram} but have $R_{\rm diff}<R_{\rm out}$ at $t_{\rm bo}$, which would imply appreciable radiative effects already during the pre-blowout evolution. Physically, however, blowout may still proceed even if the diffusion front penetrates the admixture layer before $t_{\rm bo}$, provided the radiative losses remain modest compared with the deposited energy.

Both effects become important near the boundary set by Eq.~\ref{eq:L_sp_blowout_constraint}. In this regime, weaker engines can still yield slower first peaks, but they are not captured faithfully within the current setup. Relaxing these restrictions requires evolving the system self-consistently from the explosion epoch while allowing moderate radiative leakage before $t_{\rm bo}$. We defer a comprehensive exploration of this marginal blowout regime to future work.

\subsection{\texorpdfstring{$^{56}$Ni Decay}{56Ni Decay}}
\label{subsec:Ni_decay}

We have neglected heating from $^{56}$Ni decay so far, since the central engine injects an energy comparable to the initial explosion energy by the onset of blowout, so radioactive input is typically subdominant. While, $^{56}$Ni is a key power source in supernova light curves, and its potential impact must be assessed.

For the first peak, $^{56}$Ni is expected to be synthesized predominantly in the deepest layers of the ejecta \citep{Janka2012, Imasheva2023}. Radioactive heating prior to blowout can, in principle, add energy to the swept up shell from smaller radii and reduce the energy injection required from the central engine. Nonetheless, this effect is negligible in practice.
For a pure $^{56}$Ni, the total decay energy is $\epsilon_{\rm nuc}\simeq 0.11$ MeV/baryon. Considering the realistic $^{56}$Ni yields with $X_{\rm Ni}\lesssim \mathcal{O}(0.1)$ and a blowout timescale shorter than the relevant decay timescales, the radioactive energy that can be deposited before $t_{\rm bo}$ is limited to $\lesssim \mathcal{O}(0.01)$ MeV/baryon.
This is below the characteristic energy scale required to accelerate the shell and trigger blowout, which is of order the ejecta specific kinetic energy, $\epsilon_{\rm kin}\sim 0.1$--0.5 MeV/baryon, for expansion velocities $v_{\rm ej}$ of $\sim 3{,}000$--$10{,}000\ {\rm km\ s^{-1}}$. The radiation released around the first peak is supplied mainly by two channels: internal energy injected and stored prior to blowout and shock heating during blowout. Both operate at a scale $\sim \epsilon_{\rm kin}$. Therefore, within the blowout branch considered here, $^{56}$Ni heating does not appreciably modify the shell dynamics, the blowout time $t_{\rm bo}$, or the luminosity and timescale of the first peak. The same conclusion applies to SESNe and to lower-energy explosions, including ultrastripped and AIC/MIC SNe.

The role of $^{56}$Ni during the main peak is more parameter dependent and can be evaluated with simple scalings. \citet{Kashiyama2016} estimate the peak luminosity powered by $^{56}$Ni decay as
\[
L_{\rm sn}^{\rm Ni}\sim 10^{42}\ {\rm erg\ s^{-1}}
\left(\frac{M_{^{56}{\rm Ni}}}{0.1\,M_\odot}\right)
\left(\frac{M_{\rm ej}}{10\,M_\odot}\right)^{-1}
\left(\frac{v_{\rm ej}}{10^{9}\ {\rm cm\ s^{-1}}}\right)
\left(\frac{\kappa}{0.1\ {\rm cm^{2}\ g^{-1}}}\right)^{-1},
\]
and the peak luminosity in the pulsar driven case as
\[
L_{\rm sn}^{\rm psr}\sim 3\times 10^{44}\ {\rm erg\ s^{-1}}
\left(\frac{B_{\rm dip}}{10^{14}\ {\rm G}}\right)^{-2}
\left(\frac{M_{\rm ej}}{10\,M_\odot}\right)^{-1}
\left(\frac{v_{\rm ej}}{10^{9}\ {\rm cm\ s^{-1}}}\right)
\left(\frac{\kappa}{0.1\ {\rm cm^{2}\ g^{-1}}}\right)^{-1}.
\]
For the SESN scalings adopted in this paper, these estimates give $L_{\rm sn}^{\rm psr}\gtrsim 10\,L_{\rm sn}^{\rm Ni}$ when $B_{\rm dip}\gtrsim 3\times 10^{13}\ {\rm G}$, so the radioactive contribution to the peak luminosity is limited. At lower $B_{\rm dip}$, the relative contribution from $^{56}$Ni increases and overlaps with the low $L_{\rm sp}$ parameter space associated with marginal blowout. For low mass, low energy AIC or MIC explosions, a similar estimate indicates that $B_{\rm dip}\gtrsim$ a few $\times 10^{12}\ {\rm G}$ often still gives $L_{\rm sn}^{\rm psr}\gtrsim 10\,L_{\rm sn}^{\rm Ni}$, whereas for weaker engines near marginal blowout the radioactive contribution to the main peak can become non negligible. Since we explicitly exclude marginal blowout in Section~\ref{subsec:marginal_blowout}, radioactive heating does not affect the calculations presented here. A self consistent inclusion of $^{56}$Ni decay is required when extending the model to the marginal blowout regime.

After the main peak, as the ejecta becomes progressively more transparent, $\gamma$-ray escape from the $^{56}$Ni decay chain increases and the deposition efficiency declines. Positron deposition, mainly from the $\beta^+$ branch of $^{56}$Co decay, can then become relatively more important, which changes the late time deposition channel and can modify the optical output at specific phases \citep{Milne1999, Milne2001, Seitenzahl2011}. The formation of an admixture layer may also enhance outward mixing and transport some $^{56}$Ni to larger radii, which would promote earlier $\gamma$-ray leakage and alter the late time decline, potentially introducing non thermal signatures. In summary, within our present framework that focuses on thermal emission and excludes marginal blowout, neglecting $^{56}$Ni heating is an acceptable approximation. Incorporating $^{56}$Ni decay self consistently is a natural next step for extensions that include marginal blowout and late time emission.

\subsection{Nonthermal Emission}

Our calculations adopt a minimal treatment in which the pulsar spin-down power is thermalized in the optically thick ejecta and converted into bulk kinetic energy and quasi-thermal radiation. This approximation is suitable upto the main-peak phase, but it omits nonthermal leakage as the ejecta expand.

At the same time, a luminous spin-down engine motivates complementary nonthermal diagnostics. PWN emission from radio to X-rays and $\gamma$-rays remains widely discussed \citep[e.g.,][]{Kotera2013,Metzger2014,Murase2016,Omand2018}. Although observationally, existing constraints are dominated by non-detections and upper limits, with only one tentative PWN-like candidate reported so far \citep{Margutti2018,Bhirombhakdi2018,Andreoni2022,Eftekhari2021,Murase2021}. Whether high-energy photons escape is highly sensitive to the ejecta mass and density structure, the engine luminosity history, and the evolving ionization and absorption state.

A key pathway is ionization break-out, which can release nebular photons once the effective optical depth drops sufficiently \citep{Metzger2014,Murase2016,Omand2018}. As a benchmark, \citet{Kashiyama2016} estimated that for Ic-BL--like ejecta with $M_{\rm ej}\lesssim 5\,M_\odot$, $B_{\rm dip}\sim 5\times10^{14}\,{\rm G}$, and $P_{\rm NS}\lesssim 10\,{\rm ms}$, \emph{NuSTAR} could detect hard X-ray nebular emission at $\sim100$~days for events within $\lesssim30$~Mpc (with a larger reach for SLSN-I--like parameters), illustrating that a subset of bright engines can be detectable under favourable conditions. For the ultra-stripped and AIC/MIC events, the lower ejecta masses imply earlier transparency for comparable engines, potentially advancing the emergence of PWN component.

Beyond the PWN, blowout solutions also generate a fast, high energy-to-mass outer tail, which can enhance ejecta--ambient interaction and its synchrotron and inverse-Compton emission \citep{Chevalier&Fransson2017,Chakraborti2015,Suzuki&Maeda2018}. We therefore view nonthermal channels as a natural complement to the thermal signatures developed here. These nonthermal effects haven't been included in this work to keep the model minimal and focused on the early-time phases. Incorporating these channels is a natural next step for extending the framework and for determining when they can complement, or outshine, the thermal emission highlighted in this work.

\subsection{Comparison with alternative early-peak mechanisms}
\label{subsec:compare_alternatives}

Several alternatives to blowout can also produce a double-peaked morphology, including shock cooling of extended material and CSM interaction in some configurations \citep[e.g.,][]{Nakar&Piro2014, Chevalier&Irwin2011}. To enable observational separation, we compare their predicted timescales, temperature, color evolution, and multiband behavior against those expected from a single ejecta and engine system.

First, blowout admits a stringent internal-consistency test because the first peak and the main peak originate from the same ejecta and engine system. A single parameter set must account for both peaks, with the bulk scale primarily set by $(M_{\rm ej},\,E_{\rm sn},\,E_{\rm rot},\,L_{\rm sp})$. By contrast, CSM-based explanations introduce additional degrees of freedom (e.g., CSM mass and radius) that are only weakly constrained and can vary widely across events \citep{Piro2015, Fraser2020, smith2014}.

Second, the radiated-energy budget of the first peak is a robust discriminator. In shock-cooling models in which an extended low-mass layer reprocesses the explosion energy, the radiated energy is typically \citep{Nakar&Piro2014,Piro2015}
\begin{equation}
E_{\rm ext}\sim 9\times 10^{48}\ {\rm erg}\,
\left(\frac{E_{\rm sn}}{10^{51}\ {\rm erg}}\right)
\left(\frac{M_{\rm ej}}{10\ M_\odot}\right)^{-0.7}
\left(\frac{M_{\rm ext}}{0.01\ M_\odot}\right)^{0.7},
\end{equation}
which, for the fiducial SESN parameters adopted in this paper, is naturally of order $10^{49}$~erg and can plausibly reach $\sim 10^{50}$~erg depending on the extended-material or CSM configuration \citep[e.g.,][]{Nyholm2020,Hiramatsu2024}. In blowout, the engine must deposit an energy comparable to the explosion energy before transparency and release it rapidly. The characteristic early energy scale therefore satisfies $E_{\rm bo}\gtrsim E_{\rm sn}$ (Section~\ref{sec:blowout-condition}), and is typically orders of magnitude larger than $E_{\rm ext}$ for comparable $M_{\rm ej}$ and $E_{\rm sn}$.

Third, for a fixed observed duration $t_{\rm first}$, this energy contrast maps directly into peak luminosity and hence multiband detectability. For shock cooling, a characteristic luminosity scale is $L_{\rm ext}\sim E_{\rm ext}/t_{\rm first}\sim 10^{44}\ {\rm erg\,s^{-1}}\,(E_{\rm ext}/10^{49}\ {\rm erg})\,(t_{\rm first}/{\rm day})^{-1}$. In blowout, the first-peak luminosity is set by the energy stored in the admixture layer and released over the broadening timescale, $L_{\rm first}\sim E_{\rm kin,adm}/t_{\rm bro}$, and asymptotes to the limiting regimes in Eqs.~\eqref{eq:first_peak_Lbol_1}--\eqref{eq:first_peak_Lbol_2}. For fiducial SESN parameters this corresponds to $L_{\rm first}\sim 10^{46}\ {\rm erg\,s^{-1}}$, indicating that blowout can be brighter than shock cooling by orders of magnitude at the same $t_{\rm first}$.

Finally, the temperature and color evolution provide an additional discriminator. For shock cooling, the effective temperature depends sensitively on the radius of the extended material and can therefore span a broad range. For the same observed first-peak duration $t_{\rm first}$ and a comparable opacity normalization, a commonly used scaling is \citep{Nakar&Piro2014}
\begin{equation}
T_{\rm eff}\sim 4\times 10^{4}\ {\rm K}\,
\left(\frac{\kappa}{0.1\ {\rm cm^{2}\,g^{-1}}}\right)^{0.25}
\left(\frac{t_{\rm first}}{1\ {\rm day}}\right)^{-0.5}
\left(\frac{R_{\rm ext}}{10^{13}\ {\rm cm}}\right)^{0.25}.
\end{equation}
Accordingly, CSM-involved scenarios do not necessarily guarantee a strong UV excess, and the first-peak optical colors can show substantial event-to-event diversity, largely reflecting the spread in $R_{\rm ext}$ and other CSM properties. In the blowout picture, the rapid early expansion largely suppresses sensitivity to the progenitor radius and tends to keep the first peak hot and blue. Evaluated under the same fiducial SESN parameter set and at the same $t_{\rm first}$, our analytic estimate gives $T_{\rm eff,1}\sim 10^{5}$~K (Eq.~\eqref{eq:Teff_first_numeric}), implying a prominent UV contribution and very blue optical colors.

Taken together, blowout is distinguished not merely by a double-peaked morphology, but by the combination of (i) a coupled two-peak fit under a single physically allowed parameter set, (ii) a large early energy scale set by $E_{\rm bo}\gtrsim E_{\rm sn}$, (iii) a brighter first peak, and (iv) a typically hot and UV-bright early color temperature. Early-time UV coverage and well-sampled multiband color evolution therefore provide the most direct observational tests. We emphasize that the diversity of CSM and extended-envelope configurations can lead to partial overlap in phenomenology for some parameter choices, and the separation is more robust when the two peaks provide a test of parameter consistency, with the energy scale, first-peak brightness, and color evolution jointly constrained.

\section{Conclusion}
\label{sec:conclusion}

We have developed a minimal semi-analytic, survey-ready framework for pulsar-driven SNe in the blowout regime, focusing on systems in which the central engine deposits an energy comparable to the explosion energy before the ejecta become globally transparent. Motivated by multidimensional RHD simulations, we model the post-blowout evolution with an admixture layer characterized by an approximately radius-independent kinetic-energy flux, and use this structure together with conservation laws to compute self-consistent bolometric and multi-band light curves. Within this framework, blowout generically produces a double-peaked light curve, and the model can explain the observed morphology of some double-peaked SLSNe-I, such as LSQ14bdq.

From parameter-space scans, we find that in wind bubble blowout SESNe the first peak can be extremely blue and rapidly evolving, so optical-only surveys may miss a non-negligible fraction of such early bumps, whereas UV coverage is particularly informative. For lower-mass/low-energy ultra-stripped and AIC/MIC-like explosions, the blowout condition is easier to satisfy, and soft X-ray counterparts can become more favorable over parts of the engine parameter space; in some AIC/MIC cases, the main peak may also shift into the soft X-ray band, emphasizing the diagnostic potential of X-ray time-domain observations for these channels.

Future work will relax the near-adiabatic pre-blowout assumption and add time-dependent nonthermal emission to broaden the framework to weaker engines and multi-messenger observables.

\begin{acknowledgments}
This work was supported in part by the Advanced Graduate School Research Initiative for International Scholarly Excellence Program (AGS RISE Program) at Tohoku University (M.C.), and by the Japan Society for the Promotion of Science (JSPS) KAKENHI Grant Number JP22H00130, JP23H04899, and JP24K00668 (K.K.).
\end{acknowledgments}

\bibliography{references}{}

@ARTICLE{Chevalier1982,
       author = {{Chevalier}, R.~A.},
        title = "{Self-similar solutions for the interaction of stellar ejecta with an external medium.}",
      journal = {\apj},
         year = 1982,
        month = jul,
       volume = {258},
        pages = {790-797},
          doi = {10.1086/160126},
       adsurl = {https://ui.adsabs.harvard.edu/abs/1982ApJ...258..790C}
}

@ARTICLE{Chevalier1992,
       author = {{Chevalier}, Roger A. and {Fransson}, Claes},
        title = "{Pulsar Nebulae in Supernovae}",
      journal = {\apj},
         year = 1992,
        month = aug,
       volume = {395},
        pages = {540},
          doi = {10.1086/171674},
       adsurl = {https://ui.adsabs.harvard.edu/abs/1992ApJ...395..540C}
}

@ARTICLE{Arnett1982,
       author = {{Arnett}, W.~D.},
        title = "{Type I supernovae. I - Analytic solutions for the early part of the light curve}",
      journal = {\apj},
         year = 1982,
        month = feb,
       volume = {253},
        pages = {785-797},
          doi = {10.1086/159681},
       adsurl = {https://ui.adsabs.harvard.edu/abs/1982ApJ...253..785A}
}

@ARTICLE{Suzuki&Maeda2019,
       author = {{Suzuki}, Akihiro and {Maeda}, Keiichi},
        title = "{Three-dimensional Hydrodynamic Simulations of Supernova Ejecta with a Central Energy Source}",
      journal = {\apj},
         year = 2019,
        month = aug,
       volume = {880},
       number = {2},
          eid = {150},
        pages = {150},
          doi = {10.3847/1538-4357/ab2ad3},
archivePrefix = {arXiv},
       eprint = {1906.07381},
 primaryClass = {astro-ph.HE},
       adsurl = {https://ui.adsabs.harvard.edu/abs/2019ApJ...880..150S}
}

@ARTICLE{Suzuki&Maeda2021,
       author = {{Suzuki}, Akihiro and {Maeda}, Keiichi},
        title = "{Two-dimensional Radiation-hydrodynamic Simulations of Supernova Ejecta with a Central Power Source}",
      journal = {\apj},
         year = 2021,
        month = feb,
       volume = {908},
       number = {2},
          eid = {217},
        pages = {217},
          doi = {10.3847/1538-4357/abd54c},
archivePrefix = {arXiv},
       eprint = {2012.10057},
 primaryClass = {astro-ph.HE},
}

@ARTICLE{Kashiyama2015,
       author = {{Kashiyama}, Kazumi and {Quataert}, Eliot},
        title = "{Fast luminous blue transients from newborn black holes}",
      journal = {\mnras},
         year = 2015,
        month = aug,
       volume = {451},
       number = {3},
        pages = {2656-2662},
          doi = {10.1093/mnras/stv1164},
archivePrefix = {arXiv},
       eprint = {1504.05582},
 primaryClass = {astro-ph.HE},
       adsurl = {https://ui.adsabs.harvard.edu/abs/2015MNRAS.451.2656K}
}

@article{Kashiyama2016,
       author = {{Kashiyama}, Kazumi and {Murase}, Kohta and {Bartos}, Imre and {Kiuchi}, Kenta and {Margutti}, Raffaella},
        title = "{Multi-messenger Tests for Fast-spinning Newborn Pulsars Embedded in Stripped-envelope Supernovae}",
      journal = {\apj},
         year = 2016,
        month = feb,
       volume = {818},
       number = {1},
          eid = {94},
        pages = {94},
          doi = {10.3847/0004-637X/818/1/94},
archivePrefix = {arXiv},
       eprint = {1508.04393},
 primaryClass = {astro-ph.HE},
       adsurl = {https://ui.adsabs.harvard.edu/abs/2016ApJ...818...94K}
}

@ARTICLE{Kasen2016,
       author = {{Kasen}, Daniel and {Metzger}, Brian D. and {Bildsten}, Lars},
        title = "{Magnetar-driven Shock Breakout and Double-peaked Supernova Light Curves}",
      journal = {\apj},
         year = 2016,
        month = apr,
       volume = {821},
       number = {1},
          eid = {36},
        pages = {36},
          doi = {10.3847/0004-637X/821/1/36},
archivePrefix = {arXiv},
       eprint = {1507.03645},
 primaryClass = {astro-ph.HE},
       adsurl = {https://ui.adsabs.harvard.edu/abs/2016ApJ...821...36K}
}

@article{Matzner&McKee1999,
       author = {{Matzner}, Christopher D. and {McKee}, Christopher F.},
        title = "{The Expulsion of Stellar Envelopes in Core-Collapse Supernovae}",
      journal = {\apj},
         year = 1999,
        month = jan,
       volume = {510},
       number = {1},
        pages = {379-403},
          doi = {10.1086/306571},
archivePrefix = {arXiv},
       eprint = {astro-ph/9807046},
 primaryClass = {astro-ph},
       adsurl = {https://ui.adsabs.harvard.edu/abs/1999ApJ...510..379M}
}

@article{Harding2006,
       author = {{Harding}, Alice K. and {Lai}, Dong},
        title = "{Physics of strongly magnetized neutron stars}",
      journal = {Reports on Progress in Physics},
         year = 2006,
        month = sep,
       volume = {69},
       number = {9},
        pages = {2631-2708},
          doi = {10.1088/0034-4885/69/9/R03},
archivePrefix = {arXiv},
       eprint = {astro-ph/0606674},
 primaryClass = {astro-ph},
       adsurl = {https://ui.adsabs.harvard.edu/abs/2006RPPh...69.2631H}
}

@article{Tauris2013,
       author = {{Tauris}, T.~M. and {Langer}, N. and {Moriya}, T.~J. and {Podsiadlowski}, Ph. and {Yoon}, S.-C. and {Blinnikov}, S.~I.},
        title = "{Ultra-stripped Type Ic Supernovae from Close Binary Evolution}",
      journal = {\apjl},
         year = 2013,
        month = dec,
       volume = {778},
       number = {2},
          eid = {L23},
        pages = {L23},
          doi = {10.1088/2041-8205/778/2/L23},
archivePrefix = {arXiv},
       eprint = {1310.6356},
 primaryClass = {astro-ph.SR},
       adsurl = {https://ui.adsabs.harvard.edu/abs/2013ApJ...778L..23T}
}

@article{Tauris2015,
       author = {{Tauris}, Thomas M. and {Langer}, Norbert and {Podsiadlowski}, Philipp},
        title = "{Ultra-stripped supernovae: progenitors and fate}",
      journal = {\mnras},
         year = 2015,
        month = aug,
       volume = {451},
       number = {2},
        pages = {2123-2144},
          doi = {10.1093/mnras/stv990},
archivePrefix = {arXiv},
       eprint = {1505.00270},
 primaryClass = {astro-ph.SR},
       adsurl = {https://ui.adsabs.harvard.edu/abs/2015MNRAS.451.2123T}
}

@article{Piro2012,
       author = {{Piro}, Anthony L. and {Kulkarni}, S.~R.},
        title = "{Radio Transients from the Accretion-induced Collapse of White Dwarfs}",
      journal = {\apjl},
         year = 2013,
        month = jan,
       volume = {762},
       number = {2},
          eid = {L17},
        pages = {L17},
          doi = {10.1088/2041-8205/762/2/L17},
archivePrefix = {arXiv},
       eprint = {1211.0547},
 primaryClass = {astro-ph.HE},
       adsurl = {https://ui.adsabs.harvard.edu/abs/2013ApJ...762L..17P}
}

@ARTICLE{Ablimit2022,
       author = {{Ablimit}, Iminhaji and {Podsiadlowski}, Philipp and {Hirai}, Ryosuke and {Wicker}, James},
        title = "{Stellar core-merger-induced collapse: new formation pathways for black holes, Thorne-{\.Z}ytkow objects, magnetars, and superluminous supernovae}",
      journal = {\mnras},
         year = 2022,
        month = jul,
       volume = {513},
       number = {4},
        pages = {4802-4813},
          doi = {10.1093/mnras/stac631},
archivePrefix = {arXiv},
       eprint = {2108.08430},
 primaryClass = {astro-ph.HE},
       adsurl = {https://ui.adsabs.harvard.edu/abs/2022MNRAS.513.4802A}
}

@ARTICLE{Liu2021,
       author = {{Liu}, Liang-Duan and {Gao}, He and {Wang}, Xiao-Feng and {Yang}, Sheng},
        title = "{Magnetar-driven Shock Breakout Revisited and Implications for Double-peaked Type I Superluminous Supernovae}",
      journal = {\apj},
         year = 2021,
        month = apr,
       volume = {911},
       number = {2},
          eid = {142},
        pages = {142},
          doi = {10.3847/1538-4357/abf042},
archivePrefix = {arXiv},
       eprint = {2103.09971},
 primaryClass = {astro-ph.HE},
       adsurl = {https://ui.adsabs.harvard.edu/abs/2021ApJ...911..142L}
}

@ARTICLE{Campana2006,
       author = {{Campana}, S. and {Mangano}, V. and {Blustin}, A.~J. and {Brown}, P. and {Burrows}, D.~N. and {Chincarini}, G. and {Cummings}, J.~R. and {Cusumano}, G. and {Della Valle}, M. and {Malesani}, D. and {M{\'e}sz{\'a}ros}, P. and {Nousek}, J.~A. and {Page}, M. and {Sakamoto}, T. and {Waxman}, E. and {Zhang}, B. and {Dai}, Z.~G. and {Gehrels}, N. and {Immler}, S. and {Marshall}, F.~E. and {Mason}, K.~O. and {Moretti}, A. and {O'Brien}, P.~T. and {Osborne}, J.~P. and {Page}, K.~L. and {Romano}, P. and {Roming}, P.~W.~A. and {Tagliaferri}, G. and {Cominsky}, L.~R. and {Giommi}, P. and {Godet}, O. and {Kennea}, J.~A. and {Krimm}, H. and {Angelini}, L. and {Barthelmy}, S.~D. and {Boyd}, P.~T. and {Palmer}, D.~M. and {Wells}, A.~A. and {White}, N.~E.},
        title = "{The association of GRB 060218 with a supernova and the evolution of the shock wave}",
      journal = {\nat},
         year = 2006,
        month = aug,
       volume = {442},
       number = {7106},
        pages = {1008-1010},
          doi = {10.1038/nature04892},
archivePrefix = {arXiv},
       eprint = {astro-ph/0603279},
 primaryClass = {astro-ph},
       adsurl = {https://ui.adsabs.harvard.edu/abs/2006Natur.442.1008C}
}

@ARTICLE{Blondin&Chevalier2017,
       author = {{Blondin}, John M. and {Chevalier}, Roger A.},
        title = "{Pulsar Wind Bubble Blowout from a Supernova}",
      journal = {\apj},
         year = 2017,
        month = aug,
       volume = {845},
       number = {2},
          eid = {139},
        pages = {139},
          doi = {10.3847/1538-4357/aa8267},
archivePrefix = {arXiv},
       eprint = {1707.07021},
 primaryClass = {astro-ph.HE},
       adsurl = {https://ui.adsabs.harvard.edu/abs/2017ApJ...845..139B}
}

@ARTICLE{Ho2023,
       author = {{Ho}, Anna Y.~Q. and {Perley}, Daniel A. and {Gal-Yam}, Avishay and {Lunnan}, Ragnhild and {Sollerman}, Jesper and {Schulze}, Steve and {Das}, Kaustav K. and {Dobie}, Dougal and {Yao}, Yuhan and {Fremling}, Christoffer and {Adams}, Scott and {Anand}, Shreya and {Andreoni}, Igor and {Bellm}, Eric C. and {Bruch}, Rachel J. and {Burdge}, Kevin B. and {Castro-Tirado}, Alberto J. and {Dahiwale}, Aishwarya and {De}, Kishalay and {Dekany}, Richard and {Drake}, Andrew J. and {Duev}, Dmitry A. and {Graham}, Matthew J. and {Helou}, George and {Kaplan}, David L. and {Karambelkar}, Viraj and {Kasliwal}, Mansi M. and {Kool}, Erik C. and {Kulkarni}, S.~R. and {Mahabal}, Ashish A. and {Medford}, Michael S. and {Miller}, A.~A. and {Nordin}, Jakob and {Ofek}, Eran and {Petitpas}, Glen and {Riddle}, Reed and {Sharma}, Yashvi and {Smith}, Roger and {Stewart}, Adam J. and {Taggart}, Kirsty and {Tartaglia}, Leonardo and {Tzanidakis}, Anastasios and {Winters}, Jan Martin},
        title = "{A Search for Extragalactic Fast Blue Optical Transients in ZTF and the Rate of AT2018cow-like Transients}",
      journal = {\apj},
         year = 2023,
        month = jun,
       volume = {949},
       number = {2},
          eid = {120},
        pages = {120},
          doi = {10.3847/1538-4357/acc533},
archivePrefix = {arXiv},
       eprint = {2105.08811},
 primaryClass = {astro-ph.HE},
       adsurl = {https://ui.adsabs.harvard.edu/abs/2023ApJ...949..120H}
}

@article{Nicholl2015,
       author = {{Nicholl}, M. and {Smartt}, S.~J. and {Jerkstrand}, A. and {Sim}, S.~A. and {Inserra}, C. and {Anderson}, J.~P. and {Baltay}, C. and {Benetti}, S. and {Chambers}, K. and {Chen}, T.-W. and {Elias-Rosa}, N. and {Feindt}, U. and {Flewelling}, H.~A. and {Fraser}, M. and {Gal-Yam}, A. and {Galbany}, L. and {Huber}, M.~E. and {Kangas}, T. and {Kankare}, E. and {Kotak}, R. and {Kr{\"u}hler}, T. and {Maguire}, K. and {McKinnon}, R. and {Rabinowitz}, D. and {Rostami}, S. and {Schulze}, S. and {Smith}, K.~W. and {Sullivan}, M. and {Tonry}, J.~L. and {Valenti}, S. and {Young}, D.~R.},
        title = "{LSQ14bdq: A Type Ic Super-luminous Supernova with a Double-peaked Light Curve}",
      journal = {\apjl},
         year = 2015,
        month = jul,
       volume = {807},
       number = {1},
          eid = {L18},
        pages = {L18},
          doi = {10.1088/2041-8205/807/1/L18},
archivePrefix = {arXiv},
       eprint = {1505.01078},
 primaryClass = {astro-ph.SR},
       adsurl = {https://ui.adsabs.harvard.edu/abs/2015ApJ...807L..18N}
}

@article{piro2011,
       author = {{Piro}, Anthony L. and {Ott}, Christian D.},
        title = "{Supernova Fallback onto Magnetars and Propeller-powered Supernovae}",
      journal = {\apj},
         year = 2011,
        month = aug,
       volume = {736},
       number = {2},
          eid = {108},
        pages = {108},
          doi = {10.1088/0004-637X/736/2/108},
archivePrefix = {arXiv},
       eprint = {1104.0252},
 primaryClass = {astro-ph.HE},
       adsurl = {https://ui.adsabs.harvard.edu/abs/2011ApJ...736..108P}
}

@article{nicholl2015b,
       author = {{Nicholl}, M. and {Smartt}, S.~J. and {Jerkstrand}, A. and {Inserra}, C. and {Sim}, S.~A. and {Chen}, T.-W. and {Benetti}, S. and {Fraser}, M. and {Gal-Yam}, A. and {Kankare}, E. and {Maguire}, K. and {Smith}, K. and {Sullivan}, M. and {Valenti}, S. and {Young}, D.~R. and {Baltay}, C. and {Bauer}, F.~E. and {Baumont}, S. and {Bersier}, D. and {Botticella}, M.-T. and {Childress}, M. and {Dennefeld}, M. and {Della Valle}, M. and {Elias-Rosa}, N. and {Feindt}, U. and {Galbany}, L. and {Hadjiyska}, E. and {Le Guillou}, L. and {Leloudas}, G. and {Mazzali}, P. and {McKinnon}, R. and {Polshaw}, J. and {Rabinowitz}, D. and {Rostami}, S. and {Scalzo}, R. and {Schmidt}, B.~P. and {Schulze}, S. and {Sollerman}, J. and {Taddia}, F. and {Yuan}, F.},
        title = "{On the diversity of superluminous supernovae: ejected mass as the dominant factor}",
      journal = {\mnras},
         year = 2015,
        month = oct,
       volume = {452},
       number = {4},
        pages = {3869-3893},
          doi = {10.1093/mnras/stv1522},
archivePrefix = {arXiv},
       eprint = {1503.03310},
 primaryClass = {astro-ph.SR},
       adsurl = {https://ui.adsabs.harvard.edu/abs/2015MNRAS.452.3869N}
}

@article{orellana2022,
       author = {{Orellana}, Mariana and {Bersten}, Melina C.},
        title = "{Supernova double-peaked light curves from double-nickel distribution}",
      journal = {\aap},
         year = 2022,
        month = nov,
       volume = {667},
          eid = {A92},
        pages = {A92},
          doi = {10.1051/0004-6361/202244124},
archivePrefix = {arXiv},
       eprint = {2209.03923},
 primaryClass = {astro-ph.SR},
       adsurl = {https://ui.adsabs.harvard.edu/abs/2022A&A...667A..92O}
}

@article{nicholl2016,
       author = {{Nicholl}, M. and {Smartt}, S.~J.},
        title = "{Seeing double: the frequency and detectability of double-peaked superluminous supernova light curves}",
      journal = {\mnras},
         year = 2016,
        month = mar,
       volume = {457},
       number = {1},
        pages = {L79-L83},
          doi = {10.1093/mnrasl/slv210},
archivePrefix = {arXiv},
       eprint = {1511.03740},
 primaryClass = {astro-ph.SR},
       adsurl = {https://ui.adsabs.harvard.edu/abs/2016MNRAS.457L..79N}
}

@article{Frohmaier2021,
       author = {{Frohmaier}, C. and {Angus}, C.~R. and {Vincenzi}, M. and {Sullivan}, M. and {Smith}, M. and {Nugent}, P.~E. and {Cenko}, S.~B. and {Gal-Yam}, A. and {Kulkarni}, S.~R. and {Law}, N.~M. and {Quimby}, R.~M.},
        title = "{From core collapse to superluminous: the rates of massive stellar explosions from the Palomar Transient Factory}",
      journal = {\mnras},
         year = 2021,
        month = jan,
       volume = {500},
       number = {4},
        pages = {5142-5158},
          doi = {10.1093/mnras/staa3607},
archivePrefix = {arXiv},
       eprint = {2010.15270},
 primaryClass = {astro-ph.HE},
       adsurl = {https://ui.adsabs.harvard.edu/abs/2021MNRAS.500.5142F}
}

@article{Yonetoku2025,
  title={Concept of high-z gamma-ray bursts unraveling the dark ages and extreme space-time mission—HiZ-GUNDAM},
  author={Yonetoku, Daisuke and Doi, Akihiro and Mihara, Tatehiro and Matsuhara, Hideo and Sakamoto, Takanori and Tsumura, Kohji and Ioka, Kunihito and Arimoto, Makoto and Ando, Yoshiyuki and Enoto, Teruaki and others},
  journal={Journal of Astronomical Telescopes, Instruments, and Systems},
  volume={11},
  number={4},
  pages={044002--044002},
  year={2025},
  publisher={Society of Photo-Optical Instrumentation Engineers}
}

@article{Graham2019,
       author = {{Graham}, Matthew J. and {Kulkarni}, S.~R. and {Bellm}, Eric C. and {Adams}, Scott M. and {Barbarino}, Cristina and {Blagorodnova}, Nadejda and {Bodewits}, Dennis and {Bolin}, Bryce and {Brady}, Patrick R. and {Cenko}, S. Bradley and {Chang}, Chan-Kao and {Coughlin}, Michael W. and {De}, Kishalay and {Eadie}, Gwendolyn and {Farnham}, Tony L. and {Feindt}, Ulrich and {Franckowiak}, Anna and {Fremling}, Christoffer and {Gezari}, Suvi and {Ghosh}, Shaon and {Goldstein}, Daniel A. and {Golkhou}, V. Zach and {Goobar}, Ariel and {Ho}, Anna Y.~Q. and {Huppenkothen}, Daniela and {Ivezi{\'c}}, {\v{Z}}eljko and {Jones}, R. Lynne and {Juric}, Mario and {Kaplan}, David L. and {Kasliwal}, Mansi M. and {Kelley}, Michael S.~P. and {Kupfer}, Thomas and {Lee}, Chien-De and {Lin}, Hsing Wen and {Lunnan}, Ragnhild and {Mahabal}, Ashish A. and {Miller}, Adam A. and {Ngeow}, Chow-Choong and {Nugent}, Peter and {Ofek}, Eran O. and {Prince}, Thomas A. and {Rauch}, Ludwig and {van Roestel}, Jan and {Schulze}, Steve and {Singer}, Leo P. and {Sollerman}, Jesper and {Taddia}, Francesco and {Yan}, Lin and {Ye}, Quan-Zhi and {Yu}, Po-Chieh and {Barlow}, Tom and {Bauer}, James and {Beck}, Ron and {Belicki}, Justin and {Biswas}, Rahul and {Brinnel}, Valery and {Brooke}, Tim and {Bue}, Brian and {Bulla}, Mattia and {Burruss}, Rick and {Connolly}, Andrew and {Cromer}, John and {Cunningham}, Virginia and {Dekany}, Richard and {Delacroix}, Alex and {Desai}, Vandana and {Duev}, Dmitry A. and {Feeney}, Michael and {Flynn}, David and {Frederick}, Sara and {Gal-Yam}, Avishay and {Giomi}, Matteo and {Groom}, Steven and {Hacopians}, Eugean and {Hale}, David and {Helou}, George and {Henning}, John and {Hover}, David and {Hillenbrand}, Lynne A. and {Howell}, Justin and {Hung}, Tiara and {Imel}, David and {Ip}, Wing-Huen and {Jackson}, Edward and {Kaspi}, Shai and {Kaye}, Stephen and {Kowalski}, Marek and {Kramer}, Emily and {Kuhn}, Michael and {Landry}, Walter and {Laher}, Russ R. and {Mao}, Peter and {Masci}, Frank J. and {Monkewitz}, Serge and {Murphy}, Patrick and {Nordin}, Jakob and {Patterson}, Maria T. and {Penprase}, Bryan and {Porter}, Michael and {Rebbapragada}, Umaa and {Reiley}, Dan and {Riddle}, Reed and {Rigault}, Mickael and {Rodriguez}, Hector and {Rusholme}, Ben and {van Santen}, Jakob and {Shupe}, David L. and {Smith}, Roger M. and {Soumagnac}, Maayane T. and {Stein}, Robert and {Surace}, Jason and {Szkody}, Paula and {Terek}, Scott and {Van Sistine}, Angela and {van Velzen}, Sjoert and {Vestrand}, W. Thomas and {Walters}, Richard and {Ward}, Charlotte and {Zhang}, Chaoran and {Zolkower}, Jeffry},
        title = "{The Zwicky Transient Facility: Science Objectives}",
      journal = {\pasp},
         year = 2019,
        month = jul,
       volume = {131},
       number = {1001},
        pages = {078001},
          doi = {10.1088/1538-3873/ab006c},
archivePrefix = {arXiv},
       eprint = {1902.01945},
 primaryClass = {astro-ph.IM},
       adsurl = {https://ui.adsabs.harvard.edu/abs/2019PASP..131g8001G}
}

@article{Leloudas2012,
       author = {{Leloudas}, G. and {Chatzopoulos}, E. and {Dilday}, B. and {Gorosabel}, J. and {Vinko}, J. and {Gallazzi}, A. and {Wheeler}, J.~C. and {Bassett}, B. and {Fischer}, J.~A. and {Frieman}, J.~A. and {Fynbo}, J.~P.~U. and {Goobar}, A. and {Jel{\'\i}nek}, M. and {Malesani}, D. and {Nichol}, R.~C. and {Nordin}, J. and {{\"O}stman}, L. and {Sako}, M. and {Schneider}, D.~P. and {Smith}, M. and {Sollerman}, J. and {Stritzinger}, M.~D. and {Th{\"o}ne}, C.~C. and {de Ugarte Postigo}, A.},
        title = "{SN 2006oz: rise of a super-luminous supernova observed by the SDSS-II SN Survey}",
      journal = {\aap},
         year = 2012,
        month = may,
       volume = {541},
          eid = {A129},
        pages = {A129},
          doi = {10.1051/0004-6361/201118498},
archivePrefix = {arXiv},
       eprint = {1201.5393},
 primaryClass = {astro-ph.SR},
       adsurl = {https://ui.adsabs.harvard.edu/abs/2012A&A...541A.129L}
}

@article{Smith2016,
       author = {{Smith}, M. and {Sullivan}, M. and {D'Andrea}, C.~B. and {Castander}, F.~J. and {Casas}, R. and {Prajs}, S. and {Papadopoulos}, A. and {Nichol}, R.~C. and {Karpenka}, N.~V. and {Bernard}, S.~R. and {Brown}, P. and {Cartier}, R. and {Cooke}, J. and {Curtin}, C. and {Davis}, T.~M. and {Finley}, D.~A. and {Foley}, R.~J. and {Gal-Yam}, A. and {Goldstein}, D.~A. and {Gonz{\'a}lez-Gait{\'a}n}, S. and {Gupta}, R.~R. and {Howell}, D.~A. and {Inserra}, C. and {Kessler}, R. and {Lidman}, C. and {Marriner}, J. and {Nugent}, P. and {Pritchard}, T.~A. and {Sako}, M. and {Smartt}, S. and {Smith}, R.~C. and {Spinka}, H. and {Thomas}, R.~C. and {Wolf}, R.~C. and {Zenteno}, A. and {Abbott}, T.~M.~C. and {Benoit-L{\'e}vy}, A. and {Bertin}, E. and {Brooks}, D. and {Buckley-Geer}, E. and {Carnero Rosell}, A. and {Carrasco Kind}, M. and {Carretero}, J. and {Crocce}, M. and {Cunha}, C.~E. and {da Costa}, L.~N. and {Desai}, S. and {Diehl}, H.~T. and {Doel}, P. and {Estrada}, J. and {Evrard}, A.~E. and {Flaugher}, B. and {Fosalba}, P. and {Frieman}, J. and {Gerdes}, D.~W. and {Gruen}, D. and {Gruendl}, R.~A. and {James}, D.~J. and {Kuehn}, K. and {Kuropatkin}, N. and {Lahav}, O. and {Li}, T.~S. and {Marshall}, J.~L. and {Martini}, P. and {Miller}, C.~J. and {Miquel}, R. and {Nord}, B. and {Ogando}, R. and {Plazas}, A.~A. and {Reil}, K. and {Romer}, A.~K. and {Roodman}, A. and {Rykoff}, E.~S. and {Sanchez}, E. and {Scarpine}, V. and {Schubnell}, M. and {Sevilla-Noarbe}, I. and {Soares-Santos}, M. and {Sobreira}, F. and {Suchyta}, E. and {Swanson}, M.~E.~C. and {Tarle}, G. and {Walker}, A.~R. and {Wester}, W. and {DES Collaboration}},
        title = "{DES14X3taz: A Type I Superluminous Supernova Showing a Luminous, Rapidly Cooling Initial Pre-peak Bump}",
      journal = {\apjl},
         year = 2016,
        month = feb,
       volume = {818},
       number = {1},
          eid = {L8},
        pages = {L8},
          doi = {10.3847/2041-8205/818/1/L8},
archivePrefix = {arXiv},
       eprint = {1512.06043},
 primaryClass = {astro-ph.SR},
       adsurl = {https://ui.adsabs.harvard.edu/abs/2016ApJ...818L...8S}
}

@article{Nicholl2021,
  title={Superluminous supernovae: an explosive decade},
  author={Nicholl, Matt},
  journal={arXiv preprint arXiv:2109.08697},
  year={2021}
}

@ARTICLE{Gruzinov2005,
       author = {{Gruzinov}, Andrei},
        title = "{Power of an Axisymmetric Pulsar}",
      journal = {\prl},
         year = 2005,
        month = jan,
       volume = {94},
       number = {2},
          eid = {021101},
        pages = {021101},
          doi = {10.1103/PhysRevLett.94.021101},
archivePrefix = {arXiv},
       eprint = {astro-ph/0407279},
 primaryClass = {astro-ph},
       adsurl = {https://ui.adsabs.harvard.edu/abs/2005PhRvL..94b1101G}
}

@ARTICLE{Spitkovsky2006,
       author = {{Spitkovsky}, Anatoly},
        title = "{Time-dependent Force-free Pulsar Magnetospheres: Axisymmetric and Oblique Rotators}",
      journal = {\apjl},
         year = 2006,
        month = sep,
       volume = {648},
       number = {1},
        pages = {L51-L54},
          doi = {10.1086/507518},
archivePrefix = {arXiv},
       eprint = {astro-ph/0603147},
 primaryClass = {astro-ph},
       adsurl = {https://ui.adsabs.harvard.edu/abs/2006ApJ...648L..51S}
}

@ARTICLE{Tchekhovskoy2013,
       author = {{Tchekhovskoy}, A. and {Spitkovsky}, A. and {Li}, J.~G.},
        title = "{Time-dependent 3D magnetohydrodynamic pulsar magnetospheres: oblique  rotators.}",
      journal = {\mnras},
         year = 2013,
        month = aug,
       volume = {435},
        pages = {L1-L5},
          doi = {10.1093/mnrasl/slt076},
archivePrefix = {arXiv},
       eprint = {1211.2803},
 primaryClass = {astro-ph.HE},
       adsurl = {https://ui.adsabs.harvard.edu/abs/2013MNRAS.435L...1T}
}

@ARTICLE{Blondin2001,
       author = {{Blondin}, John M. and {Chevalier}, Roger A. and {Frierson}, Dargan M.},
        title = "{Pulsar Wind Nebulae in Evolved Supernova Remnants}",
      journal = {\apj},
         year = 2001,
        month = dec,
       volume = {563},
       number = {2},
        pages = {806-815},
          doi = {10.1086/324042},
archivePrefix = {arXiv},
       eprint = {astro-ph/0107076},
 primaryClass = {astro-ph},
       adsurl = {https://ui.adsabs.harvard.edu/abs/2001ApJ...563..806B}
}

@ARTICLE{Jun1998,
       author = {{Jun}, Byung-Il},
        title = "{Interaction of a Pulsar Wind with the Expanding Supernova Remnant}",
      journal = {\apj},
         year = 1998,
        month = may,
       volume = {499},
       number = {1},
        pages = {282-293},
          doi = {10.1086/305627},
archivePrefix = {arXiv},
       eprint = {astro-ph/9712054},
 primaryClass = {astro-ph},
       adsurl = {https://ui.adsabs.harvard.edu/abs/1998ApJ...499..282J}
}

@ARTICLE{Moriya2018,
       author = {{Moriya}, Takashi J. and {Sorokina}, Elena I. and {Chevalier}, Roger A.},
        title = "{Superluminous Supernovae}",
      journal = {\ssr},
         year = 2018,
        month = mar,
       volume = {214},
       number = {2},
          eid = {59},
        pages = {59},
          doi = {10.1007/s11214-018-0493-6},
archivePrefix = {arXiv},
       eprint = {1803.01875},
 primaryClass = {astro-ph.HE},
       adsurl = {https://ui.adsabs.harvard.edu/abs/2018SSRv..214...59M}
}

@ARTICLE{Kisaka2015,
       author = {{Kisaka}, Shota and {Ioka}, Kunihito and {Takami}, Hajime},
        title = "{Energy Sources and Light Curves of Macronovae}",
      journal = {\apj},
         year = 2015,
        month = apr,
       volume = {802},
       number = {2},
          eid = {119},
        pages = {119},
          doi = {10.1088/0004-637X/802/2/119},
archivePrefix = {arXiv},
       eprint = {1410.0966},
 primaryClass = {astro-ph.HE},
       adsurl = {https://ui.adsabs.harvard.edu/abs/2015ApJ...802..119K}
}

@ARTICLE{Yuan2025,
       author = {{Yuan}, Weimin and {Dai}, Lixin and {Feng}, Hua and {Jin}, Chichuan and {Jonker}, Peter and {Kuulkers}, Erik and {Liu}, Yuan and {Nandra}, Kirpal and {O'Brien}, Paul and {Piro}, Luigi and {Rau}, Arne and {Rea}, Nanda and {Sanders}, Jeremy and {Tao}, Lian and {Wang}, Junfeng and {Wu}, Xuefeng and {Zhang}, Bing and {Zhang}, Shuangnan and {Ai}, Shunke and {Buchner}, Johannes and {Bulbul}, Esra and {Chen}, Hechao and {Chen}, Minghua and {Chen}, Yong and {Chen}, Yu-Peng and {Coleiro}, Alexis and {Coti Zelati}, Francesco and {Dai}, Zigao and {Fan}, Xilong and {Fan}, Zhou and {Friedrich}, Susanne and {Gao}, He and {Ge}, Chong and {Ge}, Mingyu and {Geng}, Jinjun and {Ghirlanda}, Giancarlo and {Gianfagna}, Giulia and {Gou}, Lijun and {Guillot}, S{\'e}bastien and {Hou}, Xian and {Hu}, Jingwei and {Huang}, Yongfeng and {Ji}, Long and {Jia}, Shumei and {Komossa}, S. and {Kong}, Albert K.~H. and {Lan}, Lin and {Li}, An and {Li}, Ang and {Li}, Chengkui and {Li}, Dongyue and {Li}, Jian and {Li}, Zhaosheng and {Ling}, Zhixing and {Liu}, Ang and {Liu}, Jinzhong and {Liu}, Liangduan and {Liu}, Zhu and {Luo}, Jiawei and {Ma}, Ruican and {Maggi}, Pierre and {Maitra}, Chandreyee and {Marino}, Alessio and {Ng}, Stephen Chi-Yung and {Pan}, Haiwu and {Rukdee}, Surangkhana and {Soria}, Roberto and {Sun}, Hui and {Tam}, Pak-Hin Thomas and {Thakur}, Aishwarya Linesh and {Tian}, Hui and {Troja}, Eleonora and {Wang}, Wei and {Wang}, Xiangyu and {Wang}, Yanan and {Wei}, Junjie and {Wen}, Sixiang and {Wu}, Jianfeng and {Wu}, Ting and {Xiao}, Di and {Xu}, Dong and {Xu}, Renxin and {Xu}, Yanjun and {Xu}, Yu and {Yang}, Haonan and {You}, Bei and {Yu}, Heng and {Yu}, Yunwei and {Zhang}, Binbin and {Zhang}, Chen and {Zhang}, Guobao and {Zhang}, Liang and {Zhang}, Wenda and {Zhang}, Yu and {Zhou}, Ping and {Zou}, Zecheng},
        title = "{Science objectives of the Einstein Probe mission}",
      journal = {Science China Physics, Mechanics, and Astronomy},
         year = 2025,
        month = mar,
       volume = {68},
       number = {3},
          eid = {239501},
        pages = {239501},
          doi = {10.1007/s11433-024-2600-3},
archivePrefix = {arXiv},
       eprint = {2501.07362},
 primaryClass = {astro-ph.HE},
       adsurl = {https://ui.adsabs.harvard.edu/abs/2025SCPMA..6839501Y}
}

@ARTICLE{De2018,
       author = {{De}, K. and {Kasliwal}, M.~M. and {Ofek}, E.~O. and {Moriya}, T.~J. and {Burke}, J. and {Cao}, Y. and {Cenko}, S.~B. and {Doran}, G.~B. and {Duggan}, G.~E. and {Fender}, R.~P. and {Fransson}, C. and {Gal-Yam}, A. and {Horesh}, A. and {Kulkarni}, S.~R. and {Laher}, R.~R. and {Lunnan}, R. and {Manulis}, I. and {Masci}, F. and {Mazzali}, P.~A. and {Nugent}, P.~E. and {Perley}, D.~A. and {Petrushevska}, T. and {Piro}, A.~L. and {Rumsey}, C. and {Sollerman}, J. and {Sullivan}, M. and {Taddia}, F.},
        title = "{A hot and fast ultra-stripped supernova that likely formed a compact neutron star binary}",
      journal = {Science},
         year = 2018,
        month = oct,
       volume = {362},
       number = {6411},
        pages = {201-206},
          doi = {10.1126/science.aas8693},
archivePrefix = {arXiv},
       eprint = {1810.05181},
 primaryClass = {astro-ph.HE},
       adsurl = {https://ui.adsabs.harvard.edu/abs/2018Sci...362..201D}
}

@ARTICLE{Yao2020,
       author = {{Yao}, Yuhan and {De}, Kishalay and {Kasliwal}, Mansi M. and {Ho}, Anna Y.~Q. and {Schulze}, Steve and {Li}, Zhihui and {Kulkarni}, S.~R. and {Fruchter}, Andrew and {Rubin}, David and {Perley}, Daniel A. and {Fuller}, Jim and {Piro}, Anthony L. and {Fremling}, C. and {Bellm}, Eric C. and {Burruss}, Rick and {Duev}, Dmitry A. and {Feeney}, Michael and {Gal-Yam}, Avishay and {Golkhou}, V. Zach and {Graham}, Matthew J. and {Helou}, George and {Kupfer}, Thomas and {Laher}, Russ R. and {Masci}, Frank J. and {Miller}, Adam A. and {Rusholme}, Ben and {Shupe}, David L. and {Smith}, Roger and {Sollerman}, Jesper and {Soumagnac}, Maayane T. and {Zolkower}, Jeffry},
        title = "{SN2019dge: A Helium-rich Ultra-stripped Envelope Supernova}",
      journal = {\apj},
         year = 2020,
        month = sep,
       volume = {900},
       number = {1},
          eid = {46},
        pages = {46},
          doi = {10.3847/1538-4357/abaa3d},
archivePrefix = {arXiv},
       eprint = {2005.12922},
 primaryClass = {astro-ph.HE},
       adsurl = {https://ui.adsabs.harvard.edu/abs/2020ApJ...900...46Y}
}

@ARTICLE{Agudo2023,
       author = {{Agudo}, I. and {Amati}, L. and {An}, T. and {Bauer}, F.~E. and {Benetti}, S. and {Bernardini}, M.~G. and {Beswick}, R. and {Bhirombhakdi}, K. and {de Boer}, T. and {Branchesi}, M. and {Brennan}, S.~J. and {Brocato}, E. and {Caballero-Garc{\'\i}a}, M.~D. and {Cappellaro}, E. and {Castro Rodr{\'\i}guez}, N. and {Castro-Tirado}, A.~J. and {Chambers}, K.~C. and {Chassande-Mottin}, E. and {Chaty}, S. and {Chen}, T.-W. and {Coleiro}, A. and {Covino}, S. and {D'Ammando}, F. and {D'Avanzo}, P. and {D'Elia}, V. and {Fiore}, A. and {Fl{\"o}rs}, A. and {Fraser}, M. and {Frey}, S. and {Frohmaier}, C. and {Fulton}, M. and {Galbany}, L. and {Gall}, C. and {Gao}, H. and {Garc{\'\i}a-Rojas}, J. and {Ghirlanda}, G. and {Giarratana}, S. and {Gillanders}, J.~H. and {Giroletti}, M. and {Gompertz}, B.~P. and {Gromadzki}, M. and {Heintz}, K.~E. and {Hjorth}, J. and {Hu}, Y.-D. and {Huber}, M.~E. and {Inkenhaag}, A. and {Izzo}, L. and {Jin}, Z.~P. and {Jonker}, P.~G. and {Kann}, D.~A. and {Kool}, E.~C. and {Kotak}, R. and {Leloudas}, G. and {Levan}, A.~J. and {Lin}, C.-C. and {Lyman}, J.~D. and {Magnier}, E.~A. and {Maguire}, K. and {Mandel}, I. and {Marcote}, B. and {Mata S{\'a}nchez}, D. and {Mattila}, S. and {Melandri}, A. and {Micha{\l}owski}, M.~J. and {Moldon}, J. and {Nicholl}, M. and {Nicuesa Guelbenzu}, A. and {Oates}, S.~R. and {Onori}, F. and {Orienti}, M. and {Paladino}, R. and {Paragi}, Z. and {Perez-Torres}, M. and {Pian}, E. and {Pignata}, G. and {Piranomonte}, S. and {Quirola-V{\'a}squez}, J. and {Ragosta}, F. and {Rau}, A. and {Ronchini}, S. and {Rossi}, A. and {S{\'a}nchez-Ram{\'\i}rez}, R. and {Salafia}, O.~S. and {Schulze}, S. and {Smartt}, S.~J. and {Smith}, K.~W. and {Sollerman}, J. and {Srivastav}, S. and {Starling}, R.~L.~C. and {Steeghs}, D. and {Stevance}, H.~F. and {Tanvir}, N.~R. and {Testa}, V. and {Torres}, M.~A.~P. and {Valeev}, A. and {Vergani}, S.~D. and {Vescovi}, D. and {Wainscost}, R. and {Watson}, D. and {Wiersema}, K. and {Wyrzykowski}, {\L}. and {Yang}, J. and {Yang}, S. and {Young}, D.~R.},
        title = "{Panning for gold, but finding helium: Discovery of the ultra-stripped supernova SN 2019wxt from gravitational-wave follow-up observations}",
      journal = {\aap},
         year = 2023,
        month = jul,
       volume = {675},
          eid = {A201},
        pages = {A201},
          doi = {10.1051/0004-6361/202244751},
archivePrefix = {arXiv},
       eprint = {2208.09000},
 primaryClass = {astro-ph.HE},
       adsurl = {https://ui.adsabs.harvard.edu/abs/2023A&A...675A.201A}
}

@ARTICLE{Sawada2022,
       author = {{Sawada}, Ryo and {Kashiyama}, Kazumi and {Suwa}, Yudai},
        title = "{On the Energy Source of Ultrastripped Supernovae}",
      journal = {\apj},
         year = 2022,
        month = mar,
       volume = {927},
       number = {2},
          eid = {223},
        pages = {223},
          doi = {10.3847/1538-4357/ac53ae},
archivePrefix = {arXiv},
       eprint = {2112.10782},
 primaryClass = {astro-ph.HE},
       adsurl = {https://ui.adsabs.harvard.edu/abs/2022ApJ...927..223S}
}

@ARTICLE{Bellm2019,
       author = {{Bellm}, Eric C. and {Kulkarni}, Shrinivas R. and {Graham}, Matthew J. and {Dekany}, Richard and {Smith}, Roger M. and {Riddle}, Reed and {Masci}, Frank J. and {Helou}, George and {Prince}, Thomas A. and {Adams}, Scott M. and {Barbarino}, C. and {Barlow}, Tom and {Bauer}, James and {Beck}, Ron and {Belicki}, Justin and {Biswas}, Rahul and {Blagorodnova}, Nadejda and {Bodewits}, Dennis and {Bolin}, Bryce and {Brinnel}, Valery and {Brooke}, Tim and {Bue}, Brian and {Bulla}, Mattia and {Burruss}, Rick and {Cenko}, S. Bradley and {Chang}, Chan-Kao and {Connolly}, Andrew and {Coughlin}, Michael and {Cromer}, John and {Cunningham}, Virginia and {De}, Kishalay and {Delacroix}, Alex and {Desai}, Vandana and {Duev}, Dmitry A. and {Eadie}, Gwendolyn and {Farnham}, Tony L. and {Feeney}, Michael and {Feindt}, Ulrich and {Flynn}, David and {Franckowiak}, Anna and {Frederick}, S. and {Fremling}, C. and {Gal-Yam}, Avishay and {Gezari}, Suvi and {Giomi}, Matteo and {Goldstein}, Daniel A. and {Golkhou}, V. Zach and {Goobar}, Ariel and {Groom}, Steven and {Hacopians}, Eugean and {Hale}, David and {Henning}, John and {Ho}, Anna Y.~Q. and {Hover}, David and {Howell}, Justin and {Hung}, Tiara and {Huppenkothen}, Daniela and {Imel}, David and {Ip}, Wing-Huen and {Ivezi{\'c}}, {\v{Z}}eljko and {Jackson}, Edward and {Jones}, Lynne and {Juric}, Mario and {Kasliwal}, Mansi M. and {Kaspi}, S. and {Kaye}, Stephen and {Kelley}, Michael S.~P. and {Kowalski}, Marek and {Kramer}, Emily and {Kupfer}, Thomas and {Landry}, Walter and {Laher}, Russ R. and {Lee}, Chien-De and {Lin}, Hsing Wen and {Lin}, Zhong-Yi and {Lunnan}, Ragnhild and {Giomi}, Matteo and {Mahabal}, Ashish and {Mao}, Peter and {Miller}, Adam A. and {Monkewitz}, Serge and {Murphy}, Patrick and {Ngeow}, Chow-Choong and {Nordin}, Jakob and {Nugent}, Peter and {Ofek}, Eran and {Patterson}, Maria T. and {Penprase}, Bryan and {Porter}, Michael and {Rauch}, Ludwig and {Rebbapragada}, Umaa and {Reiley}, Dan and {Rigault}, Mickael and {Rodriguez}, Hector and {van Roestel}, Jan and {Rusholme}, Ben and {van Santen}, Jakob and {Schulze}, S. and {Shupe}, David L. and {Singer}, Leo P. and {Soumagnac}, Maayane T. and {Stein}, Robert and {Surace}, Jason and {Sollerman}, Jesper and {Szkody}, Paula and {Taddia}, F. and {Terek}, Scott and {Van Sistine}, Angela and {van Velzen}, Sjoert and {Vestrand}, W. Thomas and {Walters}, Richard and {Ward}, Charlotte and {Ye}, Quan-Zhi and {Yu}, Po-Chieh and {Yan}, Lin and {Zolkower}, Jeffry},
        title = "{The Zwicky Transient Facility: System Overview, Performance, and First Results}",
      journal = {\pasp},
         year = 2019,
        month = jan,
       volume = {131},
       number = {995},
        pages = {018002},
          doi = {10.1088/1538-3873/aaecbe},
archivePrefix = {arXiv},
       eprint = {1902.01932},
 primaryClass = {astro-ph.IM},
       adsurl = {https://ui.adsabs.harvard.edu/abs/2019PASP..131a8002B}
}

@ARTICLE{Dessart2006,
       author = {{Dessart}, L. and {Burrows}, A. and {Ott}, C.~D. and {Livne}, E. and {Yoon}, S.-C. and {Langer}, N.},
        title = "{Multidimensional Simulations of the Accretion-induced Collapse of White Dwarfs to Neutron Stars}",
      journal = {\apj},
         year = 2006,
        month = jun,
       volume = {644},
       number = {2},
        pages = {1063-1084},
          doi = {10.1086/503626},
archivePrefix = {arXiv},
       eprint = {astro-ph/0601603},
 primaryClass = {astro-ph},
       adsurl = {https://ui.adsabs.harvard.edu/abs/2006ApJ...644.1063D}
}

@ARTICLE{Mori2025,
       author = {{Mori}, Masamitsu and {Sawada}, Ryo and {Suwa}, Yudai and {Tanikawa}, Ataru and {Kashiyama}, Kazumi and {Murase}, Kohta},
        title = "{Gravitational collapse of white dwarfs to neutron stars: From initial conditions to explosions with neutrino-radiation hydrodynamics simulations}",
      journal = {\pasj},
         year = 2025,
        month = feb,
       volume = {77},
       number = {1},
        pages = {127-138},
          doi = {10.1093/pasj/psae104},
archivePrefix = {arXiv},
       eprint = {2306.17381},
 primaryClass = {astro-ph.HE},
       adsurl = {https://ui.adsabs.harvard.edu/abs/2025PASJ...77..127M}
}

@ARTICLE{Nakar&Piro2014,
       author = {{Nakar}, Ehud and {Piro}, Anthony L.},
        title = "{Supernovae with Two Peaks in the Optical Light Curve and the Signature of Progenitors with Low-mass Extended Envelopes}",
      journal = {\apj},
         year = 2014,
        month = jun,
       volume = {788},
       number = {2},
          eid = {193},
        pages = {193},
          doi = {10.1088/0004-637X/788/2/193},
archivePrefix = {arXiv},
       eprint = {1401.7013},
 primaryClass = {astro-ph.HE},
       adsurl = {https://ui.adsabs.harvard.edu/abs/2014ApJ...788..193N}
}

@ARTICLE{Piro2015,
       author = {{Piro}, Anthony L.},
        title = "{Using Double-peaked Supernova Light Curves to Study Extended Material}",
      journal = {\apjl},
         year = 2015,
        month = aug,
       volume = {808},
       number = {2},
          eid = {L51},
        pages = {L51},
          doi = {10.1088/2041-8205/808/2/L51},
archivePrefix = {arXiv},
       eprint = {1505.07103},
 primaryClass = {astro-ph.HE},
       adsurl = {https://ui.adsabs.harvard.edu/abs/2015ApJ...808L..51P}
}

@ARTICLE{Chevalier&Irwin2011,
       author = {{Chevalier}, Roger A. and {Irwin}, Christopher M.},
        title = "{Shock Breakout in Dense Mass Loss: Luminous Supernovae}",
      journal = {\apjl},
         year = 2011,
        month = mar,
       volume = {729},
       number = {1},
          eid = {L6},
        pages = {L6},
          doi = {10.1088/2041-8205/729/1/L6},
archivePrefix = {arXiv},
       eprint = {1101.1111},
 primaryClass = {astro-ph.HE},
       adsurl = {https://ui.adsabs.harvard.edu/abs/2011ApJ...729L...6C}
}

@ARTICLE{Galama1998,
       author = {{Galama}, T.~J. and {Vreeswijk}, P.~M. and {van Paradijs}, J. and {Kouveliotou}, C. and {Augusteijn}, T. and {B{\"o}hnhardt}, H. and {Brewer}, J.~P. and {Doublier}, V. and {Gonzalez}, J.-F. and {Leibundgut}, B. and {Lidman}, C. and {Hainaut}, O.~R. and {Patat}, F. and {Heise}, J. and {in't Zand}, J. and {Hurley}, K. and {Groot}, P.~J. and {Strom}, R.~G. and {Mazzali}, P.~A. and {Iwamoto}, K. and {Nomoto}, K. and {Umeda}, H. and {Nakamura}, T. and {Young}, T.~R. and {Suzuki}, T. and {Shigeyama}, T. and {Koshut}, T. and {Kippen}, M. and {Robinson}, C. and {de Wildt}, P. and {Wijers}, R.~A.~M.~J. and {Tanvir}, N. and {Greiner}, J. and {Pian}, E. and {Palazzi}, E. and {Frontera}, F. and {Masetti}, N. and {Nicastro}, L. and {Feroci}, M. and {Costa}, E. and {Piro}, L. and {Peterson}, B.~A. and {Tinney}, C. and {Boyle}, B. and {Cannon}, R. and {Stathakis}, R. and {Sadler}, E. and {Begam}, M.~C. and {Ianna}, P.},
        title = "{An unusual supernova in the error box of the {\ensuremath{\gamma}}-ray burst of 25 April 1998}",
      journal = {\nat},
         year = 1998,
        month = oct,
       volume = {395},
       number = {6703},
        pages = {670-672},
          doi = {10.1038/27150},
archivePrefix = {arXiv},
       eprint = {astro-ph/9806175},
 primaryClass = {astro-ph},
       adsurl = {https://ui.adsabs.harvard.edu/abs/1998Natur.395..670G}
}

@ARTICLE{Pian2006,
       author = {{Pian}, E. and {Mazzali}, P.~A. and {Masetti}, N. and {Ferrero}, P. and {Klose}, S. and {Palazzi}, E. and {Ramirez-Ruiz}, E. and {Woosley}, S.~E. and {Kouveliotou}, C. and {Deng}, J. and {Filippenko}, A.~V. and {Foley}, R.~J. and {Fynbo}, J.~P.~U. and {Kann}, D.~A. and {Li}, W. and {Hjorth}, J. and {Nomoto}, K. and {Patat}, F. and {Sauer}, D.~N. and {Sollerman}, J. and {Vreeswijk}, P.~M. and {Guenther}, E.~W. and {Levan}, A. and {O'Brien}, P. and {Tanvir}, N.~R. and {Wijers}, R.~A.~M.~J. and {Dumas}, C. and {Hainaut}, O. and {Wong}, D.~S. and {Baade}, D. and {Wang}, L. and {Amati}, L. and {Cappellaro}, E. and {Castro-Tirado}, A.~J. and {Ellison}, S. and {Frontera}, F. and {Fruchter}, A.~S. and {Greiner}, J. and {Kawabata}, K. and {Ledoux}, C. and {Maeda}, K. and {M{\o}ller}, P. and {Nicastro}, L. and {Rol}, E. and {Starling}, R.},
        title = "{An optical supernova associated with the X-ray flash XRF 060218}",
      journal = {\nat},
         year = 2006,
        month = aug,
       volume = {442},
       number = {7106},
        pages = {1011-1013},
          doi = {10.1038/nature05082},
archivePrefix = {arXiv},
       eprint = {astro-ph/0603530},
 primaryClass = {astro-ph},
       adsurl = {https://ui.adsabs.harvard.edu/abs/2006Natur.442.1011P}
}

@ARTICLE{Chornock2010,
       author = {{Chornock}, Ryan and {Berger}, Edo and {Levesque}, Emily M. and {Soderberg}, Alicia M. and {Foley}, Ryan J. and {Fox}, Derek B. and {Frebel}, Anna and {Simon}, Joshua D. and {Bochanski}, John J. and {Challis}, Peter J. and {Kirshner}, Robert P. and {Podsiadlowski}, Philipp and {Roth}, Katherine and {Rutledge}, Robert E. and {Schmidt}, Brian P. and {Sheppard}, Scott S. and {Simcoe}, Robert A.},
        title = "{Spectroscopic Discovery of the Broad-Lined Type Ic Supernova 2010bh Associated with the Low-Redshift GRB 100316D}",
      journal = {arXiv e-prints},
         year = 2010,
        month = apr,
          eid = {arXiv:1004.2262},
        pages = {arXiv:1004.2262},
          doi = {10.48550/arXiv.1004.2262},
archivePrefix = {arXiv},
       eprint = {1004.2262},
 primaryClass = {astro-ph.HE},
       adsurl = {https://ui.adsabs.harvard.edu/abs/2010arXiv1004.2262C}
}

@ARTICLE{Wang2018,
       author = {{Wang}, J. and {Zhu}, Z.~P. and {Xu}, D. and {Xin}, L.~P. and {Deng}, J.~S. and {Qiu}, Y.~L. and {Qiu}, P. and {Wang}, H.~J. and {Zhang}, J.~B. and {Wei}, J.~Y.},
        title = "{Spectroscopy of the Type Ic Supernova SN 2017iuk Associated with Low-redshift GRB 171205A}",
      journal = {\apj},
         year = 2018,
        month = nov,
       volume = {867},
       number = {2},
          eid = {147},
        pages = {147},
          doi = {10.3847/1538-4357/aae6c3},
archivePrefix = {arXiv},
       eprint = {1810.03250},
 primaryClass = {astro-ph.HE},
       adsurl = {https://ui.adsabs.harvard.edu/abs/2018ApJ...867..147W}
}

@ARTICLE{Metzger2017,
       author = {{Metzger}, Brian D. and {Berger}, Edo and {Margalit}, Ben},
        title = "{Millisecond Magnetar Birth Connects FRB 121102 to Superluminous Supernovae and Long-duration Gamma-Ray Bursts}",
      journal = {\apj},
         year = 2017,
        month = may,
       volume = {841},
       number = {1},
          eid = {14},
        pages = {14},
          doi = {10.3847/1538-4357/aa633d},
archivePrefix = {arXiv},
       eprint = {1701.02370},
 primaryClass = {astro-ph.HE},
       adsurl = {https://ui.adsabs.harvard.edu/abs/2017ApJ...841...14M}
}

@ARTICLE{Murase2016,
       author = {{Murase}, Kohta and {Kashiyama}, Kazumi and {M{\'e}sz{\'a}ros}, Peter},
        title = "{A burst in a wind bubble and the impact on baryonic ejecta: high-energy gamma-ray flashes and afterglows from fast radio bursts and pulsar-driven supernova remnants}",
      journal = {\mnras},
         year = 2016,
        month = sep,
       volume = {461},
       number = {2},
        pages = {1498-1511},
          doi = {10.1093/mnras/stw1328},
archivePrefix = {arXiv},
       eprint = {1603.08875},
 primaryClass = {astro-ph.HE},
       adsurl = {https://ui.adsabs.harvard.edu/abs/2016MNRAS.461.1498M}
}

@ARTICLE{Kashiyama2017,
       author = {{Kashiyama}, Kazumi and {Murase}, Kohta},
        title = "{Testing the Young Neutron Star Scenario with Persistent Radio Emission Associated with FRB 121102}",
      journal = {\apjl},
         year = 2017,
        month = apr,
       volume = {839},
       number = {1},
          eid = {L3},
        pages = {L3},
          doi = {10.3847/2041-8213/aa68e1},
archivePrefix = {arXiv},
       eprint = {1701.04815},
 primaryClass = {astro-ph.HE},
       adsurl = {https://ui.adsabs.harvard.edu/abs/2017ApJ...839L...3K}
}

@ARTICLE{Murase2021,
       author = {{Murase}, Kohta and {Omand}, Conor M.~B. and {Coppejans}, Deanne L. and {Nagai}, Hiroshi and {Bower}, Geoffrey C. and {Chornock}, Ryan and {Fox}, Derek B. and {Kashiyama}, Kazumi and {Law}, Casey and {Margutti}, Raffaella and {M{\'e}sz{\'a}ros}, Peter},
        title = "{ALMA and NOEMA constraints on synchrotron nebular emission from embryonic superluminous supernova remnants and radio-gamma-ray connection}",
      journal = {\mnras},
         year = 2021,
        month = nov,
       volume = {508},
       number = {1},
        pages = {44-51},
          doi = {10.1093/mnras/stab2506},
archivePrefix = {arXiv},
       eprint = {2105.05239},
 primaryClass = {astro-ph.HE},
       adsurl = {https://ui.adsabs.harvard.edu/abs/2021MNRAS.508...44M}
}

@ARTICLE{Chakraborti2015,
       author = {{Chakraborti}, Sayan and {Soderberg}, Alicia and {Chomiuk}, Laura and {Kamble}, Atish and {Yadav}, Naveen and {Ray}, Alak and {Hurley}, Kevin and {Margutti}, Raffaella and {Milisavljevic}, Dan and {Bietenholz}, Michael and {Brunthaler}, Andreas and {Pignata}, Giuliano and {Pian}, Elena and {Mazzali}, Paolo and {Fransson}, Claes and {Bartel}, Norbert and {Hamuy}, Mario and {Levesque}, Emily and {MacFadyen}, Andrew and {Dittmann}, Jason and {Krauss}, Miriam and {Briggs}, M.~S. and {Connaughton}, V. and {Yamaoka}, K. and {Takahashi}, T. and {Ohno}, M. and {Fukazawa}, Y. and {Tashiro}, M. and {Terada}, Y. and {Murakami}, T. and {Goldsten}, J. and {Barthelmy}, S. and {Gehrels}, N. and {Cummings}, J. and {Krimm}, H. and {Palmer}, D. and {Golenetskii}, S. and {Aptekar}, R. and {Frederiks}, D. and {Svinkin}, D. and {Cline}, T. and {Mitrofanov}, I.~G. and {Golovin}, D. and {Litvak}, M.~L. and {Sanin}, A.~B. and {Boynton}, W. and {Fellows}, C. and {Harshman}, K. and {Enos}, H. and {von Kienlin}, A. and {Rau}, A. and {Zhang}, X. and {Savchenko}, V.},
        title = "{A Missing-link in the Supernova-GRB Connection: The Case of SN 2012ap}",
      journal = {\apj},
         year = 2015,
        month = jun,
       volume = {805},
       number = {2},
          eid = {187},
        pages = {187},
          doi = {10.1088/0004-637X/805/2/187},
archivePrefix = {arXiv},
       eprint = {1402.6336},
 primaryClass = {astro-ph.HE},
       adsurl = {https://ui.adsabs.harvard.edu/abs/2015ApJ...805..187C}
}

@ARTICLE{Suzuki&Maeda2018,
       author = {{Suzuki}, Akihiro and {Maeda}, Keiichi},
        title = "{Broad-band emission properties of central engine-powered supernova ejecta interacting with a circumstellar medium}",
      journal = {\mnras},
         year = 2018,
        month = jul,
       volume = {478},
       number = {1},
        pages = {110-125},
          doi = {10.1093/mnras/sty999},
archivePrefix = {arXiv},
       eprint = {1804.05397},
 primaryClass = {astro-ph.HE},
       adsurl = {https://ui.adsabs.harvard.edu/abs/2018MNRAS.478..110S}
}

@INCOLLECTION{Chevalier&Fransson2017,
       author = {{Chevalier}, Roger A. and {Fransson}, Claes},
        title = "{Thermal and Non-thermal Emission from Circumstellar Interaction}",
    booktitle = {Handbook of Supernovae},
         year = 2017,
       editor = {{Alsabti}, Athem W. and {Murdin}, Paul},
        pages = {875},
        publisher = {Springer},
          doi = {10.1007/978-3-319-21846-5_34},
       adsurl = {https://ui.adsabs.harvard.edu/abs/2017hsn..book..875C}
}

@ARTICLE{Tonry2018,
       author = {{Tonry}, J.~L. and {Denneau}, L. and {Heinze}, A.~N. and {Stalder}, B. and {Smith}, K.~W. and {Smartt}, S.~J. and {Stubbs}, C.~W. and {Weiland}, H.~J. and {Rest}, A.},
        title = "{ATLAS: A High-cadence All-sky Survey System}",
      journal = {\pasp},
         year = 2018,
        month = jun,
       volume = {130},
       number = {988},
        pages = {064505},
          doi = {10.1088/1538-3873/aabadf},
archivePrefix = {arXiv},
       eprint = {1802.00879},
 primaryClass = {astro-ph.IM},
       adsurl = {https://ui.adsabs.harvard.edu/abs/2018PASP..130f4505T}
}

@ARTICLE{GalYam&Avishay2019,
       author = {{Gal-Yam}, Avishay},
        title = "{The Most Luminous Supernovae}",
      journal = {\araa},
         year = 2019,
        month = aug,
       volume = {57},
        pages = {305-333},
          doi = {10.1146/annurev-astro-081817-051819},
archivePrefix = {arXiv},
       eprint = {1812.01428},
 primaryClass = {astro-ph.HE},
       adsurl = {https://ui.adsabs.harvard.edu/abs/2019ARA&A..57..305G}
}

@ARTICLE{Kasen2011,
       author = {{Kasen}, Daniel and {Woosley}, S.~E. and {Heger}, Alexander},
        title = "{Pair Instability Supernovae: Light Curves, Spectra, and Shock Breakout}",
      journal = {\apj},
         year = 2011,
        month = jun,
       volume = {734},
       number = {2},
          eid = {102},
        pages = {102},
          doi = {10.1088/0004-637X/734/2/102},
archivePrefix = {arXiv},
       eprint = {1101.3336},
 primaryClass = {astro-ph.HE},
       adsurl = {https://ui.adsabs.harvard.edu/abs/2011ApJ...734..102K}
}

@ARTICLE{Dessart2013,
       author = {{Dessart}, Luc and {Waldman}, Roni and {Livne}, Eli and {Hillier}, D. John and {Blondin}, St{\'e}phane},
        title = "{Radiative properties of pair-instability supernova explosions}",
      journal = {\mnras},
         year = 2013,
        month = feb,
       volume = {428},
       number = {4},
        pages = {3227-3251},
          doi = {10.1093/mnras/sts269},
archivePrefix = {arXiv},
       eprint = {1210.6163},
 primaryClass = {astro-ph.SR},
       adsurl = {https://ui.adsabs.harvard.edu/abs/2013MNRAS.428.3227D}
}

@ARTICLE{Ho2019,
       author = {{Ho}, Anna Y.~Q. and {Goldstein}, Daniel A. and {Schulze}, Steve and {Khatami}, David K. and {Perley}, Daniel A. and {Ergon}, Mattias and {Gal-Yam}, Avishay and {Corsi}, Alessandra and {Andreoni}, Igor and {Barbarino}, Cristina and {Bellm}, Eric C. and {Blagorodnova}, Nadia and {Bright}, Joe S. and {Burns}, E. and {Cenko}, S. Bradley and {Cunningham}, Virginia and {De}, Kishalay and {Dekany}, Richard and {Dugas}, Alison and {Fender}, Rob P. and {Fransson}, Claes and {Fremling}, Christoffer and {Goldstein}, Adam and {Graham}, Matthew J. and {Hale}, David and {Horesh}, Assaf and {Hung}, Tiara and {Kasliwal}, Mansi M. and {Kuin}, N. Paul M. and {Kulkarni}, S.~R. and {Kupfer}, Thomas and {Lunnan}, Ragnhild and {Masci}, Frank J. and {Ngeow}, Chow-Choong and {Nugent}, Peter E. and {Ofek}, Eran O. and {Patterson}, Maria T. and {Petitpas}, Glen and {Rusholme}, Ben and {Sai}, Hanna and {Sfaradi}, Itai and {Shupe}, David L. and {Sollerman}, Jesper and {Soumagnac}, Maayane T. and {Tachibana}, Yutaro and {Taddia}, Francesco and {Walters}, Richard and {Wang}, Xiaofeng and {Yao}, Yuhan and {Zhang}, Xinhan},
        title = "{Evidence for Late-stage Eruptive Mass Loss in the Progenitor to SN2018gep, a Broad-lined Ic Supernova: Pre-explosion Emission and a Rapidly Rising Luminous Transient}",
      journal = {\apj},
         year = 2019,
        month = dec,
       volume = {887},
       number = {2},
          eid = {169},
        pages = {169},
          doi = {10.3847/1538-4357/ab55ec},
archivePrefix = {arXiv},
       eprint = {1904.11009},
 primaryClass = {astro-ph.HE},
       adsurl = {https://ui.adsabs.harvard.edu/abs/2019ApJ...887..169H}
}

@ARTICLE{Mazzali2002,
       author = {{Mazzali}, P.~A. and {Deng}, J. and {Maeda}, K. and {Nomoto}, K. and {Umeda}, H. and {Hatano}, K. and {Iwamoto}, K. and {Yoshii}, Y. and {Kobayashi}, Y. and {Minezaki}, T. and {Doi}, M. and {Enya}, K. and {Tomita}, H. and {Smartt}, S.~J. and {Kinugasa}, K. and {Kawakita}, H. and {Ayani}, K. and {Kawabata}, T. and {Yamaoka}, H. and {Qiu}, Y.~L. and {Motohara}, K. and {Gerardy}, C.~L. and {Fesen}, R. and {Kawabata}, K.~S. and {Iye}, M. and {Kashikawa}, N. and {Kosugi}, G. and {Ohyama}, Y. and {Takada-Hidai}, M. and {Zhao}, G. and {Chornock}, R. and {Filippenko}, A.~V. and {Benetti}, S. and {Turatto}, M.},
        title = "{The Type Ic Hypernova SN 2002ap}",
      journal = {\apjl},
         year = 2002,
        month = jun,
       volume = {572},
       number = {1},
        pages = {L61-L65},
          doi = {10.1086/341504},
       adsurl = {https://ui.adsabs.harvard.edu/abs/2002ApJ...572L..61M}
}

@ARTICLE{Taddia2019,
       author = {{Taddia}, F. and {Sollerman}, J. and {Fremling}, C. and {Barbarino}, C. and {Karamehmetoglu}, E. and {Arcavi}, I. and {Cenko}, S.~B. and {Filippenko}, A.~V. and {Gal-Yam}, A. and {Hiramatsu}, D. and {Hosseinzadeh}, G. and {Howell}, D.~A. and {Kulkarni}, S.~R. and {Laher}, R. and {Lunnan}, R. and {Masci}, F. and {Nugent}, P.~E. and {Nyholm}, A. and {Perley}, D.~A. and {Quimby}, R. and {Silverman}, J.~M.},
        title = "{Analysis of broad-lined Type Ic supernovae from the (intermediate) Palomar Transient Factory}",
      journal = {\aap},
         year = 2019,
        month = jan,
       volume = {621},
          eid = {A71},
        pages = {A71},
          doi = {10.1051/0004-6361/201834429},
archivePrefix = {arXiv},
       eprint = {1811.09544},
 primaryClass = {astro-ph.HE},
       adsurl = {https://ui.adsabs.harvard.edu/abs/2019A&A...621A..71T}
}

@ARTICLE{Quimby2007,
       author = {{Quimby}, Robert M. and {Aldering}, Greg and {Wheeler}, J. Craig and {H{\"o}flich}, Peter and {Akerlof}, Carl W. and {Rykoff}, Eli S.},
        title = "{SN 2005ap: A Most Brilliant Explosion}",
      journal = {\apjl},
         year = 2007,
        month = oct,
       volume = {668},
       number = {2},
        pages = {L99-L102},
          doi = {10.1086/522862},
archivePrefix = {arXiv},
       eprint = {0709.0302},
 primaryClass = {astro-ph},
       adsurl = {https://ui.adsabs.harvard.edu/abs/2007ApJ...668L..99Q}
}

@ARTICLE{Pastorello2010,
       author = {{Pastorello}, A. and {Smartt}, S.~J. and {Botticella}, M.~T. and {Maguire}, K. and {Fraser}, M. and {Smith}, K. and {Kotak}, R. and {Magill}, L. and {Valenti}, S. and {Young}, D.~R. and {Gezari}, S. and {Bresolin}, F. and {Kudritzki}, R. and {Howell}, D.~A. and {Rest}, A. and {Metcalfe}, N. and {Mattila}, S. and {Kankare}, E. and {Huang}, K.~Y. and {Urata}, Y. and {Burgett}, W.~S. and {Chambers}, K.~C. and {Dombeck}, T. and {Flewelling}, H. and {Grav}, T. and {Heasley}, J.~N. and {Hodapp}, K.~W. and {Kaiser}, N. and {Luppino}, G.~A. and {Lupton}, R.~H. and {Magnier}, E.~A. and {Monet}, D.~G. and {Morgan}, J.~S. and {Onaka}, P.~M. and {Price}, P.~A. and {Rhoads}, P.~H. and {Siegmund}, W.~A. and {Stubbs}, C.~W. and {Sweeney}, W.~E. and {Tonry}, J.~L. and {Wainscoat}, R.~J. and {Waterson}, M.~F. and {Waters}, C. and {Wynn-Williams}, C.~G.},
        title = "{Ultra-bright Optical Transients are Linked with Type Ic Supernovae}",
      journal = {\apjl},
         year = 2010,
        month = nov,
       volume = {724},
       number = {1},
        pages = {L16-L21},
          doi = {10.1088/2041-8205/724/1/L16},
archivePrefix = {arXiv},
       eprint = {1008.2674},
 primaryClass = {astro-ph.SR},
       adsurl = {https://ui.adsabs.harvard.edu/abs/2010ApJ...724L..16P}
}

@ARTICLE{Andersen2020,
       author = {{CHIME/FRB Collaboration} and {Andersen}, B.~C. and {Bandura}, K.~M. and {Bhardwaj}, M. and {Bij}, A. and {Boyce}, M.~M. and {Boyle}, P.~J. and {Brar}, C. and {Cassanelli}, T. and {Chawla}, P. and {Chen}, T. and {Cliche}, J.-F. and {Cook}, A. and {Cubranic}, D. and {Curtin}, A.~P. and {Denman}, N.~T. and {Dobbs}, M. and {Dong}, F.~Q. and {Fandino}, M. and {Fonseca}, E. and {Gaensler}, B.~M. and {Giri}, U. and {Good}, D.~C. and {Halpern}, M. and {Hill}, A.~S. and {Hinshaw}, G.~F. and {H{\"o}fer}, C. and {Josephy}, A. and {Kania}, J.~W. and {Kaspi}, V.~M. and {Landecker}, T.~L. and {Leung}, C. and {Li}, D.~Z. and {Lin}, H.-H. and {Masui}, K.~W. and {McKinven}, R. and {Mena-Parra}, J. and {Merryfield}, M. and {Meyers}, B.~W. and {Michilli}, D. and {Milutinovic}, N. and {Mirhosseini}, A. and {M{\"u}nchmeyer}, M. and {Naidu}, A. and {Newburgh}, L.~B. and {Ng}, C. and {Patel}, C. and {Pen}, U.-L. and {Pinsonneault-Marotte}, T. and {Pleunis}, Z. and {Quine}, B.~M. and {Rafiei-Ravandi}, M. and {Rahman}, M. and {Ransom}, S.~M. and {Renard}, A. and {Sanghavi}, P. and {Scholz}, P. and {Shaw}, J.~R. and {Shin}, K. and {Siegel}, S.~R. and {Singh}, S. and {Smegal}, R.~J. and {Smith}, K.~M. and {Stairs}, I.~H. and {Tan}, C.~M. and {Tendulkar}, S.~P. and {Tretyakov}, I. and {Vanderlinde}, K. and {Wang}, H. and {Wulf}, D. and {Zwaniga}, A.~V.},
        title = "{A bright millisecond-duration radio burst from a Galactic magnetar}",
      journal = {\nat},
         year = 2020,
        month = nov,
       volume = {587},
       number = {7832},
        pages = {54-58},
          doi = {10.1038/s41586-020-2863-y},
archivePrefix = {arXiv},
       eprint = {2005.10324},
 primaryClass = {astro-ph.HE},
       adsurl = {https://ui.adsabs.harvard.edu/abs/2020Natur.587...54C}
}

@ARTICLE{Bochenek2020,
       author = {{Bochenek}, C.~D. and {Ravi}, V. and {Belov}, K.~V. and {Hallinan}, G. and {Kocz}, J. and {Kulkarni}, S.~R. and {McKenna}, D.~L.},
        title = "{A fast radio burst associated with a Galactic magnetar}",
      journal = {\nat},
         year = 2020,
        month = nov,
       volume = {587},
       number = {7832},
        pages = {59-62},
          doi = {10.1038/s41586-020-2872-x},
archivePrefix = {arXiv},
       eprint = {2005.10828},
 primaryClass = {astro-ph.HE},
       adsurl = {https://ui.adsabs.harvard.edu/abs/2020Natur.587...59B}
}

@ARTICLE{Kasen2010,
       author = {{Kasen}, Daniel and {Bildsten}, Lars},
        title = "{Supernova Light Curves Powered by Young Magnetars}",
      journal = {\apj},
         year = 2010,
        month = jul,
       volume = {717},
       number = {1},
        pages = {245-249},
          doi = {10.1088/0004-637X/717/1/245},
archivePrefix = {arXiv},
       eprint = {0911.0680},
 primaryClass = {astro-ph.HE},
       adsurl = {https://ui.adsabs.harvard.edu/abs/2010ApJ...717..245K}
}

@ARTICLE{Woosley2010,
       author = {{Woosley}, S.~E.},
        title = "{Bright Supernovae from Magnetar Birth}",
      journal = {\apjl},
         year = 2010,
        month = aug,
       volume = {719},
       number = {2},
        pages = {L204-L207},
          doi = {10.1088/2041-8205/719/2/L204},
archivePrefix = {arXiv},
       eprint = {0911.0698},
 primaryClass = {astro-ph.HE},
       adsurl = {https://ui.adsabs.harvard.edu/abs/2010ApJ...719L.204W}
}

@INPROCEEDINGS{Woosley2006,
       author = {{Woosley}, S.~E. and {Heger}, A.},
        title = "{The Supernova Gamma-Ray Burst Connection}",
    booktitle = {Gamma-Ray Bursts in the Swift Era},
         year = 2006,
       editor = {{Holt}, S.~S. and {Gehrels}, N. and {Nousek}, J.~A.},
       series = {American Institute of Physics Conference Series},
       volume = {836},
        month = may,
    publisher = {AIP},
        pages = {398-407},
          doi = {10.1063/1.2207927},
archivePrefix = {arXiv},
       eprint = {astro-ph/0604131},
 primaryClass = {astro-ph},
       adsurl = {https://ui.adsabs.harvard.edu/abs/2006AIPC..836..398W}
}

@ARTICLE{Metzger2011,
       author = {{Metzger}, B.~D. and {Giannios}, D. and {Thompson}, T.~A. and {Bucciantini}, N. and {Quataert}, E.},
        title = "{The protomagnetar model for gamma-ray bursts}",
      journal = {\mnras},
         year = 2011,
        month = may,
       volume = {413},
       number = {3},
        pages = {2031-2056},
          doi = {10.1111/j.1365-2966.2011.18280.x},
archivePrefix = {arXiv},
       eprint = {1012.0001},
 primaryClass = {astro-ph.HE},
       adsurl = {https://ui.adsabs.harvard.edu/abs/2011MNRAS.413.2031M}
}

@ARTICLE{Metzger2014,
       author = {{Metzger}, Brian D. and {Vurm}, Indrek and {Hasco{\"e}t}, Romain and {Beloborodov}, Andrei M.},
        title = "{Ionization break-out from millisecond pulsar wind nebulae: an X-ray probe of the origin of superluminous supernovae}",
      journal = {\mnras},
         year = 2014,
        month = jan,
       volume = {437},
       number = {1},
        pages = {703-720},
          doi = {10.1093/mnras/stt1922},
archivePrefix = {arXiv},
       eprint = {1307.8115},
 primaryClass = {astro-ph.HE},
       adsurl = {https://ui.adsabs.harvard.edu/abs/2014MNRAS.437..703M}
}

@ARTICLE{Omand2018,
       author = {{Omand}, Conor M.~B. and {Kashiyama}, Kazumi and {Murase}, Kohta},
        title = "{Radio emission from embryonic superluminous supernova remnants}",
      journal = {\mnras},
         year = 2018,
        month = feb,
       volume = {474},
       number = {1},
        pages = {573-579},
          doi = {10.1093/mnras/stx2743},
archivePrefix = {arXiv},
       eprint = {1704.00456},
 primaryClass = {astro-ph.HE},
       adsurl = {https://ui.adsabs.harvard.edu/abs/2018MNRAS.474..573O}
}

@ARTICLE{Margutti2018,
       author = {{Margutti}, R. and {Chornock}, R. and {Metzger}, B.~D. and {Coppejans}, D.~L. and {Guidorzi}, C. and {Migliori}, G. and {Milisavljevic}, D. and {Berger}, E. and {Nicholl}, M. and {Zauderer}, B.~A. and {Lunnan}, R. and {Kamble}, A. and {Drout}, M. and {Modjaz}, M.},
        title = "{Results from a Systematic Survey of X-Ray Emission from Hydrogen-poor Superluminous SNe}",
      journal = {\apj},
         year = 2018,
        month = sep,
       volume = {864},
       number = {1},
          eid = {45},
        pages = {45},
          doi = {10.3847/1538-4357/aad2df},
archivePrefix = {arXiv},
       eprint = {1704.05865},
 primaryClass = {astro-ph.HE},
       adsurl = {https://ui.adsabs.harvard.edu/abs/2018ApJ...864...45M}
}

@ARTICLE{Bhirombhakdi2018,
       author = {{Bhirombhakdi}, Kornpob and {Chornock}, Ryan and {Margutti}, Raffaella and {Nicholl}, Matt and {Metzger}, Brian D. and {Berger}, Edo and {Margalit}, Ben and {Milisavljevic}, Dan},
        title = "{Where is the Engine Hiding Its Missing Energy? Constraints from a Deep X-Ray Non-detection of the Superluminous SN 2015bn}",
      journal = {\apjl},
         year = 2018,
        month = dec,
       volume = {868},
       number = {2},
          eid = {L32},
        pages = {L32},
          doi = {10.3847/2041-8213/aaee83},
archivePrefix = {arXiv},
       eprint = {1809.02760},
 primaryClass = {astro-ph.HE},
       adsurl = {https://ui.adsabs.harvard.edu/abs/2018ApJ...868L..32B}
}

@ARTICLE{Andreoni2022,
       author = {{Andreoni}, Igor and {Lu}, Wenbin and {Grefenstette}, Brian and {Kasliwal}, Mansi and {Yan}, Lin and {Hare}, Jeremy},
        title = "{Hard X-Ray Observations of the Hydrogen-poor Superluminous Supernova SN 2018hti with NuSTAR}",
      journal = {\apjl},
         year = 2022,
        month = dec,
       volume = {941},
       number = {1},
          eid = {L16},
        pages = {L16},
          doi = {10.3847/2041-8213/aca593},
archivePrefix = {arXiv},
       eprint = {2211.15749},
 primaryClass = {astro-ph.HE},
       adsurl = {https://ui.adsabs.harvard.edu/abs/2022ApJ...941L..16A}
}

@ARTICLE{Suzuki&Maeda2017,
       author = {{Suzuki}, Akihiro and {Maeda}, Keiichi},
        title = "{Supernova ejecta with a relativistic wind from a central compact object: a unified picture for extraordinary supernovae}",
      journal = {\mnras},
         year = 2017,
        month = apr,
       volume = {466},
       number = {3},
        pages = {2633-2657},
          doi = {10.1093/mnras/stw3259},
archivePrefix = {arXiv},
       eprint = {1612.03911},
 primaryClass = {astro-ph.HE},
       adsurl = {https://ui.adsabs.harvard.edu/abs/2017MNRAS.466.2633S}
}

@ARTICLE{Chen2016,
       author = {{Chen}, Ke-Jung and {Woosley}, S.~E. and {Sukhbold}, Tuguldur},
        title = "{Magnetar-Powered Supernovae in Two Dimensions. I. Superluminous Supernovae}",
      journal = {\apj},
         year = 2016,
        month = nov,
       volume = {832},
       number = {1},
          eid = {73},
        pages = {73},
          doi = {10.3847/0004-637X/832/1/73},
archivePrefix = {arXiv},
       eprint = {1604.07989},
 primaryClass = {astro-ph.HE},
       adsurl = {https://ui.adsabs.harvard.edu/abs/2016ApJ...832...73C}
}

@ARTICLE{Cheng2025,
       author = {{Cheng}, Huaqing and {Zhang}, Chen and {Ling}, Zhixing and {Sun}, Xiaojin and {Sun}, Shengli and {Liu}, Yuan and {Dai}, Yanfeng and {Jia}, Zhenqing and {Pan}, Haiwu and {Wang}, Wenxin and {Zhao}, Donghua and {Chen}, Yifan and {Cheng}, Zhiwei and {Fu}, Wei and {Han}, Yixiao and {Li}, Junfei and {Li}, Zhengda and {Ma}, Xiaohao and {Xue}, Yulong and {Yan}, Ailiang and {Zhang}, Qiang and {Wang}, Yusa and {Yang}, Xiongtao and {Zhao}, Zijian and {Li}, Longhui and {Jin}, Ge and {Yuan}, Weimin},
        title = "{Ground calibration result of the wide-field X-ray telescope (WXT) onboard the Einstein probe}",
      journal = {Experimental Astronomy},
         year = 2025,
        month = oct,
       volume = {60},
       number = {2},
          eid = {15},
        pages = {15},
          doi = {10.1007/s10686-025-10025-9},
archivePrefix = {arXiv},
       eprint = {2505.18939},
 primaryClass = {astro-ph.IM},
       adsurl = {https://ui.adsabs.harvard.edu/abs/2025ExA....60...15C}
}

@ARTICLE{Yuan2015,
       author = {{Yuan}, Weimin and {Zhang}, C. and {Feng}, H. and {Zhang}, S.~N. and {Ling}, Z.~X. and {Zhao}, D. and {Deng}, J. and {Qiu}, Y. and {Osborne}, J.~P. and {O'Brien}, P. and {Willingale}, R. and {Lapington}, J. and {Fraser}, G.~W. and {the Einstein Probe team}},
        title = "{Einstein Probe - a small mission to monitor and explore the dynamic X-ray Universe}",
      journal = {arXiv e-prints},
         year = 2015,
        month = jun,
          eid = {arXiv:1506.07735},
        pages = {arXiv:1506.07735},
          doi = {10.48550/arXiv.1506.07735},
archivePrefix = {arXiv},
       eprint = {1506.07735},
 primaryClass = {astro-ph.HE},
       adsurl = {https://ui.adsabs.harvard.edu/abs/2015arXiv150607735Y}
}

@ARTICLE{Kotera2013,
       author = {{Kotera}, K. and {Phinney}, E.~S. and {Olinto}, A.~V.},
        title = "{Signatures of pulsars in the light curves of newly formed supernova remnants}",
      journal = {\mnras},
         year = 2013,
        month = jul,
       volume = {432},
       number = {4},
        pages = {3228-3236},
          doi = {10.1093/mnras/stt680},
archivePrefix = {arXiv},
       eprint = {1304.5326},
 primaryClass = {astro-ph.HE},
       adsurl = {https://ui.adsabs.harvard.edu/abs/2013MNRAS.432.3228K}
}

@ARTICLE{Eftekhari2021,
       author = {{Eftekhari}, T. and {Margalit}, B. and {Omand}, C.~M.~B. and {Berger}, E. and {Blanchard}, P.~K. and {Demorest}, P. and {Metzger}, B.~D. and {Murase}, K. and {Nicholl}, M. and {Villar}, V.~A. and {Williams}, P.~K.~G. and {Alexander}, K.~D. and {Chatterjee}, S. and {Coppejans}, D.~L. and {Cordes}, J.~M. and {Gomez}, S. and {Hosseinzadeh}, G. and {Hsu}, B. and {Kashiyama}, K. and {Margutti}, R. and {Yin}, Y.},
        title = "{Late-time Radio and Millimeter Observations of Superluminous Supernovae and Long Gamma-Ray Bursts: Implications for Central Engines, Fast Radio Bursts, and Obscured Star Formation}",
      journal = {\apj},
         year = 2021,
        month = may,
       volume = {912},
       number = {1},
          eid = {21},
        pages = {21},
          doi = {10.3847/1538-4357/abe9b8},
archivePrefix = {arXiv},
       eprint = {2010.06612},
 primaryClass = {astro-ph.HE},
       adsurl = {https://ui.adsabs.harvard.edu/abs/2021ApJ...912...21E}
}

@ARTICLE{Ivezic2019,
       author = {{Ivezi{\'c}}, {\v{Z}}eljko and {Kahn}, Steven M. and {Tyson}, J. Anthony and {Abel}, Bob and {Acosta}, Emily and {Allsman}, Robyn and {Alonso}, David and {AlSayyad}, Yusra and {Anderson}, Scott F. and {Andrew}, John and {Angel}, James Roger P. and {Angeli}, George Z. and {Ansari}, Reza and {Antilogus}, Pierre and {Araujo}, Constanza and {Armstrong}, Robert and {Arndt}, Kirk T. and {Astier}, Pierre and {Aubourg}, {\'E}ric and {Auza}, Nicole and {Axelrod}, Tim S. and {Bard}, Deborah J. and {Barr}, Jeff D. and {Barrau}, Aurelian and {Bartlett}, James G. and {Bauer}, Amanda E. and {Bauman}, Brian J. and {Baumont}, Sylvain and {Bechtol}, Ellen and {Bechtol}, Keith and {Becker}, Andrew C. and {Becla}, Jacek and {Beldica}, Cristina and {Bellavia}, Steve and {Bianco}, Federica B. and {Biswas}, Rahul and {Blanc}, Guillaume and {Blazek}, Jonathan and {Blandford}, Roger D. and {Bloom}, Josh S. and {Bogart}, Joanne and {Bond}, Tim W. and {Booth}, Michael T. and {Borgland}, Anders W. and {Borne}, Kirk and {Bosch}, James F. and {Boutigny}, Dominique and {Brackett}, Craig A. and {Bradshaw}, Andrew and {Brandt}, William Nielsen and {Brown}, Michael E. and {Bullock}, James S. and {Burchat}, Patricia and {Burke}, David L. and {Cagnoli}, Gianpietro and {Calabrese}, Daniel and {Callahan}, Shawn and {Callen}, Alice L. and {Carlin}, Jeffrey L. and {Carlson}, Erin L. and {Chandrasekharan}, Srinivasan and {Charles-Emerson}, Glenaver and {Chesley}, Steve and {Cheu}, Elliott C. and {Chiang}, Hsin-Fang and {Chiang}, James and {Chirino}, Carol and {Chow}, Derek and {Ciardi}, David R. and {Claver}, Charles F. and {Cohen-Tanugi}, Johann and {Cockrum}, Joseph J. and {Coles}, Rebecca and {Connolly}, Andrew J. and {Cook}, Kem H. and {Cooray}, Asantha and {Covey}, Kevin R. and {Cribbs}, Chris and {Cui}, Wei and {Cutri}, Roc and {Daly}, Philip N. and {Daniel}, Scott F. and {Daruich}, Felipe and {Daubard}, Guillaume and {Daues}, Greg and {Dawson}, William and {Delgado}, Francisco and {Dellapenna}, Alfred and {de Peyster}, Robert and {de Val-Borro}, Miguel and {Digel}, Seth W. and {Doherty}, Peter and {Dubois}, Richard and {Dubois-Felsmann}, Gregory P. and {Durech}, Josef and {Economou}, Frossie and {Eifler}, Tim and {Eracleous}, Michael and {Emmons}, Benjamin L. and {Fausti Neto}, Angelo and {Ferguson}, Henry and {Figueroa}, Enrique and {Fisher-Levine}, Merlin and {Focke}, Warren and {Foss}, Michael D. and {Frank}, James and {Freemon}, Michael D. and {Gangler}, Emmanuel and {Gawiser}, Eric and {Geary}, John C. and {Gee}, Perry and {Geha}, Marla and {Gessner}, Charles J.~B. and {Gibson}, Robert R. and {Gilmore}, D. Kirk and {Glanzman}, Thomas and {Glick}, William and {Goldina}, Tatiana and {Goldstein}, Daniel A. and {Goodenow}, Iain and {Graham}, Melissa L. and {Gressler}, William J. and {Gris}, Philippe and {Guy}, Leanne P. and {Guyonnet}, Augustin and {Haller}, Gunther and {Harris}, Ron and {Hascall}, Patrick A. and {Haupt}, Justine and {Hernandez}, Fabio and {Herrmann}, Sven and {Hileman}, Edward and {Hoblitt}, Joshua and {Hodgson}, John A. and {Hogan}, Craig and {Howard}, James D. and {Huang}, Dajun and {Huffer}, Michael E. and {Ingraham}, Patrick and {Innes}, Walter R. and {Jacoby}, Suzanne H. and {Jain}, Bhuvnesh and {Jammes}, Fabrice and {Jee}, M. James and {Jenness}, Tim and {Jernigan}, Garrett and {Jevremovi{\'c}}, Darko and {Johns}, Kenneth and {Johnson}, Anthony S. and {Johnson}, Margaret W.~G. and {Jones}, R. Lynne and {Juramy-Gilles}, Claire and {Juri{\'c}}, Mario and {Kalirai}, Jason S. and {Kallivayalil}, Nitya J. and {Kalmbach}, Bryce and {Kantor}, Jeffrey P. and {Karst}, Pierre and {Kasliwal}, Mansi M. and {Kelly}, Heather and {Kessler}, Richard and {Kinnison}, Veronica and {Kirkby}, David and {Knox}, Lloyd and {Kotov}, Ivan V. and {Krabbendam}, Victor L. and {Krughoff}, K. Simon and {Kub{\'a}nek}, Petr and {Kuczewski}, John and {Kulkarni}, Shri and {Ku}, John and {Kurita}, Nadine R. and {Lage}, Craig S. and {Lambert}, Ron and {Lange}, Travis and {Langton}, J. Brian and {Le Guillou}, Laurent and {Levine}, Deborah and {Liang}, Ming and {Lim}, Kian-Tat and {Lintott}, Chris J. and {Long}, Kevin E. and {Lopez}, Margaux and {Lotz}, Paul J. and {Lupton}, Robert H. and {Lust}, Nate B. and {MacArthur}, Lauren A. and {Mahabal}, Ashish and {Mandelbaum}, Rachel and {Markiewicz}, Thomas W. and {Marsh}, Darren S. and {Marshall}, Philip J. and {Marshall}, Stuart and {May}, Morgan and {McKercher}, Robert and {McQueen}, Michelle and {Meyers}, Joshua and {Migliore}, Myriam and {Miller}, Michelle and {Mills}, David J.},
        title = "{LSST: From Science Drivers to Reference Design and Anticipated Data Products}",
      journal = {\apj},
         year = 2019,
        month = mar,
       volume = {873},
       number = {2},
          eid = {111},
        pages = {111},
          doi = {10.3847/1538-4357/ab042c},
archivePrefix = {arXiv},
       eprint = {0805.2366},
 primaryClass = {astro-ph},
       adsurl = {https://ui.adsabs.harvard.edu/abs/2019ApJ...873..111I}
}

@ARTICLE{Bianco2022,
       author = {{Bianco}, Federica B. and {Ivezi{\'c}}, {\v{Z}}eljko and {Jones}, R. Lynne and {Graham}, Melissa L. and {Marshall}, Phil and {Saha}, Abhijit and {Strauss}, Michael A. and {Yoachim}, Peter and {Ribeiro}, Tiago and {Anguita}, Timo and {Bauer}, A.~E. and {Bauer}, Franz E. and {Bellm}, Eric C. and {Blum}, Robert D. and {Brandt}, William N. and {Brough}, Sarah and {Catelan}, M{\'a}rcio and {Clarkson}, William I. and {Connolly}, Andrew J. and {Gawiser}, Eric and {Gizis}, John E. and {Hlo{\v{z}}ek}, Ren{\'e}e and {Kaviraj}, Sugata and {Liu}, Charles T. and {Lochner}, Michelle and {Mahabal}, Ashish A. and {Mandelbaum}, Rachel and {McGehee}, Peregrine and {Neilsen}, Jr., Eric H. and {Olsen}, Knut A.~G. and {Peiris}, Hiranya V. and {Rhodes}, Jason and {Richards}, Gordon T. and {Ridgway}, Stephen and {Schwamb}, Megan E. and {Scolnic}, Dan and {Shemmer}, Ohad and {Slater}, Colin T. and {Slosar}, An{\v{z}}e and {Smartt}, Stephen J. and {Strader}, Jay and {Street}, Rachel and {Trilling}, David E. and {Verma}, Aprajita and {Vivas}, A.~K. and {Wechsler}, Risa H. and {Willman}, Beth},
        title = "{Optimization of the Observing Cadence for the Rubin Observatory Legacy Survey of Space and Time: A Pioneering Process of Community-focused Experimental Design}",
      journal = {\apjs},
         year = 2022,
        month = jan,
       volume = {258},
       number = {1},
          eid = {1},
        pages = {1},
          doi = {10.3847/1538-4365/ac3e72},
archivePrefix = {arXiv},
       eprint = {2108.01683},
 primaryClass = {astro-ph.IM},
       adsurl = {https://ui.adsabs.harvard.edu/abs/2022ApJS..258....1B}
}

@ARTICLE{Gezari2018,
       author = {{Gezari}, Suvi and {van Velzen}, Sjoert and {Hung}, Tiara and {Cenko}, Brad and {Arcavi}, Iair},
        title = "{A Smart and Colorful Cadence for the LSST Wide Fast Deep Survey: Maximizing TDE Science}",
      journal = {arXiv e-prints},
         year = 2018,
        month = dec,
          eid = {arXiv:1812.07036},
        pages = {arXiv:1812.07036},
          doi = {10.48550/arXiv.1812.07036},
archivePrefix = {arXiv},
       eprint = {1812.07036},
 primaryClass = {astro-ph.IM},
       adsurl = {https://ui.adsabs.harvard.edu/abs/2018arXiv181207036G}
}

@ARTICLE{Andreoni2024,
       author = {{Andreoni}, Igor and {Margutti}, Raffaella and {Banovetz}, John and {Greenstreet}, Sarah and {Hebert}, Claire-Alice and {Lister}, Tim and {Palmese}, Antonella and {Piranomonte}, Silvia and {Smartt}, S.~J. and {Smith}, Graham P. and {Stein}, Robert and {Ahumada}, Tomas and {Anand}, Shreya and {Auchettl}, Katie and {Bannister}, Michele T. and {Bellm}, Eric C. and {Bloom}, Joshua S. and {Bolin}, Bryce T. and {Bom}, Clecio R. and {Brethauer}, Daniel and {Brucker}, Melissa J. and {Buckley}, David A.~H. and {Chandra}, Poonam and {Chornock}, Ryan and {Christensen}, Eric and {Cooke}, Jeff and {Corsi}, Alessandra and {Coughlin}, Michael W. and {Cuevas-Otahola}, Bolivia and {Filippo}, D'Ammando and {Dai}, Biwei and {Dhawan}, S. and {Filippenko}, Alexei V. and {Foley}, Ryan J. and {Franckowiak}, Anna and {Gomboc}, Andreja and {Gompertz}, Benjamin P. and {Guy}, Leanne P. and {Hazra}, Nandini and {Hernandez}, Christopher and {Hosseinzadeh}, Griffin and {Hussaini}, Maryam and {Ibrahimzade}, Dina and {Izzo}, Luca and {Jones}, R. Lynne and {Kang}, Yijung and {Kasliwal}, Mansi M. and {Knight}, Matthew and {Kunnumkai}, Keerthi and {Lamb}, Gavin P and {LeBaron}, Natalie and {Lejoly}, Cassandra and {Levan}, Andrew J. and {MacBride}, Sean and {Mallia}, Franco and {Malz}, Alex I. and {Miller}, Adam A. and {Mora}, J.~C. and {Narayan}, Gautham and {Nayana A.}, J. and {Nicholl}, Matt and {Nichols}, Tiffany and {Oates}, S.~R. and {Panayada}, Akshay and {Ragosta}, Fabio and {Ribeiro}, Tiago and {Ryczanowski}, Dan and {Sarin}, Nikhil and {Schwamb}, Megan E. and {Sears}, Huei and {Seligman}, Darryl Z. and {Sharma}, Ritwik and {Shrestha}, Manisha and {Simran} and {Stroh}, Michael C. and {Terreran}, Giacomo and {Linesh Thakur}, Aishwarya and {Trivedi}, Aum and {Tyson}, J. Anthony and {Utsumi}, Yousuke and {Verma}, Aprajita and {Villar}, V. Ashley and {Volk}, Kathryn and {Vyas}, Meet J. and {Wasserman}, Amanda R. and {Wheeler}, J. Craig and {Yoachim}, Peter and {Zegarelli}, Angela and {Bianco}, Federica},
        title = "{Rubin ToO 2024: Envisioning the Vera C. Rubin Observatory LSST Target of Opportunity program}",
      journal = {arXiv e-prints},
         year = 2024,
        month = nov,
          eid = {arXiv:2411.04793},
        pages = {arXiv:2411.04793},
          doi = {10.48550/arXiv.2411.04793},
archivePrefix = {arXiv},
       eprint = {2411.04793},
 primaryClass = {astro-ph.IM},
       adsurl = {https://ui.adsabs.harvard.edu/abs/2024arXiv241104793A}
}

@ARTICLE{Stone&Gardiner2007,
       author = {{Stone}, James M. and {Gardiner}, Thomas},
        title = "{The Magnetic Rayleigh-Taylor Instability in Three Dimensions}",
      journal = {\apj},
         year = 2007,
        month = dec,
       volume = {671},
       number = {2},
        pages = {1726-1735},
          doi = {10.1086/523099},
archivePrefix = {arXiv},
       eprint = {0709.0452},
 primaryClass = {astro-ph},
       adsurl = {https://ui.adsabs.harvard.edu/abs/2007ApJ...671.1726S}
}

@ARTICLE{Bucciantini2008,
       author = {{Bucciantini}, N. and {Quataert}, E. and {Arons}, J. and {Metzger}, B.~D. and {Thompson}, T.~A.},
        title = "{Relativistic jets and long-duration gamma-ray bursts from the birth of magnetars}",
      journal = {\mnras},
         year = 2008,
        month = jan,
       volume = {383},
       number = {1},
        pages = {L25-L29},
          doi = {10.1111/j.1745-3933.2007.00403.x},
archivePrefix = {arXiv},
       eprint = {0707.2100},
 primaryClass = {astro-ph},
       adsurl = {https://ui.adsabs.harvard.edu/abs/2008MNRAS.383L..25B}
}

@ARTICLE{Bucciantini2009,
       author = {{Bucciantini}, N. and {Quataert}, E. and {Metzger}, B.~D. and {Thompson}, T.~A. and {Arons}, J. and {Del Zanna}, L.},
        title = "{Magnetized relativistic jets and long-duration GRBs from magnetar spin-down during core-collapse supernovae}",
      journal = {\mnras},
         year = 2009,
        month = jul,
       volume = {396},
       number = {4},
        pages = {2038-2050},
          doi = {10.1111/j.1365-2966.2009.14940.x},
archivePrefix = {arXiv},
       eprint = {0901.3801},
 primaryClass = {astro-ph.HE},
       adsurl = {https://ui.adsabs.harvard.edu/abs/2009MNRAS.396.2038B}
}

@ARTICLE{Janka2012,
       author = {{Janka}, Hans-Thomas},
        title = "{Explosion Mechanisms of Core-Collapse Supernovae}",
      journal = {Annual Review of Nuclear and Particle Science},
         year = 2012,
        month = nov,
       volume = {62},
       number = {1},
        pages = {407-451},
          doi = {10.1146/annurev-nucl-102711-094901},
archivePrefix = {arXiv},
       eprint = {1206.2503},
 primaryClass = {astro-ph.SR},
       adsurl = {https://ui.adsabs.harvard.edu/abs/2012ARNPS..62..407J}
}

@ARTICLE{Imasheva2023,
       author = {{Imasheva}, Liliya and {Janka}, Hans-Thomas and {Weiss}, Achim},
        title = "{Parametrizations of thermal bomb explosions for core-collapse supernovae and $^{56}$Ni production}",
      journal = {\mnras},
         year = 2023,
        month = jan,
       volume = {518},
       number = {2},
        pages = {1818-1839},
          doi = {10.1093/mnras/stac3239},
archivePrefix = {arXiv},
       eprint = {2209.10989},
 primaryClass = {astro-ph.HE},
       adsurl = {https://ui.adsabs.harvard.edu/abs/2023MNRAS.518.1818I}
}

@ARTICLE{Nyholm2020,
       author = {{Nyholm}, A. and {Sollerman}, J. and {Tartaglia}, L. and {Taddia}, F. and {Fremling}, C. and {Blagorodnova}, N. and {Filippenko}, A.~V. and {Gal-Yam}, A. and {Howell}, D.~A. and {Karamehmetoglu}, E. and {Kulkarni}, S.~R. and {Laher}, R. and {Leloudas}, G. and {Masci}, F. and {Kasliwal}, M.~M. and {Mor{\r{a}}}, K. and {Moriya}, T.~J. and {Ofek}, E.~O. and {Papadogiannakis}, S. and {Quimby}, R. and {Rebbapragada}, U. and {Schulze}, S.},
        title = "{Type IIn supernova light-curve properties measured from an untargeted survey sample}",
      journal = {\aap},
         year = 2020,
        month = may,
       volume = {637},
          eid = {A73},
        pages = {A73},
          doi = {10.1051/0004-6361/201936097},
archivePrefix = {arXiv},
       eprint = {1906.05812},
 primaryClass = {astro-ph.SR},
       adsurl = {https://ui.adsabs.harvard.edu/abs/2020A&A...637A..73N}
}

@ARTICLE{Smith2014,
       author = {{Smith}, Nathan},
        title = "{Mass Loss: Its Effect on the Evolution and Fate of High-Mass Stars}",
      journal = {\araa},
         year = 2014,
        month = aug,
       volume = {52},
        pages = {487-528},
          doi = {10.1146/annurev-astro-081913-040025},
archivePrefix = {arXiv},
       eprint = {1402.1237},
 primaryClass = {astro-ph.SR},
       adsurl = {https://ui.adsabs.harvard.edu/abs/2014ARA&A..52..487S}
}

@ARTICLE{Fraser2020,
       author = {{Fraser}, Morgan},
        title = "{Supernovae and transients with circumstellar interaction}",
      journal = {Royal Society Open Science},
         year = 2020,
        month = jul,
       volume = {7},
       number = {7},
          eid = {200467},
        pages = {200467},
          doi = {10.1098/rsos.200467},
       adsurl = {https://ui.adsabs.harvard.edu/abs/2020RSOS....700467F}
}

@ARTICLE{Hiramatsu2024,
       author = {{Hiramatsu}, Daichi and {Berger}, Edo and {Gomez}, Sebastian and {Blanchard}, Peter K. and {Kumar}, Harsh and {Athukoralalage}, Wasundara},
        title = "{Type IIn Supernovae. I. Uniform Light Curve Characterization and a Bimodality in the Radiated Energy Distribution}",
      journal = {arXiv e-prints},
         year = 2024,
        month = nov,
          eid = {arXiv:2411.07287},
        pages = {arXiv:2411.07287},
          doi = {10.48550/arXiv.2411.07287},
archivePrefix = {arXiv},
       eprint = {2411.07287},
 primaryClass = {astro-ph.HE},
       adsurl = {https://ui.adsabs.harvard.edu/abs/2024arXiv241107287H}
}

@ARTICLE{Seitenzahl2011,
       author = {{Seitenzahl}, I.~R.},
        title = "{Internal conversion electrons and supernova light curves}",
      journal = {Progress in Particle and Nuclear Physics},
         year = 2011,
        month = apr,
       volume = {66},
       number = {2},
        pages = {329-334},
          doi = {10.1016/j.ppnp.2011.01.028},
archivePrefix = {arXiv},
       eprint = {1012.4647},
 primaryClass = {astro-ph.SR},
       adsurl = {https://ui.adsabs.harvard.edu/abs/2011PrPNP..66..329S}
}

@ARTICLE{Milne1999,
       author = {{Milne}, P.~A. and {The}, L.-S. and {Leising}, M.~D.},
        title = "{Positron Escape from Type IA Supernovae}",
      journal = {\apjs},
         year = 1999,
        month = oct,
       volume = {124},
       number = {2},
        pages = {503-526},
          doi = {10.1086/313262},
archivePrefix = {arXiv},
       eprint = {astro-ph/9901206},
 primaryClass = {astro-ph},
       adsurl = {https://ui.adsabs.harvard.edu/abs/1999ApJS..124..503M}
}

@ARTICLE{Milne2001,
       author = {{Milne}, P.~A. and {The}, L.-S. and {Leising}, M.~D.},
        title = "{Late Light Curves of Type Ia Supernovae}",
      journal = {\apj},
         year = 2001,
        month = oct,
       volume = {559},
       number = {2},
        pages = {1019-1031},
          doi = {10.1086/322352},
archivePrefix = {arXiv},
       eprint = {astro-ph/0104185},
 primaryClass = {astro-ph},
       adsurl = {https://ui.adsabs.harvard.edu/abs/2001ApJ...559.1019M}
}
\bibliographystyle{aasjournalv7}



\end{document}